\documentclass[12pt,letterpaper]{article}
\usepackage{amsmath,amssymb,array,calc,rotating,epsfig,psfrag,amscd, cite}

\makeatletter
\renewcommand\section{\@startsection {section}{1}{\z@}%
                                   {-3.5ex \@plus -1ex \@minus -.2ex}
                                   {2.3ex \@plus.2ex}%
                                   {\normalfont\large\bfseries}}
\renewcommand\subsection{\@startsection{subsection}{2}{\z@}%
                                     {-3.25ex\@plus -1ex \@minus -.2ex}%
                                     {1.5ex \@plus .2ex}%
                                     {\normalfont\bfseries}}
\makeatother

\let\non\nonumber

\let\s=\sigma

\newcommand{\bea}{\begin{eqnarray}}
\newcommand{\eea}{\end{eqnarray}}
\newcommand{\be}{\begin{equation}}
\newcommand{\ee}{\end{equation}}

\newcommand{\p}{\partial}

\newcommand{\C}[1]{$(\ref{#1})$}

\def\IZ{\relax\ifmmode\mathchoice
{\hbox{\cmss Z\kern-.4em Z}}{\hbox{\cmss Z\kern-.4em Z}}
{\lower.9pt\hbox{\cmsss Z\kern-.4em Z}} {\lower1.2pt\hbox{\cmsss
Z\kern-.4em Z}}\else{\cmss Z\kern-.4em Z}\fi}
\def\IR{\relax{\rm I\kern-.18em R}}

\def\one{{\hbox{ 1\kern-.8mm l}}}

\newlength{\bredde}
\def\slash#1{\settowidth{\bredde}{$#1$}\ifmmode\,\raisebox{.15ex}{/}
\hspace*{-\bredde} #1\else$\,\raisebox{.15ex}{/}\hspace*{-\bredde}
#1$\fi}

\newsavebox{\zzzbar}
\sbox{\zzzbar}
  {\setlength{\unitlength}{0.9em}
  \begin{picture}(0.6,0.7)
  \thinlines
  \put(0,0){\line(1,0){0.6}}
  \put(0,0.75){\line(1,0){0.575}}
  \multiput(0,0)(0.0125,0.025){30}{\rule{0.3pt}{0.3pt}}
  \multiput(0.2,0)(0.0125,0.025){30}{\rule{0.3pt}{0.3pt}}
  \put(0,0.75){\line(0,-1){0.15}}
  \put(0.015,0.75){\line(0,-1){0.1}}
  \put(0.03,0.75){\line(0,-1){0.075}}
  \put(0.045,0.75){\line(0,-1){0.05}}
  \put(0.05,0.75){\line(0,-1){0.025}}
  \put(0.6,0){\line(0,1){0.15}}
  \put(0.585,0){\line(0,1){0.1}}
  \put(0.57,0){\line(0,1){0.075}}
  \put(0.555,0){\line(0,1){0.05}}
  \put(0.55,0){\line(0,1){0.025}}
  \end{picture}}

\newcommand{\ena}{\end{eqnarray}}
\newcommand{\beqa}{\begin{eqnarray}}
\newcommand{\eeqa}{\end{eqnarray}}

\def\s{\sigma}

\begin{document}
\begin{titlepage}

\begin{center}



\vskip 2 cm
{\Large \bf The $D^6 \mathcal{R}^4$ term from three loop maximal supergravity}\\
\vskip 1.25 cm { Anirban Basu\footnote{email address:
    anirbanbasu@hri.res.in} } \\
{\vskip 0.5cm Harish--Chandra Research Institute, Chhatnag Road, Jhusi,\\
Allahabad 211019, India\\}

\end{center}

\vskip 2 cm

\begin{abstract}
\baselineskip=18pt

We consider the $D^6\mathcal{R}^4$ term which is the leading contribution in the low momentum expansion of the three loop, four graviton amplitude in maximal supergravity. We calculate the moduli dependent coefficient of this term in 11 dimensional supergravity compactified on $T^2$. Only diagrams that involve the Mercedes skeleton contribute resulting in a compact expression involving an integral over 6 Schwinger parameters. We express this integral as an integral over the moduli of an auxiliary $T^3$. This includes an $SL(3,\mathbb{Z})$ invariant integral over the shape moduli of the $T^3$. We discuss the renormalization of the ultraviolet divergences of this amplitude that arise from the boundaries of moduli space. The renormalized amplitude is simple which is a consequence of the fact that the $D^6\mathcal{R}^4$ term is BPS.

\end{abstract}

\end{titlepage}


\section{Introduction}

Understanding the low energy effective action of string theory is important in order to analyze the various perturbative and non--perturbative aspects of the theory, in particular the duality symmetries. While this has not been done in generic cases, some of the leading terms in the $l_s^2$ expansion of the effective action have been calculated in maximally supersymmetric compactifications. These have been calculated using both spacetime and worldsheet techniques which complement each other. The moduli dependent coefficients of the various interactions are highly constrained by the U--duality symmetries of the theory, which has allowed them to be highly constrained or calculated exactly for a class of BPS interactions~\cite{Green:1997tv,Green:1997as,Kiritsis:1997em,Green:1998by,Obers:1998fb,Green:1999pu,Green:2005ba,Basu:2007ru,Basu:2007ck,Green:2010kv,Green:2010wi,Gubay:2014lwa}. When expanded at weak coupling they yield perturbative contributions which match worldsheet calculations, as well as non--perturbative contributions involving various instantons. This has provided detailed quantitative understanding of the U--duality symmetry of the theory.  

Among the class of interactions that have been studied in maximally supersymmetric compactifications, the $1/2$ BPS $\mathcal{R}^4$ interaction has been understood in great detail. The $D^4 \mathcal{R}^4$ and $D^6\mathcal{R}^4$ interactions, which are $1/4$ and $1/8$ BPS interactions respectively, have also been analyzed. While these interactions have been analyzed using various methods, we shall focus on the role played by 11 dimensional ${\mathcal{N}}=1$ supergravity in determining them. This has largely been possible because multi--loop amplitudes in maximal supergravity have been calculated~\cite{Green:1982sw,Green:1997as,Bern:1998ug,Green:1999pu,Green:2008bf,Bern:2007hh,Bern:2008pv,Bern:2009kd,Bern:2010ue}.

Eleven dimensional supergravity has played a key role in determining the $\mathcal{R}^4, D^4\mathcal{R}^4$ and $D^6\mathcal{R}^4$ interactions. They are obtained from the low momentum expansion of the four graviton amplitude in toroidal compactifications of maximal supergravity. The $\mathcal{R}^4$ interaction receives contributions only from 1 loop supergravity, while the $D^4\mathcal{R}^4$ interaction does not get contributions beyond two loops. Hence if the internal torus $T^d$ has $d \leq 2$, such that there are no non--perturbative contributions from wrapped membrane or 5--brane instantons which are not captured by supergravity, one gets the exact answer. Needless to say, supergravity by itself is ultraviolet divergent and one gets divergent answers, and they have to be regularized consistent with the symmetries of string theory. Thus this leads to finite expressions for these couplings in M theory as well as in string theory. In string theory, they give exact moduli dependent amplitudes in the type II theories in 10 and 9 dimensions. In toroidal compactifications to lower dimensions, even though the supergravity calculations do not give the complete answer, they yield perturbative results which are interesting in their own right. The one and two loop supergravity amplitudes have also been expanded to higher orders in the momentum expansion leading to information about non--BPS interactions.          

Our aim is to determine the structure of the $D^6\mathcal{R}^4$ interaction from regularized supergravity. This interaction does not receive contributions beyond three loops in supergravity. The one and two loop contributions have been obtained before, and together with the three loop contribution this provides the complete answer from supergravity. Though our calculation can be generalized to compactifications on higher dimensional torii, we shall consider compactifying on $T^2$ which yields the exact answer. Also for the class of diagrams we analyze, the method can be applied to interactions at higher orders in the momentum expansion, which are the non--BPS interactions.       

We begin with a brief discussion relating M theory on $T^2$ to the type II theories, and the regime of validity of the supergravity calculations. We next summarize the one and two loop supergravity calculations of the $\mathcal{R}^4, D^4 \mathcal{R}^4$ and $D^6\mathcal{R}^4$ amplitudes. Apart from setting up the notations, we also finish some calculations in literature, as these details are crucial for the three loop analysis. In the next section, we consider the three loop four graviton supergravity amplitude, which yields the $D^6 \mathcal{R}^4$ interaction at leading order in the low momentum expansion. This interaction receives contributions only from loop diagrams that arise from the Mercedes skeleton diagram, and not from the ladder skeleton diagram. We next discuss the symmetries of the Mercedes skeleton and the dual regular tetrahedron which is crucial in expressing the integrals in the amplitude in a manner invariant under the reparametrization symmetry of the various internal loop momenta. The relevant symmetry group is the permutation group $S_4$ and we express the integrals in an $S_4$ invariant way. The final answer involves an integral over the 6 Schwinger parameters associated with the 6 links in the Mercedes skeleton or the 6 edges of the dual regular tetrahedron. We next parametrize these 6 Schwinger parameters in terms of the 6 moduli of an auxiliary $T^3$, and express the amplitude as an integral over this moduli space. The final answer turns out to be extremely simple: the integral involves integrating over the volume modulus of the $T^3$ with a  precise power of the volume in the measure, as well as an $SL(3,\mathbb{Z})$ invariant integral with the appropriate measure over the 5 shape moduli of the $T^3$ which parametrizes the coset space $SO(3) \backslash SL(3,\mathbb{R})$. The integrand is simply the $SL(3,\mathbb{Z})$ invariant lattice factor which depends on the volume and shape moduli  of the $T^3$ and includes the infinite sum over the winding momenta running in the loops. These winding momenta are obtained from the KK momenta after performing Poisson resummation. This amplitude, however, has the divergences of supergravity which have to be regularized. To do so, we first isolate the one and two loop sub--divergences and construct moduli dependent counterterms to cancel them. Based on the structure of these counterterms, we argue how these divergences arise from the boundary of moduli space in the three loop amplitude. Finally, these divergences as well as the three loop primitive divergence are regularized using string theory leading to finite answers which give the three loop amplitude for quantum supergravity. Putting together the various regularized contributions from one, two and three loops, we get that the $D^6\mathcal{R}^4$ interaction leads to a term in the 9 dimensional effective action of the form
\be \label{eqn1}l_{11}^5 \int d^9 x \sqrt{-G^{(9)}} \mathcal{V}_2 D^6 \mathcal{R}^4 F(\Omega,\bar\Omega)\ee          
where
\be F(\Omega,\bar\Omega) = \frac{4}{21}\zeta (2) E_{5/2} (\Omega,\bar\Omega) \mathcal{V}_2^{3/2} + \mathcal{E} (\Omega,\bar\Omega) \mathcal{V}_2^{-3}+ 4\zeta (2) E_{3/2} (\Omega,\bar\Omega)\mathcal{V}_2^{-3/2} + 24\zeta (4) + \frac{8}{21}\zeta (2) \zeta (5) \mathcal{V}_2^{-6}.\ee
The $SL(2,\mathbb{Z})$ invariant modular form $\mathcal{E}$ satisfies 
\be 4\Omega_2^2 \frac{\p^2 \mathcal{E}}{\p\Omega \p\bar\Omega} = 12 \mathcal{E} - 6 E_{3/2}^2.\ee 
In \C{eqn1}, $\mathcal{V}_2$ is the dimensionless volume of the $T^2$ measured with the M theory metric, and $\Omega$ is its complex structure. Also $E_s (\Omega,\bar\Omega)$ is the non--holomorphic Eisenstein series of $SL(2,\mathbb{Z})$ defined by
\be E_s (\Omega,\bar\Omega) = \sum_{(m,n) \neq (0,0)} \frac{\Omega_2^s}{\vert m+n\Omega\vert^{2s}}.\ee
We have dropped an overall irrelevant numerical factor in \C{eqn1}. Finally, we very briefly discuss the issue of calculating the three loop amplitude for non--BPS interactions, where diagrams arising from both the Mercedes skeleton and the ladder skeleton contribute, and thus the structure of the amplitude is qualitatively different.   

\section{Relating M theory on $T^2$ to the type IIA and type IIB theories}

We shall need the various relations expressing quantities in M theory compactified on $T^2$ in terms of the moduli of the type IIA and type IIB theories~\cite{Hull:1994ys,Witten:1995ex,Aspinwall:1995fw,Schwarz:1995jq}, which we briefly review. The 11 dimensional Planck length is related to the string length by the relation $l_{11} = e^{\phi^A/3} l_s$, where $\phi^A$ ($\phi^B$) is the type IIA (IIB) dilaton.  

Keeping only the scalars and the graviton obtained from the $T^2$ compactification of M theory where $R_{11}$ and $R_{10}$ are the dimensionless radii (in units of $l_{11}$) of the two circles and dropping the 1 form gauge potentials for simplicity, the line element in M theory is given by
\be \label{metric} ds^2 = G_{MN} dx^M dx^N = G^{(9)}_{\mu\nu} dx^\mu dx^\nu + R_{11}^2 l_{11}^2(dx^{11} - C dx^{10})^2 + R_{10}^2 l_{11}^2dx_{10}^2,\ee
where $x^{11}$ and $x^{10}$ are dimensionless angular coordinates.
The 9 dimensional metric $G^{(9)}_{\mu\nu} = g_{\mu\nu}^{A/B}$ where $g_{\mu\nu}^A$ ($g_{\mu\nu}^B$) is the type IIA (IIB) metric in the string frame. The dimensionless volume $\mathcal{V}_2$ (in units of $4\pi^2 l_{11}^2$) and the complex structure $\Omega$ of the $T^2$ are related to the type IIA and type IIB moduli by the relations
\be \label{rel1}\mathcal{V}_2 = R_{10} R_{11} = e^{\phi^B/3} r_B^{-4/3} = e^{\phi_A/3} r_A, \quad \Omega_1 = C , \quad \Omega_2 = \frac{R_{10}}{R_{11}} = e^{-\phi^B} = r_A e^{-\phi^A}, \ee 
while the T duality relation is 
\be \label{rel2} r_B = r_A^{-1} = \frac{1}{R_{10} \sqrt{R_{11}}}.\ee
Here $r_A$ ($r_B$) is the dimensionless radius of the tenth dimension (in units of $l_s$) in the type IIA (IIB) string frame metric, and 
\be C = C_0 = C_1,\ee 
where $C_0$ ($C_1$) is the type IIB (IIA) 0 (1) form potential. This enables us to express the interactions in M theory on $T^2$ in terms of type IIA or type IIB interactions on $S^1$, which are related by T duality.

From \C{rel1}, it follows that the 10 dimensional type IIB theory is obtained by taking $\mathcal{V}_2 \rightarrow 0$ keeping $\Omega$ fixed and setting $\Omega= \tau$, where $\tau$ is the complex coupling of the IIB theory. The 10 dimensional type IIA theory is obtained by keeping $\phi_A$ fixed, taking $\mathcal{V}_2 \rightarrow \infty$ and recovering the 1 form from $C$.  

Let us now analyze the regime of validity of our supergravity calculations. We are calculating the four graviton amplitude in a fixed classical background. This is valid only when the volume of the various internal dimensions are much larger than the Planck length. Thus our calculation is valid for $R_{10} >> 1$ and $R_{11} >> 1$. From \C{rel1} and \C{rel2} in the type IIA theory, we get that
\be R_{10} = e^{-\phi_A/3} r_A, \quad R_{11} = e^{2\phi_A/3}.\ee
Thus our calculation is valid when $e^{2\phi_A}$ is large and $r_A$ is large. This yields results in the strongly coupled IIA theory, and on sending $r_A \rightarrow \infty$ produces the 10 dimensional answer. In the IIB theory, we get that
\be R_{10} = e^{-\phi_B/3} r_B^{-2/3}, \quad R_{11} = e^{2\phi_B/3} r_B^{-2/3},\ee
thus our calculation is valid when $r_B \rightarrow 0$ keeping $e^{2\phi_B}$ arbitrary but fixed.

\section{The four graviton supergravity amplitude at one and two loops}

We review various details of the four graviton amplitude in $N=1$ supergravity in 11 dimensions compactified on $\mathbb{R}^{8,1} \times T^2$ at one and two loops~\cite{Green:1982sw,Green:1997as,Russo:1997mk,Green:1997ud,Bern:1998ug,Green:1999pu,Green:2008bf}. Apart from setting up the notations and conventions  we shall use throughout, we also finish some incomplete calculations in the literature, as they shall be needed for our three loop analysis. The external on--shell gravitons have momenta and polarizations only in the uncompactified directions transverse to the $T^2$ (this is also true for our three loop analysis).

\subsection{The one loop four graviton amplitude}

To start with, the one loop four graviton amplitude in 11 uncompactified dimensions is given by
\be \label{1loop} \mathcal{A}_4^{(1)} = \frac{4\pi^2 \kappa_{11}^4}{(2\pi)^{11}}\Big[ I(S,T) + I(S,U) + I(T,U) \Big] \mathcal{K}. \ee

\begin{figure}[ht]
\begin{center}
\[
\mbox{\begin{picture}(150,105)(0,0)
\includegraphics[scale=.8]{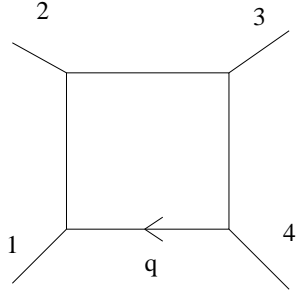}
\end{picture}}
\]
\caption{The one loop diagram $I(S,T)$ }
\end{center}
\end{figure}

In \C{1loop}, $\mathcal{K}$ is the linearized approximation to $\mathcal{R}^4$ in momentum space, and $2\kappa_{11}^2 = (2\pi)^8 l_{11}^9$. Also $S, T$ and $U$ are the Mandelstam variables defined by $S = - G^{MN} (k_1+ k_2)_M (k_1 + k_2)_N, T = -G^{MN} (k_1+ k_4)_M(k_1 + k_4)_N$ and $U = - G^{MN} (k_1 + k_3)_M (k_1 + k_3)_N$, where $G_{MN}$ is the M theory metric, and the external momenta are labelled by $k_{iM} (i=1,\ldots,4)$ and satisfy $\sum_i k_{iM} =0$ and $k_i^2 =0$. The box function $I(S,T)$ depicted in figure 1, is symmetric in $S$ and $T$ and is given by the one loop $\varphi^3$ massless four point integral
\be \label{1exp} I(S,T)= \int d^{11} q \frac{1}{q^2 (q+ k_1)^2(q+k_1+k_2)^2 (q-k_4)^2}\ee  
where the external momenta are all directed inwards in the loop diagram,
and $q_M$ is the loop momentum. We now evaluate \C{1exp} compactified on $\mathbb{R}^{8,1} \times T^2$. The dimensionless volume of $T^2$ (in units of $4\pi^2 l_{11}^2$) is $\mathcal{V}_2$ in the M theory metric, while its complex structure is $\Omega$. The KK momenta $q_I$ in the $T^2$ directions are given by $l_I/l_{11}$ for two integers $l_I$.

Finally, to express \C{1exp} compactified on $\mathbb{R}^{8,1} \times T^2$ in a form convenient for us, we introduce four Schwinger parameters for the four propagators in \C{1exp}, perform the loop integral for the 9 dimensional momentum $q_\mu$ and make a change of variables to put $I(S,T)$ in the form
\be \label{def1} 4\pi^2 I(S,T) = \frac{\pi^{9/2}}{l_{11}^2 \mathcal{V}_2} \int_0^\infty d\s \s^{-3/2} \int_0^1 d\omega_3\int_0^{\omega_3} d\omega_2\int_0^{\omega_2} d \omega_1 \sum_{l_I} e^{-\s G^{IJ} l_I l_J /l_{11}^2 - \s Q(S,T;\omega_i)} \ee
where
\be Q(S,T;\omega_i) = -S \omega_1 (\omega_3 - \omega_2) -T (\omega_2 -\omega_1)(1-\omega_3).\ee
In \C{def1}, the metric of $T^2$ is given by
\be \label{defmet}
G_{IJ} = \frac{\mathcal{V}_2}{\Omega_2} \left( \begin{array}{cc} \vert \Omega \vert^2 & -\Omega_1 \\ -\Omega_1 & 1 \end{array} \right),\ee
and thus
\be G^{IJ} l_I l_J = \frac{1}{\mathcal{V}_2\Omega_2} \vert l_1 + l_2 \Omega\vert^2.\ee

First let us consider the $\mathcal{R}^4$ interaction in \C{1loop} that results from \C{def1}. This is obtained by setting the Mandelstam variables to zero in the box integrals leading to
\be \label{easy1}4\pi^2 I(0,0) = \frac{\pi^{11/2}}{6} \sum_{\hat{l}_I} \int_0^\infty d\hat\s \sqrt{\hat\s} e^{-\pi^2 l_{11}^2 \hat\s G_{IJ} \hat{l}_I \hat{l}_J},\ee 
where we have Poisson resummed to go from a sum over momentum modes to go to a sum over winding modes, and substituted $\hat\s = \s^{-1}$. We get that
\be \label{psum} 3  I(0,0) = \frac{1}{4\pi^2} \cdot \frac{\pi^3}{4 l_{11}^3} \Big[ \frac{4\pi^{5/2}}{3} (\Lambda l_{11})^3 + \mathcal{V}^{-3/2}_2 E_{3/2} (\Omega,\bar\Omega)\Big],\ee
where we have cutoff the UV divergence at $\s = \hat\s^{-1} = \Lambda^{-2}$.

To cancel the UV divergence, we add a one loop local counterterm depicted in figure 2 to the action given by
\be \delta \mathcal{A}_4^{(1)} = \frac{\kappa_{11}^4}{(2\pi)^{11}} \mathcal{K} \cdot  \frac{\pi^{3}}{4 l_{11}^3} c_1.\ee
 
\begin{figure}[ht]
\begin{center}
\[
\mbox{\begin{picture}(50,60)(0,0)
\includegraphics[scale=1]{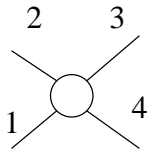}
\end{picture}}
\]
\caption{The one loop counterterm }
\end{center}
\end{figure}

Adding the two contributions gives a local contribution to the effective action of M theory on $\mathbb{R}^{8,1} \times T^2$ given by
\be \label{M9}l_{11}^{-1} \int d^9 x \sqrt{-G^{(9)}} \mathcal{V}_2 \mathcal{R}^4 \Big[ \frac{4\pi^{5/2}}{3} (\Lambda l_{11})^3 + c_1 + \mathcal{V}^{-3/2}_2 E_{3/2} (\Omega,\bar\Omega)\Big] ,\ee
where $G^{(9)}_{\mu\nu}$ is the 9 dimensional M theory metric \C{metric} and we have neglected an overall irrelevant numerical factor in the normalization of the action. Expanding
\be E_{3/2} (\Omega, \bar\Omega) = 2\zeta (3) \Omega_2^{3/2} + 4\zeta (2) \Omega_2^{-1/2} + O(e^{-\Omega_2}),\ee
for large $\Omega_2$, and dropping the exponentially suppressed terms we get that the expression in \C{M9} is given in the type IIA and type IIB theories by
\bea \label{AB}&&l_s^{-1} \int d^9 x \sqrt{-g^{A}} r_A \mathcal{R}^4 \Big[ 2\zeta (3) e^{-2\phi^A} + \frac{2\pi^2}{3 r_A^2} + \frac{4\pi^{5/2}}{3} (\Lambda l_{11})^3 + c_1+\ldots\Big]\non \\ &&= l_s^{-1} \int d^9 x \sqrt{-g^{B}} r_B \mathcal{R}^4 \Big[  2\zeta (3) e^{-2\phi^B} + \frac{2\pi^2}{3} +\frac{4\pi^{5/2}(\Lambda l_{11})^3/3 + c_1}{r_B^2}+\ldots\Big]\eea
on using \C{rel1} and \C{rel2}.
Thus using the genus one equality of the four graviton IIA and IIB amplitudes we get that\footnote{The $\mathcal{R}^4$ term does not receive any contribution beyond one loop in supergravity, so there is no issue of any additional corrections.}
\be \label{valc1}\frac{4\pi^{5/2}}{3} (\Lambda l_{11})^3 + c_1 = \frac{2\pi^2}{3}.\ee
Hence use of string duality removes the UV divergence unambiguously, and leads to a finite term in the M theory effective action. 

The remaining terms from \C{def1}, and $I(S,U)$ and $I(T,U)$ for non-vanishing Mandelstam variables give local as well as non--local contributions in 9 dimensions. The local contributions are given by 
\bea \label{lnl}4\pi^2 \Big[I(S,T)+ I(S,U)+I(T,U)\Big] =  \pi^{9/2} \sum_{n=2}^\infty \frac{\mathcal{W}^n}{n!} (l_{11}^2 \mathcal{V}_2)^{n-3/2} \Gamma(n-1/2) E_{n-1/2} (\Omega, \bar\Omega)\eea
where
\be \mathcal{W}^n = {G}_{ST}^n +{G}_{SU}^n + {G}_{TU}^n,\ee
where
\be {G}_{ST}^n = \int_0^1 d\omega_3 \int_0^{\omega_3} d\omega_2 \int_0^{\omega_2} d\omega_1 \Big(-Q(S,T;\omega_i)\Big)^n.\ee

Thus, upto exponentially suppressed terms, the renormalized one loop amplitude is given by
\bea \label{1loopfin}
&&\mathcal{A}_4^{(1)}= (2\pi^8 l_{11}^{15} r_B) \mathcal{K} r_B \Big[ 2\zeta (3) e^{-2\phi^B} + \frac{2\pi^2}{3} ( 1+r_B^{-2})+ \frac{2\pi^2 l_s^4}{6!r_B^4} \s_2 \Big( \zeta (3) + 2 \zeta (2) e^{2\phi^B} \Big) \non \\&&+ \frac{l_s^6}{2\cdot 4! r_B^6}\s_3\Big( \frac{1}{21} \zeta (2)\zeta (5)+ \frac{1}{9} \zeta (6) e^{4\phi^B} \Big)+ O(k^8)\Big]\non \\ &&= (2\pi^8 l_{11}^{15} r_A^{-1}) \mathcal{K} r_A \Big[ 2\zeta (3) e^{-2\phi^A} + \frac{2\pi^2}{3} (1+r_A^{-2}) +\frac{2\pi^2 l_s^4}{6!} \s_2\Big(  \zeta (3)r_A^2  + 2 \zeta (2) e^{2\phi^A}\Big) \non \\ &&+ \frac{l_s^6}{2\cdot 4!}\s_3\Big( \frac{1}{21} \zeta (2)\zeta (5)r_A^4 + \frac{1}{9} \zeta (6) e^{4\phi^A}   \Big)+O(k^8) \Big],\eea
where
\be \s_n \equiv S^n + T^n + U^n.\ee
The overall factor of $2\pi^8 l_{11}^{15} r_B = 2\pi^8 l_{11}^{15} r_A^{-1}$ is needed to correctly normalize the action and is common to multiloop amplitudes.

\subsection{The two loop four graviton amplitude}

\begin{figure}[ht]
\begin{center}
\[
\mbox{\begin{picture}(310,80)(0,0)
\includegraphics[scale=.65]{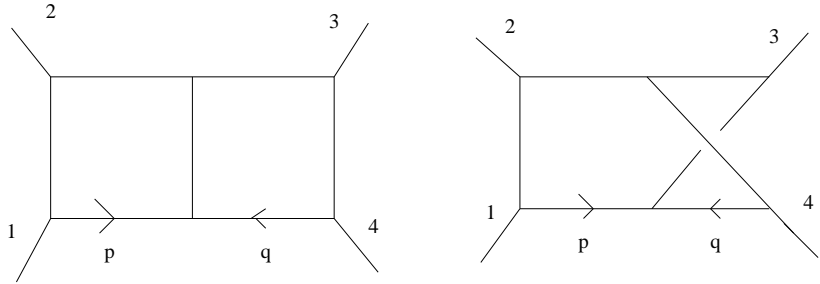}
\end{picture}}
\]
\caption{The two loop diagrams $I^P (S,T)$ and $I^{NP} (S,T)$}
\end{center}
\end{figure}

The two loop four graviton amplitude in 11 uncompactified dimensions is given by
\bea \label{2loop}\mathcal{A}_4^{(2)} = \frac{(4\pi^2)^2 \kappa_{11}^6}{(2\pi)^{22}} \Big[ S^2 \Big( I_P (S,T) + I_P (S,U) +I_{NP} (S,T) + I_{NP} (S,U)\Big)\non \\ + T^2 \Big( I_P (T,S) + I_P (T,U) + I_{NP} (T,S) + I_{NP} (T,U)\Big)\non \\ + U^2 \Big( I_P (U,S) + I_P (U,T) + I_{NP} (U,S) + I_{NP} (U,T)\Big)\Big] \mathcal{K}, \eea 
where the planar contribution $I_P (S,T)$ depicted in figure 3 is given by
\be \label{2planar}I_P (S,T) = \int d^{11} p \int d^{11}q \frac{1}{p^2 (p-k_1)^2 (p-k_1 - k_2)^2 (p+q)^2 q^2 (q-k_4)^2 (q-k_3 - k_4)^2}\ee
while the non--planar contribution $I_{NP} (S,T)$ also depicted in figure 3 is given by
\be \label{2nplanar}I_{NP} (S,T) = \int d^{11} p \int d^{11}q \frac{1}{p^2 (p-k_1)^2 (p-k_1 - k_2)^2 (p+q)^2 q^2 (q-k_4)^2 (p+q+k_3)^2}\ee
where the momenta are directed inwards in the loop diagrams and $p_M$ and $q_M$ are the loop momenta. The remaining terms in \C{2loop} are obtained by simply interchanging the Mandelstam variables. Note that the various contributions are given by massless $\varphi^3$ field theory Feynman diagrams. 

The planar integral \C{2planar} can be simplified in the $T^2$ compactification  by introducing Schwinger parameters and performing the loop integrals to give
\bea \label{2PLANAR}&&(4\pi^2)^2 I_P (S,T) = \frac{\pi^9}{(l_{11}^2 \mathcal{V}_2)^2} \sum_{m_I,n_I}\int_0^\infty d\lambda d\s d\rho \frac{\lambda^2 \s^2}{\Delta^{9/2}} e^{-G^{IJ} \Big(\s m_I m_J + \lambda n_I n_J +\rho(m+n)_I(m+n)_J\Big)/l_{11}^2} \non \\  &&\int_0^1 d\omega_2 \int_0^{\omega_2} d\omega_1 \int_0^1 dv_2\int_0^{v_2} d v_1 e^{\rho\s\lambda (v_2 - v_1)(\omega_2 - \omega_1)T/\Delta + \Big(\rho\s\lambda(v_1 - \omega_1)(v_2 -\omega_2)/\Delta + \s v_1 (1-v_2) + \lambda \omega_1 (1-\omega_2)\Big)S},\non \\ \eea
while the non--planar integral \C{2nplanar} simplifies and becomes
\bea \label{2NPLANAR}&&(4\pi^2)^2 I_{NP} (S,T) = \frac{\pi^9}{(l_{11}^2\mathcal{V}_2)^2}  \sum_{m_I,n_I} \int_0^\infty d\lambda d\s d\rho \frac{\lambda^2 \s \rho}{\Delta^{9/2}} e^{-G^{IJ} \Big(\s m_I m_J + \lambda n_I n_J +\rho(m+n)_I(m+n)_J\Big)/l_{11}^2} \non \\  &&\int_0^1 du_1 d v_1 d\omega_2 \int_0^{\omega_2} d\omega_1 e^{\s\rho\lambda (\omega_2 -\omega_1) (u_1 - v_1)T/\Delta +\Big( (\s+\rho)\lambda^2 \omega_1 (1-\omega_2) +\s\lambda\rho (\omega_1 (1-u_1) + v_1 (u_1 - \omega_2))\Big)S/\Delta} .  \non \\ \eea
In \C{2PLANAR} and \C{2NPLANAR}, $\Delta$ is given by
\be \label{Delta} \Delta = \s\rho +\s\lambda +\rho\lambda,\ee 
while $m_I$ ($I=1,2$) and $n_I$ ($I=1,2$) are integers representing the KK momenta along $T^2$ in the two supergravity loops. 

First consider the $D^4 \mathcal{R}^4$ interaction which involves
\bea (4\pi^2)^2 [ I_P(0,0) + I_{NP} (0,0)]=\frac{\pi^{11}}{12} \sum_{\hat{m}_I ,\hat{n}_I} \int_0^\infty d\hat\lambda d\hat\s d\hat\rho \hat\Delta^{1/2} e^{-\pi^2 l_{11}^2 G_{IJ} \Big( \hat\lambda \hat{m}_I \hat{m}_J + \hat\s \hat{n}_I \hat{n}_J +\hat\rho{(\hat{m}+\hat{n})}_I{(\hat{m}+\hat{n})}_J\Big)},\non \\ \eea
where we have Poisson resummed to go from momentum modes to winding modes and defined
\be \hat\rho = \frac{\rho}{\Delta}, \quad \hat\s = \frac{\s}{\Delta}, \quad \hat\lambda = \frac{\lambda}{\Delta},\ee
and
\be \hat\Delta = \hat\s\hat\rho +\hat\s\hat\lambda + \hat\rho\hat\lambda= \Delta^{-1}.\ee  
Further defining
\be \tau_1 = \frac{\hat\rho}{\hat\rho +\hat\lambda}, \quad \tau_2 = \frac{\sqrt{\hat\Delta}}{\hat\rho+\hat\lambda}, \quad V_2 = l_{11}^2 \sqrt{\hat\Delta},\ee
we get that
\be (4\pi^2)^2 [I_P(0,0) + I_{NP} (0,0)] = \frac{\pi^{11}}{2 l_{11}^8} \sum_{\hat{m}_I ,\hat{n}_I} \int_0^\infty dV_2 V_2^3 \int_{\mathcal{F}_2} \frac{d^2 \tau}{\tau_2^2} e^{-\pi^2 G_{IJ} (\hat{m}+\hat{n}\tau)_I (\hat{m}+\hat{n}\bar\tau)_J V_2/\tau_2}, \ee
where $d^2 \tau= d\tau_1 d\tau_2$ and $\mathcal{F}_2$ is the fundamental domain of $SL(2,\mathbb{Z})$ defined by
\be \mathcal{F}_2 = \{ -\frac{1}{2} \leq \tau_1 \leq \frac{1}{2}, \tau_2 \geq 0, \vert \tau \vert^2 \geq 1\}.\ee 
Thus the amplitude boils down to an integral over the moduli space of an auxiliary $T^2$ parametrized by volume $V_2$ and complex structure $\tau$, where the integral of the complex structure $\tau$ is over $\mathcal{F}_2$. The integrand is an $SL(2,\mathbb{Z})$ invariant lattice factor. This structure will provide a crucial hint for our three loop calculation, as we shall see later.

Performing the integral, we get that
\be \label{ren1} (4\pi^2)^2 [I_P(0,0) + I_{NP} (0,0)] = a \Lambda^8 + \frac{\pi^{13/2} \Lambda^3}{8 l_{11}^5 \mathcal{V}_2^{5/2}} E_{5/2} (\Omega, \bar\Omega) +\frac{\pi^4 \zeta (3)\zeta (4)}{2l_{11}^8 \mathcal{V}_2^4}   ,\ee
where $a$ is an arbitrary constant\footnote{We leave some constants undetermined as they are not going to be relevant for our purposes.}. The various ultraviolet divergent contributions come from the boundaries of moduli space as explained in detail in~\cite{Green:1999pu}, which is used in calculating them.   

The ultraviolet divergences in \C{ren1} have to be renormalized. First consider the one loop counterterms. These are obtained in the standard way by replacing each single loop in figure 3 by a local $\mathcal{R}^4$ counterterm vertex, where the structure and factors of the counterterm amplitude are determined by \C{2loop}. Note that each planar diagram receives two contributions of the type given in figure 4 coming from either loop. However every non--planar diagram receives only one such contribution coming from one of the loops, as the other contribution involves the $\mathcal{R}^5$ counterterm vertex diverging linearly in $\Lambda$ which leaves no finite remainder after regularizing using an off--shell $\mathcal{R}^5$ counterterm vertex. This is because the $\mathcal{R}^5$ term vanishes on--shell as it is in the same supermultiplet as the $D^2 \mathcal{R}^4$ interaction, which we drop from now onwards (see figure 5). Every loop diagram is of the form given in figure 4, which has to be expanded to the required order in the momentum expansion, and all the contributions have to be added.  

\begin{figure}[ht]
\begin{center}
\[
\mbox{\begin{picture}(140,80)(0,0)
\includegraphics[scale=.65]{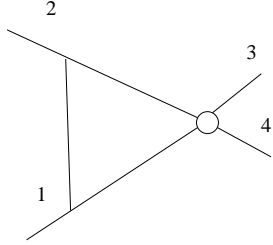}
\end{picture}}
\]
\caption{A one loop counterterm contribution to the two loop amplitude}
\end{center}
\end{figure}

\begin{figure}[ht]
\begin{center}
\[
\mbox{\begin{picture}(180,50)(0,0)
\includegraphics[scale=.65]{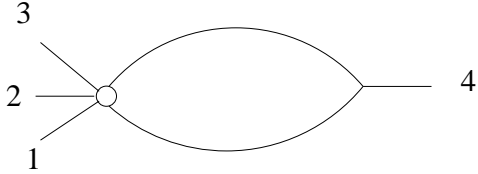}
\end{picture}}
\]
\caption{An off--shell counterterm contribution to the two loop amplitude}
\end{center}
\end{figure}

For example, let us consider the diagram given in figure 4. Upto the overall factor involving the $c_1$ vertex, the integral involved in 11 uncompactified dimensions is given by 
\be \label{dimint}\int d^{11}q \frac{1}{q^2 (q+ k_1)^2 (q-k_2)^2}.\ee
On compactifying on $T^2$, this is equal to
\be \label{Int1}\frac{\pi^{9/2}}{4\pi^2 l_{11}^2 \mathcal{V}_2} \int_0^\infty d\s_1 d\s_2 d\s_3 \s^{-9/2} e^{\s_2 \s_3 S/\s}\sum_{l_I} e^{-\s G^{IJ} l_I l_J/l_{11}^2}\ee
where 
\be \s = \s_1 +\s_2 +\s_3.\ee
Defining 
\be \omega_1 = \frac{\s_2}{\s}, \quad \omega_2 = \frac{\s_2 +\s_3}{\s}, \ee
and hence
\be 0 \leq \omega_1 \leq \omega_2 \leq 1\ee
we get that \C{Int1} is equal to
\be \frac{\pi^{9/2}}{4\pi^2 l_{11}^2 \mathcal{V}_2} \int_0^\infty d\s \int_0^1 d\omega_2 \int_0^{\omega_2} d\omega_1 \s^{-5/2} e^{\omega_1 (\omega_2 -\omega_1) \s S}\sum_{l_I} e^{-\s G^{IJ} l_I l_J/l_{11}^2}.\ee

Thus, at $O(D^4 \mathcal{R}^4)$, we get the counterterm contribution
\be \label{ren2}\delta \mathcal{A}_4^{(2)} = \frac{\pi^{11/2} \kappa_{11}^6}{(2\pi)^{22} l_{11}^5}\mathcal{K} \s_2 \cdot \frac{\pi^3}{4l_{11}^3} c_1 \cdot \Big[ \frac{2}{5} (\Lambda l_{11})^5 + \frac{3}{4\pi^{9/2}} \mathcal{V}_2^{-5/2} E_{5/2} (\Omega, \bar\Omega)\Big].\ee   
Finally adding \C{ren1} and \C{ren2} we get that
\be \mathcal{A}_4^{(2)} +\delta \mathcal{A}_4^{(2)} = \frac{\kappa_{11}^6}{(2\pi)^{22} l_{11}^8}\mathcal{K} \s_2 \Big[ \frac{\pi^6}{8} \mathcal{V}_2^{-5/2} E_{5/2} (\Omega, \bar\Omega) + \pi^4 \zeta (3) \zeta (4) \mathcal{V}_2^{-4} + \hat{a} (\Lambda l_{11})^8  + \hat{b} (\Lambda l_{11})^5\Big]\ee
on using \C{valc1}, where $\hat{a}, \hat{b}$ are constants. The $\Lambda^8$ term yields a primitive two loop divergence. The renormalized value of the last two terms has to be set to zero as they yield a contribution proportional to $e^{4\phi_B/3}$ in the effective action in the string frame\footnote{In fact, apart from the finite term we set $\Lambda l_{11} \rightarrow \infty$, so the term involving $\hat{b}$ can be dropped. Hence the two loop primitive divergence is completely cancelled by the counterterm.}.    

Finally we consider the $D^6 \mathcal{R}^4$ interaction. 
At linear order in the external momenta, we get a contribution
\bea \label{easy4}&&(4\pi^2)^2 [I_P (S,T) + I_P (S,U) + I_{NP} (S,T) + I_{NP} (S,U)] \non \\&&= \frac{\pi^{11}S}{72} \sum_{\hat{m}_I ,\hat{n}_I} \int_0^\infty d\hat\lambda d\hat\s d\hat\rho \hat\Delta^{-1/2} e^{-\pi^2 l_{11}^2 G_{IJ} \Big(\hat\lambda \hat{m}_I \hat{m}_J + \hat\s \hat{n}_I \hat{n}_J +\hat\rho{(\hat{m}+\hat{n})}_I{(\hat{m}+\hat{n})}_J\Big)}  \Big[ \hat\lambda + \hat\rho + \hat\s- 5\frac{\hat\lambda \hat\s \hat\rho}{\hat\Delta}\Big]\non \\ &&= \frac{\pi^{11} S}{12 l_{11}^6} 
\sum_{\hat{m}_I ,\hat{n}_I} \int_0^\infty dV_2 V_2^2 \int_{\mathcal{F}_2} \frac{d^2 \tau}{\tau_2^2} e^{-\pi^2 G_{IJ} (\hat{m}+\hat{n}\tau)_I (\hat{m}+\hat{n}\bar\tau)_J V_2/\tau_2} A(\tau,\bar\tau),\eea
where
\be \label{imptoo}A(\tau,\bar\tau) = \frac{\vert \tau \vert^2 - \vert \tau_1 \vert +1}{\tau_2} +\frac{5}{\tau_2^3}(\tau_1^2 - \vert \tau_1 \vert) (\vert \tau \vert^2 - \vert \tau_1 \vert),\ee
which satisfies
\be \label{impA} 4\tau_2^2 \frac{\p^2 A}{\p \tau \p \bar\tau} = 12 A - 12 \tau_2 \delta (\tau_1). \ee
Thus at $O(D^6\mathcal{R}^4)$, we have that
\bea \label{d6r4}
&&(4\pi^2)^2 \Big[S^2 \Big(I_P (S,T) + I_P (S,U) + I_{NP} (S,T) + I_{NP} (S,U)\Big)+\ldots \Big] \non \\ &&= \frac{\pi^{11} }{12 l_{11}^6} \s_3
\sum_{\hat{m}_I ,\hat{n}_I} \int_0^\infty dV_2 V_2^2 \int_{\mathcal{F}_2} \frac{d^2 \tau}{\tau_2^2} e^{-\pi^2 G_{IJ} (\hat{m}+\hat{n}\tau)_I (\hat{m}+\hat{n}\bar\tau)_J V_2/\tau_2} A(\tau,\bar\tau). \eea

For the $D^6\mathcal{R}^4$ interaction, the leading ultraviolet divergence which results from \C{d6r4} arises from a primitive two loop divergence when $\hat{m}_I =\hat{n}_I =0$ and is of the form $\Lambda^6$ coming from the boundary of the $V_2$ integral cutoff at $V_2 \sim (l_{11} \Lambda)^2$. The finite piece has been evaluated in~\cite{Green:2005ba} and is given by
\be \frac{\pi^6}{96 l_{11}^6} \s_3\mathcal{V}_2^{-3}\mathcal{E} (\Omega,\bar\Omega)\ee  
where $\mathcal{E} (\Omega,\bar\Omega)$ satisfies the Poisson equation
\be 4\Omega_2^2 \frac{\p^2 \mathcal{E}}{\p \Omega \p \bar\Omega} = 12 \mathcal{E} - 6 E_{3/2}^2,\ee
the structure of which follows from supersymmetry~\cite{Basu:2008cf}. The perturbative contributions to $\mathcal{E}$ are given by
\be \mathcal{E} (\Omega,\bar\Omega) = 4 \zeta(3)^2 \Omega_2^3 + 8\zeta (2) \zeta (3) \Omega_2 + 24 \zeta (4) \Omega_2^{-1} + \frac{8}{9} \zeta (6) \Omega_2^{-3}.\ee

The remaining ultraviolet divergence comes from the boundary of moduli space when $\tau_2 \rightarrow \infty$ keeping $V_2$ fixed~\cite{Green:1999pu}. Hence we only need to isolate the boundary contribution, which is done along the lines of~\cite{Green:1999pu} and so we only mention the results. The divergent part is given by
\be 4\Omega_2^2 \frac{\p^2 I^{div}}{\p\Omega \p \bar\Omega}  = \frac{\pi^{11}}{12 l_{11}^6}\s_3\sum_{\hat{m}_I, \hat{n}_I}\int_0^\infty d V_2 V_2^2 \int_{-1/2}^{1/2} d\tau_1 \Big[ A \frac{\p F_L }{\p\tau_2} - F_L \frac{\p A}{\p \tau_2}\Big]\Big\vert_{\tau_2 = (\Lambda l_{11})^2/V_2}\ee
where $F_L$ is the exponential involving the lattice momenta 
\be F_L = \sum_{\hat{m}_I, \hat{n}_I} e^{-\pi^2 G_{IJ} (\hat{m}+\hat{n}\tau)_I (\hat{m}+\hat{n}\bar\tau)_J V_2/\tau_2} \ee
in \C{easy4}. This structure comes from integrating by parts and considering only the boundary contribution, as all the other contributions are finite (apart from the two loop primitive divergence). Noting that as $\tau_2 \rightarrow \infty$,
\be A \rightarrow \tau_2, \quad \frac{\p A}{\p \tau_2} \rightarrow 1,\ee
we get that
\be 4\Omega_2^2 \frac{\p^2 I^{div}}{\p\Omega \p \bar\Omega}  = \frac{\pi^{17/2}}{96 l_{11}^3}\s_3\Lambda^3 \mathcal{V}_2^{-3/2} E_{3/2} (\Omega, \bar\Omega)\ee
leading to
\be I^{div} = \frac{\pi^{17/2}}{72 l_{11}^3} \s_3 \Lambda^3 \mathcal{V}_2^{-3/2} E_{3/2} (\Omega, \bar\Omega).\ee

Thus at this order
\be \label{findiv3}\mathcal{A}_4^{(2)}  = \frac{\kappa_{11}^6}{(2\pi)^{22} l_{11}^6} \mathcal{K} \s_3 \Big[ c (\Lambda l_{11})^6 + \frac{\pi^{17/2}}{72} (\Lambda l_{11})^3 \mathcal{V}_2^{-3/2} E_{3/2} (\Omega, \bar\Omega) + \frac{\pi^6}{96} \mathcal{V}_2^{-3} \mathcal{E} (\Omega,\bar\Omega)\Big], \ee
where $c$ is an undetermined constant. 

From \C{2loop} and \C{dimint} the one loop counterterm at $O(D^6 \mathcal{R}^4)$ is given by
\be \label{ren3}\delta \mathcal{A}_4^{(2)} = \frac{\pi^{11/2} \kappa_{11}^6}{12(2\pi)^{22} l_{11}^3}\mathcal{K} \s_3 \cdot \frac{\pi^3}{4l_{11}^3} c_1 \cdot \Big[ \frac{2}{3} (\Lambda l_{11})^3 + \frac{1}{2\pi^{5/2}} \mathcal{V}_2^{-3/2} E_{3/2} (\Omega, \bar\Omega)\Big].\ee
Thus adding the contributions \C{findiv3} and \C{ren3}, we get that
\be \mathcal{A}_4^{(2)} +\delta \mathcal{A}_4^{(2)} = \frac{\kappa_{11}^6}{(2\pi)^{22} l_{11}^6} \mathcal{K} \s_3 \Big[ \frac{\pi^8}{144} \mathcal{V}_2^{-3/2} E_{3/2} (\Omega,\bar\Omega) + \frac{\pi^6}{96} \Big(\mathcal{V}_2^{-3} \mathcal{E} (\Omega,\bar\Omega)+\hat{c} (\Lambda l_{11})^6 + \hat{d} (\Lambda l_{11})^{3} \Big)\Big],\ee
where $\hat{c}$ and $\hat{d}$ are undetermined constants. This leads to terms in the effective action given by
\be l_{11}^5 \int d^9 x \sqrt{-G^{(9)}} \mathcal{V}_2 D^6 \mathcal{R}^4 \Big[ \mathcal{V}_2^{-3} \mathcal{E} (\Omega,\bar\Omega) + 4\zeta (2) \mathcal{V}_2^{-3/2} E_{3/2} (\Omega,\bar\Omega) + \hat{c}(\Lambda l_{11})^6 + \hat{d} (\Lambda l_{11})^{3} \Big]\ee
where we have dropped an overall irrelevant constant. In the type IIA and IIB theories, this leads to terms in the effective action given by
\bea \label{twolooprenmore}
&& l_s^5 \int d^9 x \sqrt{-g^{A}} r_A D^6\mathcal{R}^4 \Big[ 4\zeta (3)^2 e^{-2\phi_A} + 8\zeta (2) \zeta (3) (1+ r_A^{-2}) \non \\ && + e^{2\phi_A} \Big(24\zeta (4) r_A^{-4}+ 16 \zeta(2)^2 r_A^{-2}  + \hat{c}(\Lambda l_{11})^6  \Big)+ \frac{8}{9} \zeta (6) e^{4\phi_A} r_A^{-6}\Big] \non \\ && = l_s^5 \int d^9 x \sqrt{-g^{B}} r_B D^6\mathcal{R}^4 \Big[ 4\zeta (3)^2 e^{-2\phi_B} + 8\zeta (2) \zeta (3) (1+ r_B^{-2}) \non \\ &&+ e^{2\phi_B} \Big(24\zeta (4) + 16 \zeta(2)^2 r_B^{-2}  + \frac{\hat{c}(\Lambda l_{11})^6 }{ r_B^4} \Big)+ \frac{8}{9} \zeta (6) e^{4\phi_B}\Big]\eea
where we have dropped exponentially suppressed terms, and also the divergent term involving $\hat{d}$ which is subdominant in the $\Lambda l_{11} \rightarrow \infty$ limit. Thus the remaining ultraviolet divergence $\sim \Lambda^6$ which is the two loop primitive divergence. To cancel this we finally add the primitive two loop counterterm to the amplitude given by
\be \delta \mathcal{A}_4^{(2)} = \frac{\kappa_{11}^6}{(2\pi)^{22} l_{11}^6} \mathcal{K}\s_3 \cdot \frac{\pi^6 e}{96}\ee 
thus sending
\be \hat{c}(\Lambda l_{11})^6  \rightarrow e + \hat{c}(\Lambda l_{11})^6 \ee
in \C{twolooprenmore}. 
The genus two equality of the four graviton amplitude in the type IIA and type IIB theories~\cite{Berkovits:2006vc} gives us that
\be \label{2loopren} e + \hat{c}(\Lambda l_{11})^6  = 24\zeta(4) (1-\eta),\ee 
where $\eta$ is an undetermined parameter which is not fixed by the two loop analysis. This is because we shall later see that the three loop primitive divergence also contributes to the genus two amplitude, and it is the sum of these two primitive divergences that fixes the finite part of the amplitude. We shall fix $\eta$ later.

Thus, upto exponentially suppressed terms the renormalized two loop amplitude is given by
\bea \label{2loopfin}
&&\mathcal{A}_4^{(2)}= (2\pi^8 l_{11}^{15} r_B) \mathcal{K} r_B \Big[\frac{l_s^4}{6!} \s_2 \Big(45 \zeta (5) e^{-2\phi^B} + 2\pi^2 \zeta (3) r_B^2  + 4\pi^2 \zeta (2) e^{2\phi^B} \Big) \non \\ &&+ \frac{l_s^6}{16\cdot 4!}\s_3\Big( 4\zeta (3)^2 e^{-2\phi^B} +8\zeta (2) \zeta (3) (1+ r_B^{-2})+ 24\zeta (4) e^{2\phi^B} (1+(1-\eta)r_B^{-4}) \non \\ &&+ 16\zeta(2)^2 e^{2\phi_B} r_B^{-2}+ \frac{8}{9} \zeta (6) e^{4\phi^B} \Big)+ O(k^8)\Big]\non \\ &&= (2\pi^8 l_{11}^{15} r_A^{-1}) \mathcal{K} r_A \Big[ \frac{l_s^4}{6!} \s_2\Big( 45 \zeta (5) e^{-2\phi^A} + \frac{2\pi^2}{r_A^4} \zeta (3) + \frac{4\pi^2}{r_A^4} \zeta (2) e^{2\phi^A}\Big) \non \\ &&+ \frac{l_s^6}{16\cdot 4!}\s_3\Big( 4\zeta (3)^2 e^{-2\phi^A} +8 \zeta (2) \zeta (3) (1 + r_A^{-2}) + 24\zeta (4) e^{2\phi^A} ((1-\eta)+r_A^{-4}) \non \\ &&+ 16 \zeta(2)^2 e^{2\phi_A} r_A^{-2}+ \frac{8}{9 r_A^6} \zeta (6) e^{4\phi^A}  \Big)+O(k^8) \Big].\eea

The complete perturbative part of the $\mathcal{R}^4$ and $D^4\mathcal{R}^4$ amplitudes is obtained by adding \C{1loopfin} and \C{2loopfin}. While the $\mathcal{R}^4$ term does not get contributions beyond genus one, the $D^4\mathcal{R}^4$ term does not get contributions beyond genus two. This is a consequence of the fact that these interactions are BPS. Note that these amplitudes manifestly exhibit T duality, as well as the equality of the type IIA and IIB coefficients at genus zero, one and two. This is also in agreement with the U--duality invariant results obtained in 8 dimensions on decompactifying to 9 dimensions~\cite{Kiritsis:1997em,Basu:2007ru}.  

However, the features of the $D^6\mathcal{R}^4$ interaction obtained from \C{1loopfin} and \C{2loopfin} are quite different. While the T duality of the IIA and IIB theories is manifest, the genus one amplitude is not the same in the IIA and IIB theories. This is because from \C{1loopfin} we see that the IIA and IIB theories have terms of the form $\zeta (2) \zeta (5) r_A^4$ and $\zeta (2)\zeta (5)r_B^{-6}$ respectively. Hence the $D^6\mathcal{R}^4$ interaction must receive contributions beyond two loops in supergravity, which we now consider\footnote{This is true even if $\eta =0$ and the three loop primitive divergence cancels completely.}.

\section{The four graviton supergravity amplitude at three loops}

The three loop four graviton amplitude in maximal supergravity in arbitrary dimensions has been considered in~\cite{Bern:2007hh}. The leading contribution is of the form $D^6 \mathcal{R}^4$, because the coefficient of the $D^4 \mathcal{R}^4$ contribution arising from the various diagrams vanishes identically. This suggests that it should be possible to express the amplitude in a form where the leading contribution is manifestly $D^6\mathcal{R}^4$, and hence the amplitude has better manifest ultraviolet behavior, which has been obtained in~\cite{Bern:2008pv}. In either case, the total amplitude is expressed as a sum over nine basic loop diagrams. However, the higher loop generalization of the KLT relation~\cite{Kawai:1985xq} expressing the gravity amplitude as square of the gauge theory amplitude is not manifest in these expressions for the amplitude. This is because though the denominator in the integrand of each loop diagram in the supergravity amplitude is the same as the denominator of the corresponding diagram of the three loop four gluon amplitude in $\mathcal{N}=4$ Yang--Mills, the numerator of the gravity amplitude is not the square of the numerator of the color stripped Yang--Mills amplitude. This has been remedied in~\cite{Bern:2010ue}, where the KLT structure is manifest. However, this has been achieved at the expense of introducing three new basic loop diagrams for the amplitude\footnote{These three new loop diagrams are reducible.}. We find it most convenient to use the representation of the amplitude given in~\cite{Bern:2008pv}.        

The general structure of the three loop amplitude can be understood in terms of the skeleton diagrams of the amplitude, which can be deduced using the first quantized worldline formalism for the superparticle~\cite{Berkovits:2001rb,Anguelova:2004pg,Dai:2006vj,Green:2008bf,Bjornsson:2010wm,Bjornsson:2010wu}. There are two kinds of irreducible distinct skeleton diagrams that are relevant for the four graviton three loop amplitude--the ladder diagram and the Mercedes diagram depicted in figure 6. The various loop diagrams are obtained by attaching external on--shell graviton vertex operators to these skeleton diagrams such that there are no one loop bubble or triangle sub--diagrams in the resultant diagrams. 

\begin{figure}[ht]
\begin{center}
\[
\mbox{\begin{picture}(180,100)(0,0)
\includegraphics[scale=.45]{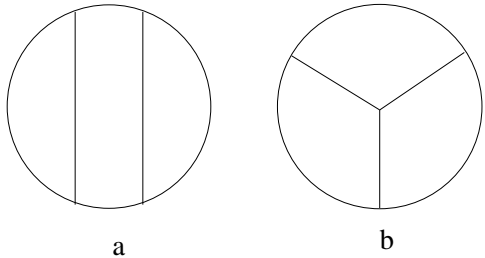}
\end{picture}}
\]
\caption{The three loop skeleton diagrams: (a) ladder and (b) Mercedes}
\end{center}
\end{figure}

The nine basic loop diagrams have been obtained in~\cite{Bern:2007hh,Bern:2008pv} using unitarity cut techniques and are depicted in figures 7 and 8, where we have used the notation of~\cite{Bern:2007hh,Bern:2008pv}.

\begin{figure}[ht]
\begin{center}
\[
\mbox{\begin{picture}(400,80)(0,0)
\includegraphics[scale=.55]{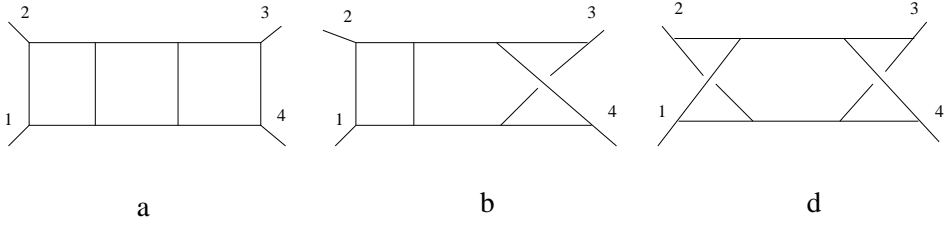}
\end{picture}}
\]
\caption{Three loop diagrams from the ladder skeleton}
\end{center}
\end{figure}

\begin{figure}[ht]
\begin{center}
\[
\mbox{\begin{picture}(350,190)(0,0)
\includegraphics[scale=.6]{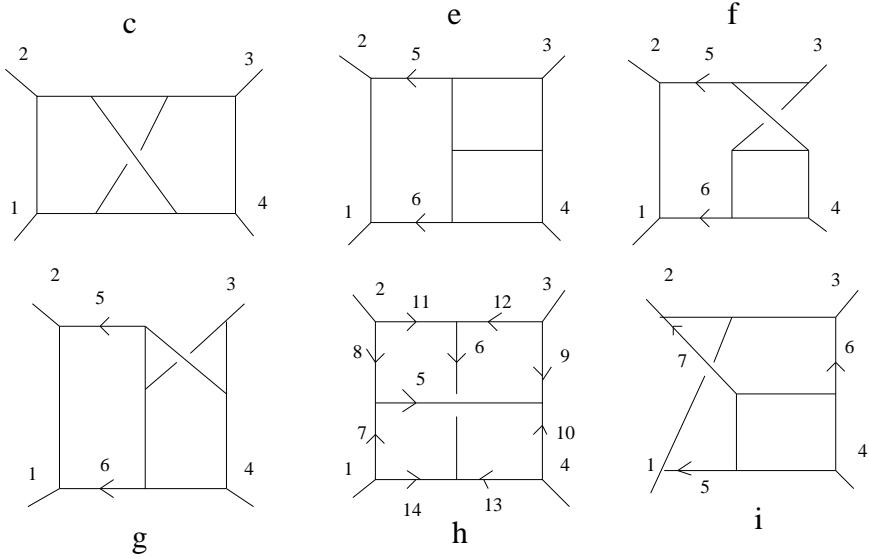}
\end{picture}}
\]
\caption{Three loop diagrams from the Mercedes skeleton}
\end{center}
\end{figure}

 Of them, diagrams $a,b$ and $d$ are obtained from the ladder skeleton diagram, while the rest are obtained from the Mercedes skeleton diagram. Unlike the one and two loop amplitudes, all the loop diagrams are not given in terms of massless $\varphi^3$ scalar field theory, even though the denominators of the integrands involve scalar propagators, and there are only cubic vertices. While the integrands for the loop diagrams $a,b,c$ and $d$ have numerator 1 and hence are given by massless $\varphi^3$ scalar field theory, the others are not as they have non--trivial dependence on the loop momenta in the numerators as given below. The total three loop amplitude is given by
\bea \label{totcont}
\mathcal{A}_4^{(3)} &=& \frac{(4\pi^2)^3 \kappa_{11}^8}{(2\pi)^{33}}\sum_{S_3} \Big[ I^{(a)} + I^{(b)} + \frac{1}{2} I^{(c)} + \frac{1}{4} I^{(d)} + 2 I^{(e)} + 2 I^{(f)} + 4 I^{(g)} + \frac{1}{2} I^{(h)} + 2 I^{(i)}\Big] \mathcal{K} \non \\ &\equiv& \frac{(4\pi^2)^3 \kappa_{11}^8}{(2\pi)^{33}} I_3 \mathcal{K}. \eea
where $S_3$ represents the 6 independent permutations of the external legs marked $\{1,2,3\}$ keeping the external leg $\{4\}$ fixed. The external momenta are directed inwards in all the loop diagrams.

\subsection{Evaluating the loop diagrams at $O(D^6\mathcal{R}^4)$}

The numerators $N^{(x)}$ for the various integrands in the loop diagrams are given by
\bea \label{num}
N^{(a)} &=& N^{(b)} = N^{(c)} = N^{(d)} = S^4 , \non \\ N^{(e)} &=& N^{(f)} = N^{(g)} = S^2 \tau_{35} \tau_{46}, \non \\ N^{(h)} &=& \Big(S(\tau_{26} +\tau_{36}) +T(\tau_{15} +\tau_{25}) +ST\Big)^2 \non \\ &&+ \Big(S^2 (\tau_{26} +\tau_{36}) - T^2 (\tau_{15} +\tau_{25}) \Big) \Big(\tau_{17} +\tau_{28} +\tau_{39} +\tau_{4,10} \Big) \non \\&& +S^2 (\tau_{17}  \tau_{28} +\tau_{39} \tau_{4,10}) +T^2 (\tau_{28} \tau_{39}+ \tau_{17} \tau_{4,10}) + U^2 (\tau_{17} \tau_{39} + \tau_{28} \tau_{4,10}), \non \\ N^{(i)} &=& (S\tau_{45} - T\tau_{46})^2 -\tau_{27} (S^2 \tau_{45} + T^2 \tau_{46}) - \tau_{15} (S^2 \tau_{47}+ U^2 \tau_{46})\non \\ &&-  \tau_{36} (T^2 \tau_{47} +U^2 \tau_{45}) - l_5^2 S^2 T - l_6^2 S T^2 +\frac{l_7^2}{3}STU,\eea
where
\be \tau_{ij} = -2 k_i \cdot l_j ~(i \leq 4, j \geq 5).\ee
The momenta $l_i$ are denoted in figure 8.

We now calculate the $D^6 \mathcal{R}^4$ term from the three loop amplitude. From \C{num} we see that the diagrams $a,b,c$ and $d$ have a leading contribution which is $O(D^8 \mathcal{R}^4)$ and are dropped. Hence the $D^6 \mathcal{R}^4$ term does not receive any contributions from loop diagrams that result from the ladder skeleton. Also all the contributions from the loop diagrams $e,f,g,h$ and $i$ are already $O(D^6 \mathcal{R}^4)$ from the numerators of the integrands, and hence the denominators which involve massless scalar propagators can be evaluated at zero external momenta. We now evaluate each of these loop diagrams, keeping only terms at $O(D^6 \mathcal{R}^4)$. While we refer to each contribution mentioned in~\C{totcont} as $I^{(x)}$, the total contribution after the sum over $S_3$ is referred to as $I^{(X)}$. The loop momenta labels for the various diagrams are given in figure 9.

\begin{figure}[ht]
\begin{center}
\[
\mbox{\begin{picture}(350,190)(0,0)
\includegraphics[scale=.6]{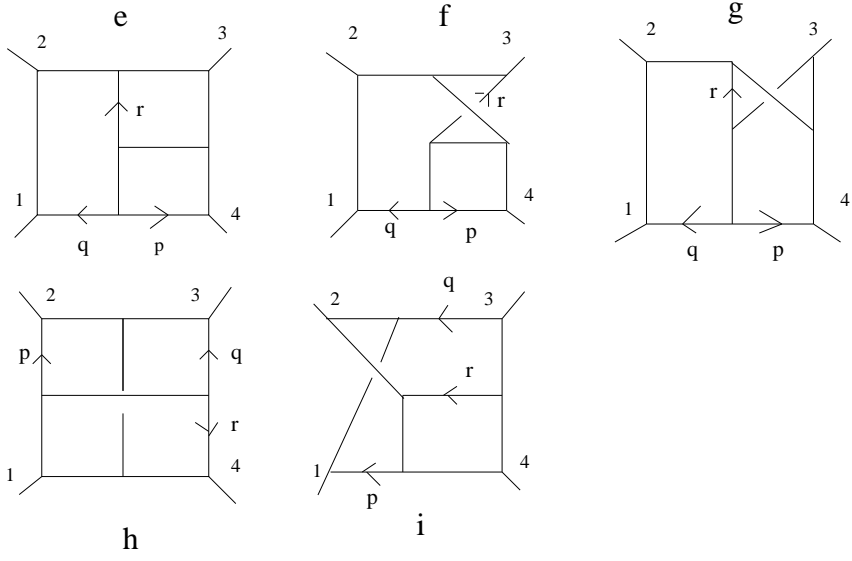}
\end{picture}}
\]
\caption{Momentum labels for the relevant three loop diagrams}
\end{center}
\end{figure}

\subsubsection{The contribution from diagram $e$}

In 11 uncompactified dimensions, the diagram $e$ contributes
\be \label{fige}
I^{(e)} = -4S^2\int d^{11}p \int d^{11} q \int d^{11}r \frac{(k_3 \cdot q)(k_4 \cdot q)}{q^6 p^4 r^2 (p+q)^2 (q+r)^4 (p+q+r)^2}.\ee
We now evaluate this integral, as well as the others, in the background $T^2 \times \mathbb{R}^{8,1}$, though it is easy to consider the most general case of compactification on $T^d$ for arbitrary $d$. Thus the 11 dimensional loop momenta $p_M$, $q_M$ and $r_M$ decompose as $\{p_\mu,l_I/l_{11}\}$, $\{q_\mu, m_I/l_{11}\}$ and $\{r_\mu, n_I/l_{11}\}$ respectively where $p_\mu,q_\mu$ and $r_\mu$ are the 9 dimensional momenta and $l_I,m_I$ and $n_I$ ($I=1,2$) are the KK momenta along $T^2$. We introduce 10 Schwinger parameters $\s^i$ for the 10 propagators. Thus the product of the propagators in the compactified theory coming from the denominator of \C{fige} is given by   
\be \int_0^\infty \prod_{i=1}^{10} d\s^ie^{-\sum_{j=1}^{10}\s^j q_j^2} e^{-\Big((\s_1 +\s_2 + \s_3){\bf{m}}^2 +(\s_4 +\s_5){\bf{l^2}}  +\s_6 {\bf{n^2}} + \s_7 {\bf{(l+m)^2}}+ (\s_8 + \s_9){\bf{(m+n)^2}} +\s_{10}{\bf{(l+m+n)^2}} \Big)/l_{11}^2}\ee
where
\be q_j = \{q,q,q,p,p,r,p+q,q+r,q+r,p+q+r\} \ee
and
\be {\bf{m}^2}\equiv G^{IJ} m_I m_J.\ee
Thus, compactifying on $T^2$ we have that
\bea \label{terme}I^{(e)} &=& -\frac{4S^2}{(4\pi^2 l_{11}^2 \mathcal{V}_2)^3} \int d^9 p \int d^9 q \int d^9 r (k_3 \cdot q)(k_4 \cdot q)\int_0^\infty \prod_{i=1}^{10} d\s^i  \non \\ && \times f_P (\lambda, \s,\s_6, \s_7,\rho,\s_{10}) F_L (\lambda, \s,\s_6, \s_7,\rho,\s_{10}),\eea  
where we have defined
\be \s = \s_1 +\s_2 +\s_3, \quad \lambda = \s_4 +\s_5, \quad \rho = \s_8 +\s_9.\ee
In \C{terme} the unintegrated momentum factor $f_P$ depends on 6 independent parameters and is given by
\be f_P (\s,\lambda,\mu,\rho,\nu,\theta) = e^{-\s p^2 -\lambda q^2 -\mu r^2 -\rho (p+q)^2 -\nu (q+r)^2 -\theta(p+q+r)^2}.\ee
The lattice factor $F_L$ also depends on 6 independent parameters and is given by
\bea \label{defF} &&F_L (\s,\lambda,\mu,\rho,\nu,\theta) \non \\ &&= \sum_{l_I, m_I, n_I} e^{-G^{IJ}\Big( \s l_I l_J +\lambda m_I m_J +\mu n_I n_J +\rho (l+m)_I (l+m)_J +\nu (m+n)_I(m+n)_J +\theta (l+m+n)_I(l+m+n)_J \Big)/l_{11}^2}.\non \\ \eea
Using the fact that $(k_3 \cdot q)(k_4 \cdot q) \rightarrow -Sq^2/18$ in the integral, we get that
\bea I^{(e)} = \frac{2S^3}{9(4\pi^2 l_{11}^2 \mathcal{V}_2)^3} \int d^9 p \int d^9 q \int d^9 r q^2 \int_0^\infty \prod_{i=1}^{10} d\s^i  f_P (\lambda, \s,\s_6, \s_7,\rho,\s_{10}) \non \\ \times F_L (\lambda, \s,\s_6, \s_7,\rho,\s_{10}). \eea
We now define
\be w_1 = \frac{\s_1}{\s}, \quad w_2 = \frac{\s_1 +\s_2}{\s}, \quad u=\frac{\s_4}{\lambda}, \quad v = \frac{\s_8}{\rho},\ee
and hence
\be 0 \leq w_1 \leq w_2 \leq 1, \quad 0 \leq u \leq 1, \quad 0 \leq v \leq 1.\ee
This simplifies the expression for $I^{(e)}$ leading to
\bea I^{(e)} = -\frac{S^3}{9(4\pi^2 l_{11}^2 \mathcal{V}_2)^3} \int_0^\infty d\s d\lambda d\rho d\s_6 d\s_7 d\s_{10} (\s^2 \lambda \rho) F_L (\lambda, \s,\s_6, \s_7,\rho,\s_{10}) \non \\ \times \frac{\p}{\p\s}\mathcal{J}(\lambda, \s,\s_6, \s_7,\rho,\s_{10}),\eea 
where
\bea \label{defJ}\mathcal{J} (\s,\lambda,\mu,\rho,\nu,\theta) &=& \int d^9 p\int d^9 q\int d^9 r e^{-\s p^2-\lambda q^2 -\mu r^2 -\rho (p+q)^2-\nu (q+r)^2-\theta (p+q+r)^2} \non \\  &=& \pi^{27/2} \Delta_3^{-9/2}(\s,\lambda,\mu,\rho,\nu,\theta).\eea
In \C{defJ}, $\Delta_3$ is defined by
\bea \label{defD}\Delta_3 (\s,\lambda,\mu,\rho,\nu,\theta) &=& \s \lambda \mu +\rho\nu\theta + \s\mu (\rho+\nu+\theta) +\lambda\mu (\rho+\theta) +\s\lambda (\nu+\theta)\non \\ &&+ \mu\nu(\rho+\theta) +\s\rho(\nu+\theta) +\lambda(\rho\nu+\nu\theta+\rho\theta).\eea

Finally, this leads to
\bea I^{(E)} = -\frac{2 \s_3}{9(4\pi^2 l_{11}^2 \mathcal{V}_2)^3} \int_0^\infty d\Upsilon (\lambda^2 \s \nu) F_L (\s,\lambda, \mu,\rho, \nu,\theta)  \frac{\p}{\p\lambda}\mathcal{J}(\s,\lambda, \mu,\rho, \nu,\theta),\eea
where we denote
\be  d\Upsilon \equiv d\s d\lambda d\mu d\rho d\nu d\theta\ee  
for brevity.

\subsubsection{The contribution from diagram $f$}

The calculations for $I^{(f)}$ and $I^{(g)}$ are very similar to the calculation for $I^{(e)}$ and so we skip the details. 
In 11 uncompactified dimensions, the loop diagram $f$ contributes
\be \label{figf}I^{(f)} = -4S^2\int d^{11}p \int d^{11} q \int d^{11}r \frac{(k_3 \cdot q)(k_4 \cdot q)}{q^6 p^4 r^4 (p+q)^2 (q+r)^2 (p+q+r)^2}.\ee
Compactifying on $T^2$ and proceeding as above by introducing 10 Schwinger parameters $\s^i$ for the momenta in the propagators in \C{figf}
\be q_j = \{ q,q,q,p,p,r,r,p+q,q+r,p+q+r\}\ee
we get that
\bea I^{(f)} &=& \frac{2S^2}{9(4\pi^2 l_{11}^2 \mathcal{V}_2)^3} \int d^9 p \int d^9 q \int d^9 r q^2 \int_0^\infty \prod_{i=1}^{10} d\s^i  \non \\ && \times f_P (\lambda, \s,\rho, \s_8,\s_9,\s_{10}) F_L (\lambda, \s,\rho, \s_8,\s_9,\s_{10}),\eea
where
\be \s = \s_1 +\s_2 +\s_3, \quad \lambda = \s_4 +\s_5, \quad \rho = \s_6 +\s_7.\ee
Thus we get that
\bea I^{(F)} = -\frac{2\s_3}{9(4\pi^2 l_{11}^2 \mathcal{V}_2)^3} \int_0^\infty d\Upsilon (\lambda^2 \s \mu) F_L (\s,\lambda, \mu,\rho, \nu,\theta)  \frac{\p}{\p\lambda}\mathcal{J}(\s,\lambda, \mu,\rho, \nu,\theta).\eea

\subsubsection{The contribution from diagram $g$}

In 11 uncompactified dimensions, the diagram $g$ contributes
\be \label{figg}I^{(g)} = -4S^2\int d^{11}p \int d^{11} q \int d^{11}r \frac{(k_3 \cdot q)(k_4 \cdot q)}{q^6 p^4 r^2 (p+q)^2 (q+r)^2 (p+q+r)^4}.\ee
Once again compactifying on $T^2$ and introducing 10 Schwinger parameters $\s^i$ for the momenta in the propagators in \C{figg}
\be q_j = \{ q,q,q,p,p,r,p+q,q+r,p+q+r,p+q+r\}\ee
we get that
\bea I^{(g)} &=& \frac{2S^2}{9(4\pi^2 l_{11}^2 \mathcal{V}_2)^3} \int d^9 p \int d^9 q \int d^9 r q^2 \int_0^\infty \prod_{i=1}^{10} d\s^i  \non \\ && \times f_P (\lambda, \s,\s_6, \s_7,\s_8,\rho) F_L (\lambda, \s,\s_6, \s_7,\s_8,\rho),\eea
where
\be \s = \s_1 +\s_2 +\s_3, \quad \lambda = \s_4 +\s_5, \quad \rho = \s_9 +\s_{10}.\ee
This leads to
\bea \label{valg}I^{(G)} = -\frac{2\s_3}{9(4\pi^2 l_{11}^2 \mathcal{V}_2)^3} \int_0^\infty d\Upsilon (\lambda^2 \s \theta) F_L (\s,\lambda, \mu,\rho, \nu,\theta) \frac{\p}{\p\lambda}\mathcal{J}(\s,\lambda, \mu,\rho, \nu,\theta).\eea

Note that there are several ways in which we could write each of these integrals, as well as the ones below. For example, we could write 
\bea I^{(G)} = -\frac{2\s_3}{9(4\pi^2 l_{11}^2 \mathcal{V}_2)^3} \int_0^\infty d\Upsilon (\s^2 \lambda \mu) F_L (\s,\lambda, \mu,\rho, \nu,\theta) \frac{\p}{\p\s}\mathcal{J}(\s,\lambda, \mu,\rho, \nu,\theta),\eea
which is equal to \C{valg} on using\footnote{In fact, the equality \C{equal} follows from $P_{18}$ in \C{listtrans}.}
\be \label{equal}F_L(\s,\lambda, \mu,\rho, \nu,\theta) = F_L (\lambda,\s, \theta,\rho, \nu,\mu), \quad \mathcal{J}(\s,\lambda, \mu,\rho, \nu,\theta) = \mathcal{J} (\lambda,\s, \theta,\rho, \nu,\mu) \ee
and trivially renaming variables in the integral. We shall discuss this issue in detail in the next section when we discuss the symmetries of the Mercedes skeleton.

\subsubsection{The contribution from diagram $h$}

In 11 uncompactified dimensions, the diagram $h$ contributes
\be \label{figh}I^{(h)} = \int d^{11}p \int d^{11} q \int d^{11}r \frac{N^{(h)}}{p^4 q^4 r^4 (p+q)^2 (q+r)^2 (p+q+r)^4}.\ee

First consider the contribution from the first two lines in $N^{(h)}$ in \C{num} after compactifying on $T^2$. Using that $q^\mu_i q^\nu_j \rightarrow \eta^{\mu\nu} (q_i \cdot q_j)/9$ in the integrals where $q_i = (p,q,r)$, at order $O(D^6 \mathcal{R}^4)$ we see that the first line for $N^{(h)}$ in \C{num} contributes a factor of 
\be -\frac{4}{9} S^2 T (p+q)^2 - \frac{4}{9} S T^2 (q+r)^2\ee
inside the integral, which is exactly cancelled by contributions coming from the second line of $N^{(h)}$ (this is true in all dimensions). The next term involving $S^2 (\tau_{17}\tau_{28} + \tau_{39} \tau_{4,10})$ gives a contribution 
\bea &=& \frac{2S^3}{9(4\pi^2 l_{11}^2 \mathcal{V}_2)^3} \int_0^\infty d\Upsilon(\s\lambda\mu\theta)F_L (\s,\lambda, \mu,\rho, \nu,\theta) \non \\  &&\times \int d^9 p \int d^9 q \int d^9 r f_P(\s,\lambda, \mu,\rho, \nu,\theta) \Big(p\cdot (p+q+r) - q\cdot r\Big)\eea
exactly along the lines of the calculations we have done before.
Similarly, the last two terms involving $T^2(\tau_{28}\tau_{39}+ \tau_{17} \tau_{4,10})$ and $U^2(\tau_{17} \tau_{39}+\tau_{28}\tau_{4,10})$ give contributions
\bea &=& \frac{2T^3}{9(4\pi^2 l_{11}^2 \mathcal{V}_2)^3} \int_0^\infty d\Upsilon(\s\lambda\mu\theta)F_L (\s,\lambda, \mu,\rho, \nu,\theta) \non \\  &&\times \int d^9 p \int d^9 q \int d^9 r f_P(\s,\lambda, \mu,\rho, \nu,\theta) \Big(r\cdot (p+q+r) - p\cdot q\Big)\eea
and
\bea &=& \frac{2U^3}{9(4\pi^2 l_{11}^2 \mathcal{V}_2)^3} \int_0^\infty d\Upsilon(\s\lambda\mu\theta)F_L (\s,\lambda, \mu,\rho, \nu,\theta) \non \\  &&\times \int d^9 p \int d^9 q \int d^9 r f_P(\s,\lambda, \mu,\rho, \nu,\theta) \Big(q\cdot (p+q+r) - p\cdot r\Big)\eea
respectively. Thus adding all the contributions, we have that
\bea I^{(H)} &=&  \frac{2\s_3}{9(4\pi^2 l_{11}^2 \mathcal{V}_2)^3} \int_0^\infty d\Upsilon(\s\lambda\mu\theta)F_L (\s,\lambda, \mu,\rho, \nu,\theta)\non \\ &&\times \int d^9 p \int d^9 q \int d^9 r f_P(\s,\lambda, \mu,\rho, \nu,\theta) (p^2 + q^2 + r^2 +(p+q+r)^2) \non \\ &=& -\frac{2\s_3}{9(4\pi^2 l_{11}^2 \mathcal{V}_2)^3} \int_0^\infty d\Upsilon(\s\lambda\mu\theta)F_L (\s,\lambda, \mu,\rho, \nu,\theta)\non \\ &&\times \Big(\frac{\p}{\p\s} +\frac{\p}{\p\lambda} +\frac{\p}{\p\mu} +\frac{\p}{\p\theta}\Big)\mathcal{J} (\s,\lambda, \mu,\rho, \nu,\theta) \non \\ &=& -\frac{8\s_3}{9(4\pi^2 l_{11}^2 \mathcal{V}_2)^3} \int_0^\infty d\Upsilon(\s\lambda\mu\theta)F_L (\s,\lambda, \mu,\rho, \nu,\theta) \frac{\p}{\p\theta}\mathcal{J} (\s,\lambda, \mu,\rho, \nu,\theta), \eea
on using
\bea F_L (\s,\lambda, \mu,\rho, \nu,\theta) = F_L(\theta,\mu,\lambda,\rho,\nu,\s),\quad \mathcal{J} (\s,\lambda, \mu,\rho, \nu,\theta) = \mathcal{J}(\theta,\mu,\lambda,\rho,\nu,\s),\non \\ F_L (\s,\lambda, \mu,\rho, \nu,\theta) = F_L (\mu,\theta,\s,\rho,\nu,\lambda), \quad \mathcal{J}(\s,\lambda, \mu,\rho, \nu,\theta) = \mathcal{J} (\mu,\theta,\s,\rho,\nu,\lambda), \non \\ F_L (\s,\lambda, \mu,\rho, \nu,\theta) = F_L (\lambda,\s,\theta,\rho,\nu,\mu), \quad \mathcal{J}(\s,\lambda, \mu,\rho, \nu,\theta) = \mathcal{J} (\lambda,\s,\theta,\rho,\nu,\mu),\eea
which follow from $P_{15}, P_{10}$ and $P_{18}$ in \C{listtrans} respectively, and trivially renaming variables in the integrals. 

\subsubsection{The contribution from diagram $i$}

Finally in 11 uncompactified dimensions, the diagram $i$ contributes
\be \label{figi}I^{(i)} = \int d^{11}p \int d^{11} q \int d^{11}r \frac{N^{(i)}}{p^4 q^4 r^2 (p+q)^4 (q+r)^4 (p+q+r)^2}.\ee

Compactifying on $T^2$ and proceeding as above, we write
\be I^{(I)} = I^{(I)}_1 + I^{(I)}_2,\ee
where $I^{(I)}_1$ includes the contribution of all but the last three terms in $N^{(i)}$ in \C{num}, and $I^{(I)}_2$ is the contribution of the last three terms.

We get that
\bea I^{(I)}_1 &=&  \frac{2\s_3}{9(4\pi^2 l_{11}^2 \mathcal{V}_2)^3} \int_0^\infty d\Upsilon(\s\lambda\rho\nu)F_L (\s,\lambda, \mu,\rho, \nu,\theta)\non \\ &&\times \int d^9 p \int d^9 q \int d^9 r f_P(\s,\lambda, \mu,\rho, \nu,\theta) (p^2 + q^2 + (p+q)^2 ) \non \\ &=& -\frac{2\s_3}{9(4\pi^2 l_{11}^2 \mathcal{V}_2)^3} \int_0^\infty d\Upsilon(\s\lambda\rho\nu)F_L (\s,\lambda, \mu,\rho, \nu,\theta)\non \\ &&\times \Big[\frac{\p}{\p\s} + \frac{\p}{\p\lambda} + \frac{\p}{\p\rho}\Big] \mathcal{J} (\s,\lambda, \mu,\rho, \nu,\theta) \non \\ &=&-\frac{2\s_3}{9(4\pi^2 l_{11}^2 \mathcal{V}_2)^3} \int_0^\infty d\Upsilon(\s\lambda\rho\nu)F_L (\s,\lambda, \mu,\rho, \nu,\theta)\non \\ &&\times \Big[2\frac{\p}{\p\s}  + \frac{\p}{\p\rho}\Big] \mathcal{J} (\s,\lambda, \mu,\rho, \nu,\theta), \eea
on using $P_{18}$ in \C{listtrans} and trivially renaming integration variables, while $I^{(I)}_2$ is given by
\bea \label{easy}I^{(I)}_2 &=& \frac{\s_3}{(4\pi^2 l_{11}^2 \mathcal{V}_2)^3} \int_0^\infty d\Upsilon \Big(2\lambda\rho\nu + \frac{2}{3} \s\lambda\nu\Big)F_L (\s,\lambda, \mu,\rho, \nu,\theta) \mathcal{J} (\s,\lambda, \mu,\rho, \nu,\theta)\non \\ &=& \frac{\s_3}{6(4\pi^2 l_{11}^2 \mathcal{V}_2)^3} \int_0^\infty d\Upsilon \Delta_3 F_L (\s,\lambda, \mu,\rho, \nu,\theta) \mathcal{J} (\s,\lambda, \mu,\rho, \nu,\theta)\eea
on using the symmetries of $F_L$ and $\mathcal{J}$ discussed in the next section.

\subsection{The total contribution from the loop diagrams at $O(D^6\mathcal{R}^4)$}

Apart from \C{easy} which is quite simple, it is illuminating to consider the pictorial representation of the various loop diagrams we have evaluated as shown in figure 10. 

\begin{figure}[ht]
\begin{center}
\[
\mbox{\begin{picture}(320,210)(0,0)
\includegraphics[scale=.55]{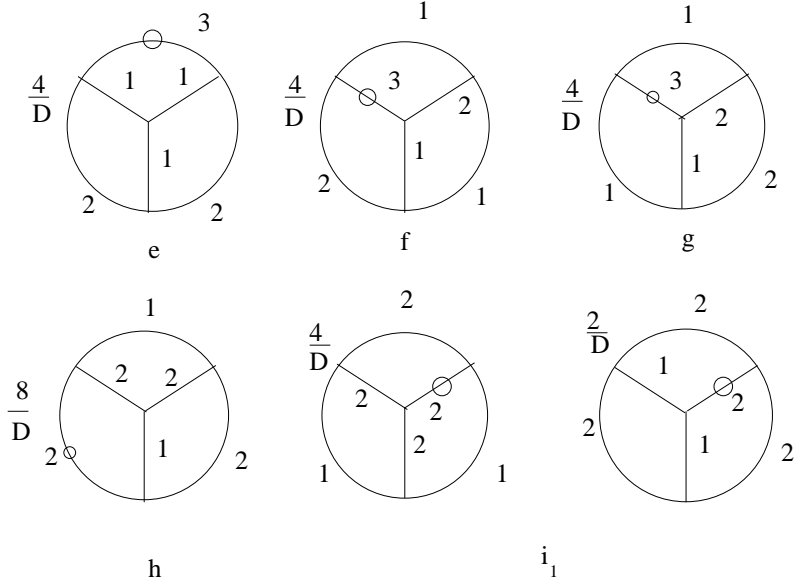}
\end{picture}}
\]
\caption{Diagrammatic representation of the loop diagrams}
\end{center}
\end{figure}

\begin{figure}[ht]
\begin{center}
\[
\mbox{\begin{picture}(320,210)(0,0)
\includegraphics[scale=.55]{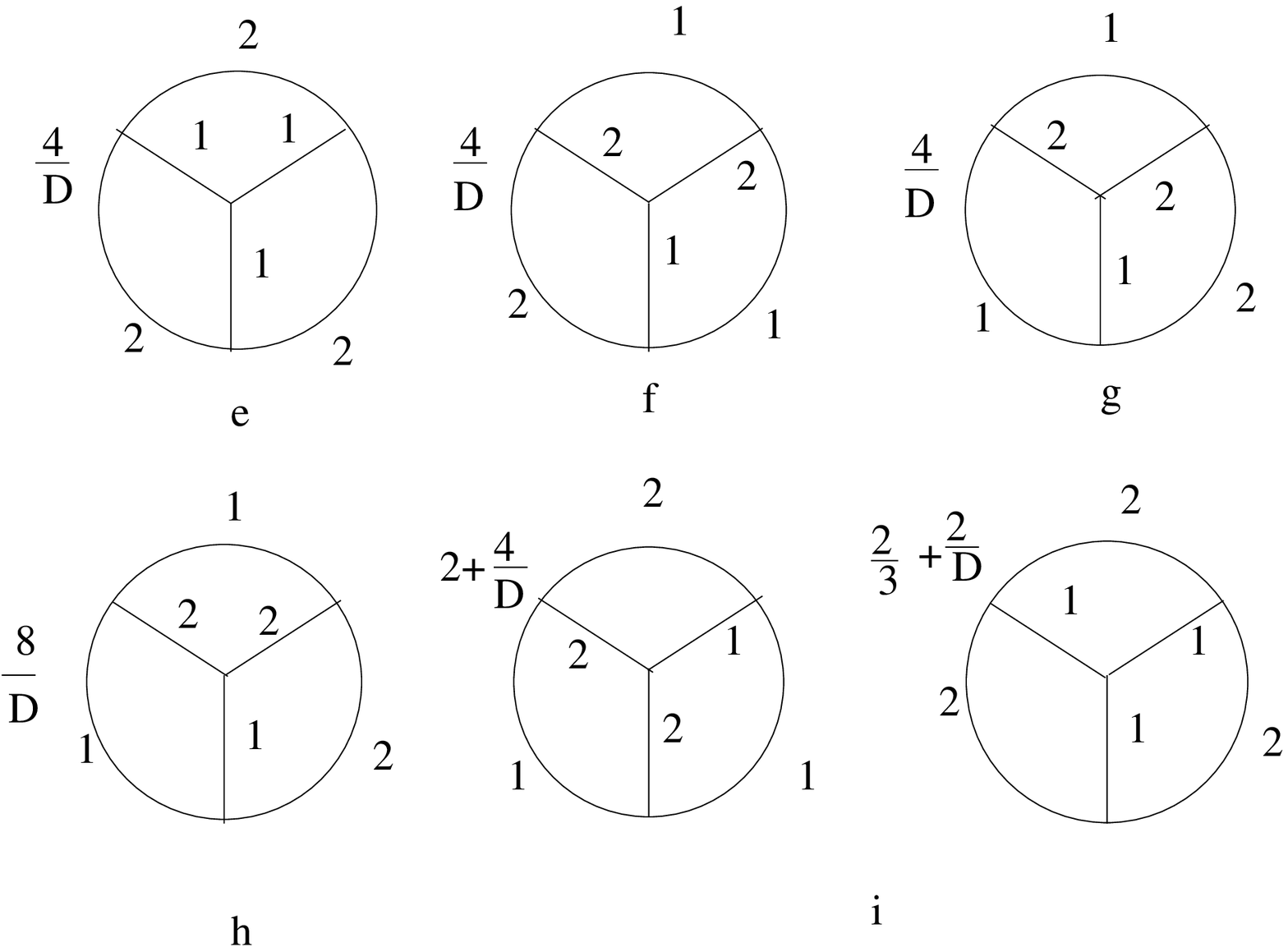}
\end{picture}}
\]
\caption{Diagrammatic representation of the field theory loop diagrams}
\end{center}
\end{figure}

We depict the contribution from each loop diagram in terms of the underlying Mercedes skeleton, where the integers (1,2,3) on each link stand for the number of Schwinger parameters $\s_i$ $(i=1,\cdots,10)$ corresponding to the momentum depicted by that link. The circle on one of the links stands for the Schwinger parameter with respect to which a derivative has been taken. The overall numerical prefactor for compactifying to $D$ non--compact dimensions is also mentioned. Each diagram is topologically distinct from the others, and hence they all yield different contributions.

What happens in the field theory limit? This is the limit when one does a simple dimensional reduction, and neglects the contributions of the massive KK modes. Then the derivative with respect to a Schwinger parameter removes a factor of the corresponding propagator in the denominator of the integral. Then the contribution of the loop diagrams look as in figure 11~\footnote{Here we have also included the contribution from \C{easy}.}. In this limit, there are only two topologically distinct loop diagrams and the prefactors precisely match the analysis in~\cite{Bern:2008pv} where the $D=6$ case has been considered in detail.

Thus adding all the contributions, the total contribution at three loops $I_3$ in the background $T^2 \times \mathbb{R}^{8,1}$ is given by
\bea \label{impexp}
I_3 &=& \frac{2\pi^{27/2}}{(4\pi^2 l_{11}^2 \mathcal{V}_2)^3} \s_3 \int_0^\infty \frac{d\Upsilon}{\Delta_3^{11/2}} F_L \Big[\lambda^2 \s\nu\frac{\p}{\p\lambda} + \lambda^2 \s\mu \frac{\p}{\p\lambda} +  2\s^2\lambda\mu \frac{\p}{\p\s} +  \s\lambda\mu\theta\frac{\p}{\p\theta} \non \\ &&+ \s\lambda\rho\nu \Big(2 \frac{\p}{\p\s}+ \frac{\p}{\p\rho}\Big)\Big]\Delta_3 + \frac{\pi^{27/2}}{3(4\pi^2 l_{11}^2 \mathcal{V}_2)^3} \s_3 \int_0^\infty d\Upsilon \Delta_3^{-7/2} F_L .\eea
From now onwards, $F_L$ and $\Delta_3$ will involve the sequence $(\s,\lambda,\mu,\rho,\nu,\theta)$.
We now use the various symmetries of the Mercedes skeleton to simplify our calculations. 

\section{The symmetries of the Mercedes skeleton}

In the calculations above, it is clear that the integrals can be written in various ways based on the symmetries of $F_L$ and $\mathcal{J}$. We now consider the basis for these relations. 

Consider the loop diagrams that contribute to the three loop four graviton amplitude that arise from adding vertex operators to the Mercedes skeleton. For the $D^6\mathcal{R}^4$ interaction, these are the only contributions, while at higher orders in the momentum expansion, these are only  part of the total contribution to the amplitude\footnote{The diagrams $a,b,c$ and $d$ contribute only from $O(D^8 \mathcal{R}^4)$ onwards. Of them, diagram $c$ arises from the Mercedes skeleton and involves 6 Schwinger parameters. However, diagrams $a,b$ and $d$ which arise from the ladder skeleton involve 5 Schwinger parameters, and hence have an entirely different structure. }. For such loop diagrams at an arbitrary order in the momentum expansion, after introducing the 6 Schwinger parameters and integrating over the non--compact momenta and summing over the compact ones, one is left with an integral over the Schwinger parameters, where the integrand always contains the lattice factor $F_L$ and a factor of $\mathcal{J}$, possibly with derivatives of Schwinger parameters acting on it. Apart from these, the integrand also contains various factors of the Schwinger parameters. Now we can express the integrand such that it respects the symmetries  of $F_L$ and $\mathcal{J}$, which is the symmetry of the underlying Mercedes skeleton. Then the integral is put in a manifestly symmetric form. This analysis is true at all orders in the momentum expansion.            
 
We would like to find this symmetry group, which will be extremely useful for our analysis.
Now $F_L$ and $\mathcal{J}$ depend on 6 Schwinger parameters, corresponding to the 6 momenta $p,q,r,p+q,q+r$ and $p+q+r$ which run along the 6 links of the Mercedes skeleton (this is also true for the dimensionless KK momenta $l,m,n,l+m,m+n,l+m+n$). Since all the continuous momenta are integrated over and the discrete momenta are summed over, the symmetries are the ones that are associated with various interchanges of the 6 parameters associated with relabelling the various momenta. To see the symmetry, we draw the Mercedes diagram with a specific choice of the 6 parameters as shown in figure 12. 

\begin{figure}[ht]
\begin{center}
\[
\mbox{\begin{picture}(300,110)(0,0)
\includegraphics[scale=.5]{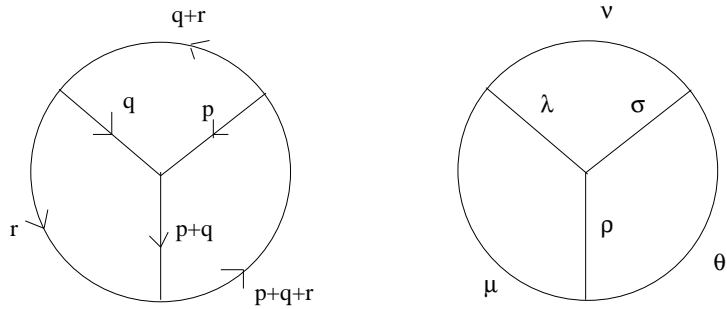}
\end{picture}}
\]
\caption{Parametrizing the Mercedes skeleton}
\end{center}
\end{figure}

The symmetry can also be seen from the dual regular tetrahedron which follows from replacing each face of the Mercedes skeleton (which is the wheel graph of order 4) with a vertex of the regular tetrahedron, such that every edge of the regular tetrahedron is parametrized by the link of the Mercedes skeleton it cuts as depicted in figure 13. Thus the symmetry group is the set of discrete transformations which interchange the 4 vertices of either diagram, keeping the links between the vertices intact. This is the symmetric group $S_4$, the automorphism group of the wheel graph of order 4, as well as the regular tetrahedron. It is now easy to see how the Schwinger parameters transform under the action of every element of $S_4$.

\begin{figure}[ht]
\begin{center}
\[
\mbox{\begin{picture}(300,120)(0,0)
\includegraphics[scale=.55]{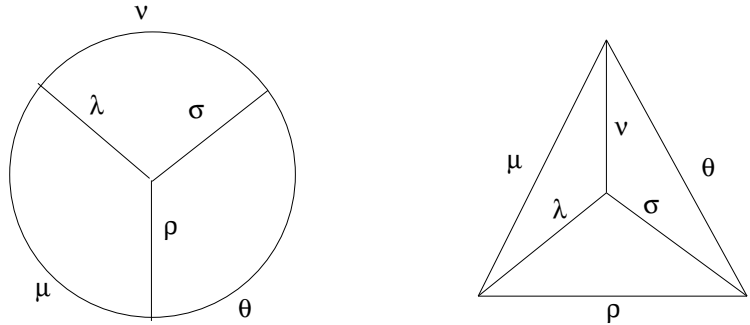}
\end{picture}}
\]
\caption{The Mercedes skeleton and the dual regular tetrahedron}
\end{center}
\end{figure}

\subsection{Transformations of the Schwinger parameters under $S_4$}

We now write down the action of the transformations of $S_4$ on the 6 Schwinger parameters. We shall use the notation

\be P_i = \begin{pmatrix}
 \s &\lambda &\mu &\rho &\nu &\theta \\
* & * & *& *& *& *
\end{pmatrix}\ee

where $i=1,\cdots, 24$ to indicate each transformation, where the lower row stands for the parameters $\s_i (\s ~\lambda ~\mu ~\rho ~\nu ~\theta)$, where $\s_i \in S_4$. Thus the 24 transformations are given by

\bea \label{listtrans}P_1 = \begin{pmatrix}
\s &\lambda &\mu &\rho &\nu &\theta \\
\s &\lambda &\mu &\rho &\nu &\theta 
\end{pmatrix} \quad
P_2 = \begin{pmatrix}
\s &\lambda &\mu &\rho &\nu &\theta \\
\theta & \rho & \lambda& \mu& \s& \nu
\end{pmatrix}  \quad
P_3 = \begin{pmatrix}
\s &\lambda &\mu &\rho &\nu &\theta \\
\nu & \mu & \rho& \lambda& \theta& \s
\end{pmatrix}\non \\
P_4 = \begin{pmatrix}
\s &\lambda &\mu &\rho &\nu &\theta \\
\lambda & \nu & \theta& \mu& \s& \rho
\end{pmatrix} \quad
P_5 = \begin{pmatrix}
\s &\lambda &\mu &\rho &\nu &\theta \\
\nu & \s & \rho& \theta& \lambda& \mu
\end{pmatrix}  \quad
P_6 = \begin{pmatrix}
\s &\lambda &\mu &\rho &\nu &\theta \\
\theta & \nu & \lambda & \s& \mu& \rho
\end{pmatrix}\non \\
P_7 = \begin{pmatrix}
\s &\lambda &\mu &\rho &\nu &\theta \\
\rho & \mu & \nu& \theta& \lambda& \s
\end{pmatrix} \quad
P_8 = \begin{pmatrix}
\s &\lambda &\mu &\rho &\nu &\theta \\
\lambda & \rho & \theta & \s & \mu & \nu
\end{pmatrix}  \quad
P_9 = \begin{pmatrix}
\s &\lambda &\mu &\rho &\nu &\theta \\
\rho & \s & \nu & \lambda & \theta & \mu
\end{pmatrix}\non \\
P_{10} = \begin{pmatrix}
\s &\lambda &\mu &\rho &\nu &\theta \\
\mu & \theta & \s& \rho& \nu& \lambda
\end{pmatrix} \quad
P_{11} = \begin{pmatrix}
\s &\lambda &\mu &\rho &\nu &\theta \\
\mu & \lambda & \s& \nu & \rho& \theta
\end{pmatrix}  \quad
P_{12} = \begin{pmatrix}
\s &\lambda &\mu &\rho &\nu &\theta \\
\s & \theta & \mu& \nu & \rho & \lambda
\end{pmatrix}\non \\
P_{13} = \begin{pmatrix}
\s &\lambda &\mu &\rho &\nu &\theta \\
\nu & \lambda & \rho& \mu& \s& \theta
\end{pmatrix} \quad
P_{14} = \begin{pmatrix}
\s &\lambda &\mu &\rho &\nu &\theta \\
\s & \rho & \mu& \lambda& \theta& \nu
\end{pmatrix}  \quad
P_{15} = \begin{pmatrix}
\s &\lambda &\mu &\rho &\nu &\theta \\
\theta & \mu & \lambda & \rho & \nu & \s
\end{pmatrix}\non \\
P_{16} = \begin{pmatrix}
\s &\lambda &\mu &\rho &\nu &\theta \\
\rho & \lambda & \nu & \s & \mu & \theta
\end{pmatrix} \quad
P_{17} = \begin{pmatrix}
\s &\lambda &\mu &\rho &\nu &\theta \\
\s & \nu & \mu & \theta & \lambda & \rho
\end{pmatrix}  \quad
P_{18} =\begin{pmatrix}
\s &\lambda &\mu &\rho &\nu &\theta \\
\lambda &\s & \theta& \rho& \nu & \mu
\end{pmatrix}\non \\
P_{19} = \begin{pmatrix}
\s &\lambda &\mu &\rho &\nu &\theta \\
\theta & \s & \lambda& \nu& \rho & \mu
\end{pmatrix} \quad
P_{20}= \begin{pmatrix}
\s &\lambda &\mu &\rho &\nu &\theta \\
\mu & \nu & \s & \lambda & \theta & \rho
\end{pmatrix}  \quad
P_{21} = \begin{pmatrix}
\s &\lambda &\mu &\rho &\nu &\theta \\
\mu &\rho &\s&\theta & \lambda &\nu
\end{pmatrix}\non \\
P_{22} = \begin{pmatrix}
\s &\lambda &\mu &\rho &\nu &\theta \\
\nu & \theta & \rho& \s& \mu& \lambda
\end{pmatrix} \quad
P_{23} = \begin{pmatrix}
\s &\lambda &\mu &\rho &\nu &\theta \\
\rho & \theta & \nu& \mu& \s& \lambda
\end{pmatrix}  \quad
P_{24}=\begin{pmatrix}
\s &\lambda &\mu &\rho &\nu &\theta \\
\lambda & \mu & \theta& \nu& \rho& \s
\end{pmatrix} .\non \\ 
\eea
In \C{listtrans}, $P_1, \cdots, P_{12}$ form $A_4$, the alternating subgroup of $S_4$, while $P_{13}, \cdots, P_{24}$ involve parity.

Note that this symmetry structure also exists for the two loop supergravity amplitude. In that case, the unique skeleton diagram is the two loop ladder skeleton and its dual is the equilateral triangle. The symmetry group is $S_3$. 

Now consider the combinations cubic in the Schwinger parameters
\bea 
\Delta^{(1)} (\s,\lambda,\mu,\rho,\nu,\theta) &=& \s\lambda\mu +\s\mu\theta+ \mu\nu\rho +\s\rho\nu+\lambda\nu\theta+\lambda\theta\rho, \non \\
\Delta^{(2)} (\s,\lambda,\mu,\rho,\nu,\theta) &=& \rho\nu\theta+ \s\mu\rho+\s\mu\nu+\lambda\mu\theta+\s\lambda\theta+\lambda\rho\nu, \non \\ 
\Delta^{(3)} (\s,\lambda,\mu,\rho,\nu,\theta) &=& \mu\nu\theta+\s\rho\theta+\rho\lambda\mu+\lambda\s\nu.
\eea 
Under $P_1, \cdots, P_{12}$, we see that $\Delta^{(1)},\Delta^{(2)}$ and $\Delta^{(3)}$ are individually invariant. Under $P_{13}, \cdots, P_{24}$, on the other hand we have that $\Delta^{(1)} \leftrightarrow \Delta^{(2)}$ while $\Delta^{(3)}$ is invariant.  
Thus from \C{defD}, we see that 
\be \Delta_3 = \Delta^{(1)} +\Delta^{(2)} +\Delta^{(3)}\ee 
is $S_4$ invariant, and hence from \C{defJ} we see that $\mathcal{J}$ is $S_4$ invariant. 
We also see that $F_L$ is invariant from \C{defF} by appropriately renaming the KK momenta, which is determined by drawing the Mercedes skeleton for each case in \C{listtrans}.

\subsection{Expressing the three loop integral in an $S_4$ invariant way}

Having understood the general symmetry structure underlying the Mercedes skeleton, 
 let us consider the expression for $I_3$ in \C{impexp}, which we shall write in a manifestly $S_4$ invariant way. This is a natural way to express the integral, since the 6 Schwinger parameters which are integrated over, enjoy the $S_4$ symmetry as discussed above. Right now it makes the expression cumbersome, but this will be very helpful for us later on. This is because we shall later describe the integrals in terms of integrals over the moduli space of an auxiliary geometry, and we want to parametrize the integrand in terms of the moduli of the auxiliary geometry in an $S_4$ invariant way, which corresponds to the freedom to rename the loop momenta in the Mercedes skeleton. Note that the measure $d\Upsilon$, $F_L$ and $\Delta_3$ in \C{impexp} are $S_4$ invariant, and so we simply need to symmetrize under the $S_4$ action the remaining factors. This is done by looking at the transformations given in \C{listtrans}.   

Thus in \C{impexp}, we make the following replacements inside the integral:

\bea  \lambda^2 \s\nu \frac{\p}{\p\lambda} + \lambda^2 \s\mu\frac{\p}{\p\lambda} +2 \s^2 \lambda\mu \frac{\p}{\p\s} \rightarrow \frac{D_1}{12}, \eea
where
\bea D_1 =
\lambda^2 \Big((\s+\rho)(\mu+\nu)+\theta(\s+\rho+\mu+\nu)\Big) \frac{\p}{\p\lambda}\non \\ +\rho^2 \Big((\s+\lambda)(\theta+\mu)+\nu(\s+\lambda+\theta+\mu)\Big) \frac{\p}{\p\rho}\non \\ +\mu^2 \Big((\nu+\lambda)(\theta+\rho)+\s(\nu+\lambda+\theta+\rho)\Big) \frac{\p}{\p\mu}\non \\ +\s^2 \Big((\nu+\theta)(\lambda+\rho)+\mu(\nu+\theta+\lambda+\rho)\Big) \frac{\p}{\p\s} \non \\ +\nu^2 \Big((\lambda+\mu)(\s+\theta)+\rho(\lambda+\mu+\s+\theta)\Big) \frac{\p}{\p\nu}\non \\ +\theta^2 \Big((\rho+\mu)(\s+\nu)+\lambda(\rho+\mu+\s+\nu)\Big) \frac{\p}{\p\theta},\non \\ \eea

\be \s\lambda\mu\theta \frac{\p}{\p\theta}\rightarrow \frac{D_2}{12}, \ee
where
\bea D_2 &=&\s\lambda\mu\theta \Big(\frac{\p}{\p\s} +\frac{\p}{\p\lambda} +\frac{\p}{\p\mu} +\frac{\p}{\p\theta}\Big) +\theta\rho\lambda\nu \Big(\frac{\p}{\p\theta} +\frac{\p}{\p\rho} +\frac{\p}{\p\lambda} +\frac{\p}{\p\nu}\Big) \non \\ &&+\mu\nu\rho\s\Big(\frac{\p}{\p\mu} +\frac{\p}{\p\nu} +\frac{\p}{\p\rho} +\frac{\p}{\p\s}\Big), \eea

\be \s\lambda\rho\nu \frac{\p}{\p\s} \rightarrow \frac{D_3}{24},\ee
where
\bea D_3 = \s\lambda\rho \Big( (\nu+\theta)\frac{\p}{\p\s} + (\mu+\nu) \frac{\p}{\p\lambda} +(\theta+\mu) \frac{\p}{\p\rho}  \Big) \non \\ + \lambda\mu\nu \Big( (\s+\theta) \frac{\p}{\p\nu} + (\rho+\theta)\frac{\p}{\p\mu} + (\rho+\s)\frac{\p}{\p\lambda}\Big) \non \\ + \nu\theta\s \Big((\mu+\rho) \frac{\p}{\p\theta}+ (\rho+\lambda)\frac{\p}{\p\s}+(\lambda+\mu)\frac{\p}{\p\nu}\Big) \non \\ +\mu\rho\theta \Big((\lambda+\s)\frac{\p}{\p\rho} +(\s+\nu)\frac{\p}{\p\theta} +(\nu+\lambda)\frac{\p}{\p\mu}\Big),\eea 

and
\be \s\lambda\rho\nu \frac{\p}{\p\rho} \rightarrow \frac{D_4}{12},\ee

where
\bea D_4 = \s\lambda\rho \Big(\nu\frac{\p}{\p\rho} +\mu\frac{\p}{\p\s} +\theta \frac{\p}{\p\lambda}\Big) + \nu\theta\s \Big(\rho\frac{\p}{\p\nu} +\mu \frac{\p}{\p\s} + \lambda\frac{\p}{\p\theta}\Big)\non \\ + \lambda\mu\nu \Big(\s\frac{\p}{\p\mu}+\theta\frac{\p}{\p\lambda} +\rho\frac{\p}{\p\nu}\Big) +\mu\rho\theta \Big(\lambda\frac{\p}{\p\theta} +\s\frac{\p}{\p\mu} +\nu \frac{\p}{\p\rho}\Big).\eea

Here $D_1, D_2,D_3$ and $D_4$ are $S_4$ invariant by construction. Now
\be \Big(D_1 + D_2 +D_3 +D_4\Big)\Delta_3 = 3 \Delta_3^2,\ee
resulting in an enormous simplification. 
Thus from \C{impexp}, we get that
\bea \label{impexp2} I_3 &=& \frac{5\pi^{27/2}}{6(4\pi^2 l_{11}^2 \mathcal{V}_2)^3} \s_3 \int_0^\infty d\Upsilon \Delta_3^{-7/2} F_L .\eea
The final answer for the three loop amplitude takes a very simple form. 

\section{The unrenormalized three loop amplitude at $O(D^6\mathcal{R}^4)$}

We now want to simplify the three loop integral given by \C{impexp2} without worrying about the ultraviolet divergences by bringing it to a tractable form. 
As in the case of the one and two loop amplitudes, it is very convenient to perform a Poisson resummation in $F_L$ to go from KK modes to winding modes. In particular, the ultraviolet divergent structure is manifest in the Poisson resummed variables. 

First let us consider the lattice factor $F_L$. We write \C{defF} in a compact way as
\be \label{newdefF} F_L = \sum_{k_{\alpha I}} e^{-G^{IJ}G^{\alpha\beta} k_{\alpha I}k_{\beta J}/l_{11}^2},\ee
where the KK integers $k_{\alpha I}$ are defined by $k_{\alpha I} = \{l_I,m_I,n_I\}$ for $\alpha =1,2,3$. In \C{newdefF}, the symmetric matrix $G^{\alpha\beta}$ ($\alpha,\beta =1,2,3$) has entries (of dimension $l_{11}^2$)
\be \label{inverse}G^{\alpha\beta} = \begin{pmatrix}
\s +\rho+\theta &\rho+\theta &\theta  \\
\rho +\theta & \lambda+\rho+\nu+\theta & \nu+\theta \\
\theta & \nu+\theta & \mu+\nu+\theta
\end{pmatrix}.\ee    
Note that ${\rm det} G^{\alpha\beta} = \Delta_3$. After Poisson resummation, we get that
\be \label{defF2} F_L = \frac{(\pi l_{11}^2 \mathcal{V}_2)^3}{\Delta_3}\sum_{\hat{k}^{\alpha I}} e^{-\pi^2 l_{11}^2 G_{IJ}G_{\alpha\beta} \hat{k}^{\alpha I} \hat{k}^{\beta J}}, \ee
where the winding mode integers $\hat{k}^{\alpha I}$ are defined by $\hat{k}^{\alpha I} = \{\hat{l}^I,\hat{m}^I,\hat{n}^I\}$ for $\alpha =1,2,3$. Also the matrix $G_{\alpha\beta}$ is the inverse of the matrix $G^{\alpha\beta}$, and has entries of dimension $l_{11}^{-2}$. 

We also want to redefine the Schwinger parameters after performing the Poisson resummation. So we define
\be \label{newS1}\hat\s = \frac{\s}{\Delta_3^{2/3}}, \quad \hat\lambda  = \frac{\lambda}{\Delta_3^{2/3}}, \quad \hat\mu  = \frac{\mu}{\Delta_3^{2/3}}, \quad \hat\rho  = \frac{\rho}{\Delta_3^{2/3}}, \quad \hat\nu  = \frac{\nu}{\Delta_3^{2/3}}, \quad \hat\theta  = \frac{\theta}{\Delta_3^{2/3}}, \ee
and so
\bea \label{newS2}\hat\Delta_3 (\hat\s,\hat\lambda,\hat\mu,\hat\rho,\hat\nu,\hat\theta) &=& \hat\s \hat\lambda \hat\mu +\hat\rho\hat\nu\hat\theta + \hat\s\hat\mu (\hat\rho+\hat\nu+\hat\theta) +\hat\lambda\hat\mu (\hat\rho+\hat\theta) +\hat\s\hat\lambda (\hat\nu+\hat\theta)\non \\ &&+ \hat\mu\hat\nu(\hat\rho+\hat\theta) +\hat\s\hat\rho(\hat\nu+\hat\theta) +\hat\lambda(\hat\rho\hat\nu+\hat\nu\hat\theta+\hat\rho\hat\theta) \non \\ &=& \Delta_3^{-1} (\s,\lambda,\mu,\rho,\nu,\theta).\eea
 Thus the variables $\hat\s,\hat\lambda,\hat\mu,\hat\rho,\hat\nu$ and $\hat\theta$ have dimensions of $l_{11}^{-2}$. We can now express the matrix $G_{\alpha\beta}$ in terms of the variables defined in \C{newS1} and \C{newS2}\footnote{The entries are messy. For example, 
\be G_{13} = \frac{\hat\rho\hat\nu-\hat\lambda\hat\theta}{\hat{\Delta}_3^{1/3}}.\ee}, however we shall not need the explicit details in this form. Finally we note that the measure transforms in a very simple way as
\be d\Upsilon = \hat{\Delta}_3^{-4} d\hat\s d\hat\lambda d\hat\mu d\hat\rho d\hat\nu d\hat\theta \equiv \hat{\Delta}_3^{-4} d\hat\Upsilon,\ee 
on using the relation
\be \Big( \s\frac{\p}{\p\s} + \lambda \frac{\p}{\p\lambda} +\mu \frac{\p}{\p\mu} +\rho \frac{\p}{\p\rho} +\nu \frac{\p}{\p\nu} +\theta \frac{\p}{\p\theta}\Big) \Delta_3 = 3 \Delta_3.\ee
Thus from \C{impexp2}, it follows that
\be \label{impexp3}
I_3 = \frac{5 \pi^{21/2}}{6\cdot 64}   \s_3 \int_0^\infty d\hat\Upsilon \hat{\Delta}_3^{1/2} \sum_{\hat{k}^{\alpha I}} e^{-\pi^2 l_{11}^2 G_{IJ}G_{\alpha\beta} \hat{k}^{\alpha I} \hat{k}^{\beta J}}.\ee

\subsection{An underlying auxiliary geometry}

The expression for the three loop amplitude given in \C{impexp3} is a rather complicated one. However, proceeding by analogy with the two loop amplitude, there is a natural guess to simplify the expression. This heuristic counting of numbers as explained below proves to be a very helpful guide for our analysis.

\begin{figure}[ht]
\begin{center}
\[
\mbox{\begin{picture}(120,90)(0,0)
\includegraphics[scale=.5]{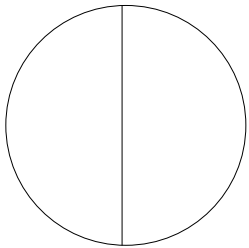}
\end{picture}}
\]
\caption{The two loop ladder skeleton}
\end{center}
\end{figure}

At two loops, there is only one skeleton diagram--the ladder skeleton as shown in figure 14, and hence there are 3 Schwinger parameters.  
The lattice factor $F_L$ in \C{2PLANAR} and \C{2NPLANAR} involves a sum over two sets of integers $m_I$ and $n_I$. In the two loop analysis, the 3 Schwinger parameters were traded off for the parameters of an auxiliary $T^2$, having moduli its volume, and complex structure. The Poisson resummed integers $\hat{m}_I$ and $\hat{n}_I$ now correspond to non--trivial winding numbers along the two non--contractible cycles of $T^2$. This non--trivial fact which was extremely useful for the two loop analysis has an obvious generalization for our case. In fact, what was crucial was that the integral over the Schwinger parameters included an integral over the complex structure moduli of $T^2$, which had precisely the $SL(2,\mathbb{Z})$ invariant measure and was integrated over 3 copies of the fundamental domain of $SL(2,\mathbb{Z})$. Renormalizing the UV divergences came from the boundary contributions of moduli space.    

We have a suggestive similar structure in our case. The Mercedes skeleton has 6 Schwinger parameters, while $F_L$ in \C{defF} involves a sum over three sets of integers $l_I$, $m_I$ and $n_I$. It is natural to assume that the 6 parameters can be traded off for the 6 moduli of an auxiliary $T^3$, of which one is the volume modulus and the rest are shape moduli. The Poisson resummed integers $\hat{l}_I$, $\hat{m}_I$ and $\hat{n}_I$ would then correspond to the winding numbers along the three non--contractible cycles of $T^3$. 

The lattice factors in \C{newdefF} and \C{defF2} support this assumption, where we identify $G_{\alpha\beta}$ with the (dimensionful) metric of the $T^3$. We now explicitly map the Schwinger parameters to the moduli of $T^3$ and show that the $SL(3,\mathbb{Z})$ invariant measure on the moduli space of the shape moduli of $T^3$ arises from the measure in \C{impexp3}.   

\subsection{Mapping the Schwinger parameters to the auxiliary $T^3$}

The three dimensional torus $T^3$ is parametrized by 6 parameters. One of them is its volume $V_3$. The remaining 5 are its shape moduli which we denote $L,T_1,T_2,A_1$ and $A_2$, which parametrize the maximally symmetric coset space $SO(3)\backslash SL(3,\mathbb{R})$ with the inverse vielbein given by \C{T1}. Thus the inverse metric on $T^3$ of volume $V_3$ is given by $G^{\alpha\beta} = l_{11}^2 V_3^{-2/3}g^{\alpha\beta}$, where $g^{\alpha\beta}$ is given by \C{T2}. We set the entries of $G^{\alpha\beta}$ to have dimension $l_{11}^2$ to match the entries in \C{inverse} as $L,T_1,T_2,A_1,A_2$ and $V_3$ are dimensionless. Thus
\be G^{\alpha\beta} = l_{11}^2 V_3^{-2/3}\begin{pmatrix}
1/L^2 & A_1/L^2 & A_2/L^2  \\
A_1/L^2 & A_1^2/L^2 +L/T_2 & A_1A_2/L^2 +LT_1/T_2 \\
A_2/L^2 & A_1A_2/L^2 +LT_1/T_2 & A_2^2/L^2 + L\vert T \vert^2/T_2,
\end{pmatrix}\ee  
which on matching with \C{inverse}, using \C{newS1} and \C{newS2} gives us that
\bea \label{invert}
l_{11}^2 \hat\s &=& \frac{1-A_1}{L^2} V_3^{2/3} ,\non \\ l_{11}^2 \hat\lambda &=& \Big[(1-T_1)\frac{L}{T_2} + \frac{(A_1 -1)(A_1 - A_2)}{L^2} \Big]V_3^{2/3}, \non \\ l_{11}^2 \hat\mu &=& \Big[(\vert T \vert^2 - T_1) \frac{L}{T_2} + \frac{A_2 (A_2 - A_1)}{L^2} \Big]V_3^{2/3},\non \\ l_{11}^2 \hat\rho &=& \frac{A_1 - A_2}{L^2} V_3^{2/3}, \non \\ l_{11}^2 \hat\nu &=& \Big[\frac{LT_1}{T_2} + \frac{A_2 (A_1 -1)}{L^2}\Big] V_3^{2/3}, \non \\  l_{11}^2 \hat\theta &=& \frac{A_2}{L^2} V_3^{2/3},\eea
and
\be \label{invert2} {\hat\Delta}_3 = \frac{V_3^2}{l_{11}^6}.\ee
Thus from \C{impexp3} we see that 
\be \label{measure}d\hat\Upsilon \hat{\Delta}_3^{1/2}  \sum_{\hat{k}^{\alpha I}} e^{-\pi^2 l_{11}^2 G_{IJ}G_{\alpha\beta} \hat{k}^{\alpha I} \hat{k}^{\beta J}} = \frac{4V_3^4}{l_{11}^{15}L^4 T_2^2} dV_3 dL dT_1 dT_2 dA_1 dA_2 \sum_{\hat{k}^{\alpha I}} e^{-\pi^2 \mathcal{V}_2 V_3^{2/3} \hat{G}_{IJ} \hat{G}_{\alpha\beta} \hat{k}^{\alpha I} \hat{k}^{\beta J}},\ee
where ${G}_{IJ} = \mathcal{V}_2 \hat{G}_{IJ}$ from \C{defmet}, and 
\be \label{needdef}
\hat{G}_{\alpha\beta} = \begin{pmatrix}
L^2 +\vert A_1 T - A_2 \vert^2/LT_2 & (A_2 T_1 - A_1 \vert T\vert^2)/LT_2 & (A_1 T_1 -A_2)/LT_2  \\
(A_2 T_1 - A_1 \vert T\vert^2)/LT_2 & \vert T\vert^2/LT_2 & -T_1/LT_2 \\
 (A_1 T_1 -A_2)/LT_2  & -T_1/LT_2& 1/LT_2,
\end{pmatrix}.\ee 
While the renaming of the Schwinger parameters in terms of the $T^3$ moduli automatically puts $F_L$ in the desired form, the relation \C{measure} for the transformation of the measure $d\hat\Upsilon$ is crucial for our purposes. Note that in \C{measure} the lattice factor has manifest $SL(3,\mathbb{Z})$ symmetry.

The dependence of the measure on the shape moduli of the $T^3$ is 
\be d\mu \equiv \frac{1}{L^4 T_2^2} dL dT_1 dT_2 dA_1 dA_2\ee  
which is precisely equal to \C{Measure}, hence making the auxiliary $T^3$ manifest. Thus the three loop amplitude \C{impexp3} is manifestly $SL(3,\mathbb{Z})$ invariant\footnote{This has also been argued in~\cite{Pioline:2014bra}.}.
This is analogous to the two loop calculation, where the leading contribution involving the $D^4 \mathcal{R}^4$ term is given by an $SL(2,\mathbb{Z})$ invariant integral over the moduli space of the auxiliary $T^2$. 

Thus the three loop amplitude \C{impexp3} involves a map from the auxiliary $T^3$ to the target space $T^2$. We would now like to understand the integral over the moduli space. In the case of the two loop amplitude, the measure was mapped to 3 copies of the fundamental domain of $SL(2,\mathbb{Z})$.
We would like to perform the analogous calculation for our case.    

\subsection{Maps from different regions of the space of Schwinger parameters to the fundamental domain of $SL(3,\mathbb{Z})$}

From \C{invert} and \C{invert2}, we have that
\bea \label{map2}&&T_1 = \frac{\hat\theta(\hat\s +\hat\nu) +\hat\nu(\hat\s+\hat\rho)}{\hat\s(\hat\rho+\hat\theta)+(\hat\lambda+\hat\nu)(\hat\s +\hat\rho+\hat\theta)}, \quad T_2 = \frac{{\hat\Delta}_3^{1/2} \sqrt{\hat\s+\hat\rho+\hat\theta}}{\hat\s(\hat\rho+\hat\theta)+(\hat\lambda+\hat\nu)(\hat\s +\hat\rho+\hat\theta)}, \non \\ &&A_1 = \frac{\hat\rho+\hat\theta}{\hat\s+\hat\rho+\hat\theta}, \quad A_2 = \frac{\hat\theta}{\hat\s+\hat\rho+\hat\theta}, \quad L= \frac{{\hat\Delta}_3^{1/6}}{\sqrt{\hat\s+\hat\rho+\hat\theta}}, \quad V_3 = l_{11}^3 \hat{\Delta}_3^{1/2},\eea
leading to the constraints
\bea \label{sl3}
&&0 \leq A_2 \leq A_1 \leq 1, \quad 0 \leq T_1 \leq 1, \non \\&& \Big\vert T -\frac{1}{2}\Big\vert^2  = \frac{1}{4} + \frac{\hat\mu(\hat\s +\hat\rho) +\hat\theta(\hat\mu+\hat\rho)}{\hat\s(\hat\rho+\hat\theta)+(\hat\lambda+\hat\nu)(\hat\s +\hat\rho+\hat\theta)} \geq \frac{1}{4}, \non \\ &&\Big(A_1 - \frac{1}{2}\Big)^2 + \frac{L^3}{T_2} = \frac{1}{4} + \frac{\hat\lambda+\hat\nu}{\hat\s+\hat\rho+\hat\theta} \geq \frac{1}{4} , \non \\ &&\Big(A_2 -\frac{1}{2}\Big)^2 + \frac{L^3}{T_2} \vert T \vert^2 = \frac{1}{4} + \frac{\hat\mu +\hat\nu}{\hat\s+\hat\rho+\hat\theta} \geq \frac{1}{4} , \non \\ &&\Big(A_1 -A_2 -\frac{1}{2}\Big)^2 +\frac{L^3}{T_2}\vert T-1\vert^2 =\frac{1}{4} + \frac{\hat\lambda+\hat\mu}{\hat\s+\hat\rho+\hat\theta} \geq \frac{1}{4}.\eea
Before proceeding, notice the similarity between \C{sl3} and \C{SL3}. This gives us a strong hint about how to proceed. 

\begin{figure}[ht]
\begin{center}
\[
\mbox{\begin{picture}(150,100)(0,0)
\includegraphics[scale=.55]{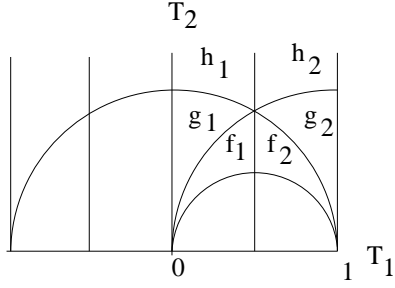}
\end{picture}}
\]
\caption{The $T$ plane}
\end{center}
\end{figure}

In \C{sl3}, $T$ lies in the region depicted in figure 15. Thus the relevant region is $f_1 \oplus f_2 \oplus g_1 \oplus g_2 \oplus h_1 \oplus h_2$. Each of these regions can be mapped to the standard fundamental domain of $GL(2,\mathbb{Z})$ given by $\vert T \vert^2 \geq 1, 0 \leq T_1 \leq 1/2$ which is the region $h_1$. This is done using the $S$ and $T$ transformations which send $T\rightarrow -1/T$ and $T\rightarrow T+1$ respectively, along with $ T \rightarrow -\bar{T}$, which is implemented by the $GL(2,\mathbb{Z})$ matrix
\be K = \begin{pmatrix}
-1 &  0 \\
0 & 1  \\
\end{pmatrix}.\ee 

\begin{figure}[ht]
\begin{center}
\[
\mbox{\begin{picture}(150,100)(0,0)
\includegraphics[scale=.55]{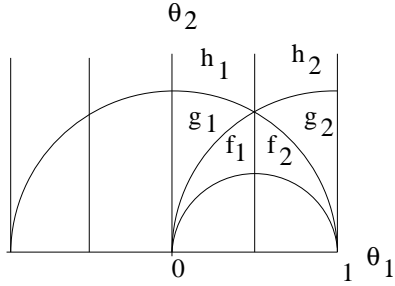}
\end{picture}}
\]
\caption{The $\Theta$ plane}
\end{center}
\end{figure}

It is also very useful for our purposes to consider the entries in the 1--2 and 1--3 blocks in \C{T2}.
Defining
\be \Theta = A_1 + i\sqrt{L^3/T_2},\quad \lambda^{-1} = \sqrt{L T_2}\ee  
we see that the entries in the 1--2 block give the inverse metric 
\be \frac{\lambda}{\Theta_2} \begin{pmatrix}
1 &  \Theta_1 \\
\Theta_1& \vert \Theta \vert^2  \\
\end{pmatrix} \ee 
of an auxiliary $T^2$ in the auxiliary $T^3$ of volume $\lambda^{-1}$. From \C{sl3}, we get that $\Theta$ satisfies
\be 0 \leq \Theta_1 \leq 1, \quad \Big\vert \Theta - \frac{1}{2} \Big\vert^2 \geq \frac{1}{4},\ee
depicted by figure 16. Thus the relevant region is $f_1 \oplus f_2 \oplus g_1 \oplus g_2 \oplus h_1 \oplus h_2$, which can be mapped into the fundamental domain of $GL(2,\mathbb{Z})_\Theta$ using $S, T$ and $K$.

\begin{figure}[ht]
\begin{center}
\[
\mbox{\begin{picture}(150,100)(0,0)
\includegraphics[scale=.55]{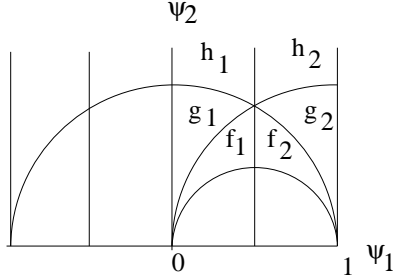}
\end{picture}}
\]
\caption{The $\Psi$ plane}
\end{center}
\end{figure}

Similarly consider the entries in the 1--3 block in \C{T2}. Defining
\be \Psi = A_2 + i\sqrt{L^3/T_2} \vert T\vert,\quad \lambda^{-1} = \sqrt{L T_2}\ee  
we see that this part gives the inverse metric 
\be \frac{\lambda}{\Psi_2} \begin{pmatrix}
1 &  \Psi_1 \\
\Psi_1& \vert \Psi \vert^2  \\
\end{pmatrix}, \ee 
of another auxiliary $T^2$ in the auxiliary $T^3$ of volume $\lambda^{-1}$. From \C{sl3}, we get that $\Psi$ satisfies
\be 0 \leq \Psi_1 \leq 1, \quad \Big\vert \Psi - \frac{1}{2} \Big\vert^2 \geq \frac{1}{4},\ee
depicted by figure 17. Again the relevant region is $f_1 \oplus f_2 \oplus g_1 \oplus g_2 \oplus h_1 \oplus h_2$, which can be mapped into the fundamental domain of $GL(2,\mathbb{Z})_\Psi$ using $S, T$ and $K$.
Note that \C{sl3} implies that $\Theta_1 \geq \Psi_1$.

\begin{figure}[ht]
\begin{center}
\[
\mbox{\begin{picture}(150,120)(0,0)
\includegraphics[scale=.55]{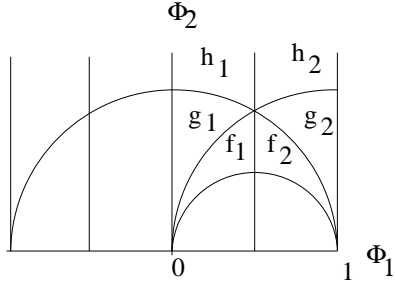}
\end{picture}}
\]
\caption{The $\Phi$ plane}
\end{center}
\end{figure}

It is also convenient for our purposes to define
\be \Phi = (A_1 - A_2) + i\sqrt{L^3/T_2} \vert T-1\vert.\ee
Thus from \C{sl3} we get that 
\be  0\leq \Phi_1 \leq 1, \quad \Big\vert \Phi - \frac{1}{2} \Big\vert^2 \geq \frac{1}{4}, \ee 
which is the region in the figure 18. Thus the relevant region is $f_1 \oplus f_2 \oplus g_1 \oplus g_2 \oplus h_1 \oplus h_2$. Clearly, the transformations among these various regions in the various figures are not independent, but are constrained by \C{sl3}.

\begin{figure}[ht]
\begin{center}
\[
\mbox{\begin{picture}(150,100)(0,0)
\includegraphics[scale=.55]{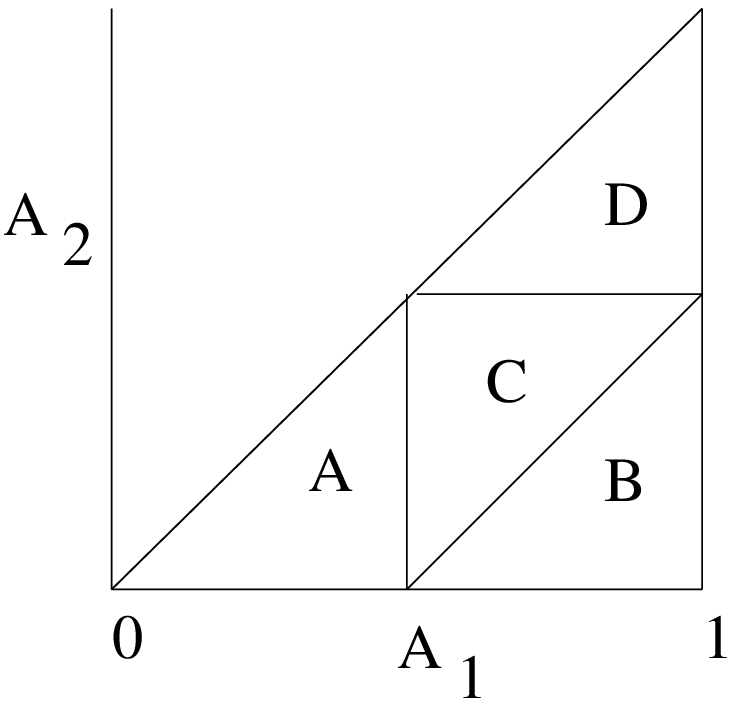}
\end{picture}}
\]
\caption{The $A_1$--$A_2$ region in \C{sl3}}
\end{center}
\end{figure}

First let us show that a part of the fundamental domain of $SL(3,\mathbb{Z})$ is trivially included in the regions we have described. Consider the region $A$ in fig 19, and the region $h_1$ in figures 15, 16, 17 and 18. This is given by
\bea \label{rule1} &&0 \leq A_1 , A_2 ,T_1 \leq \frac{1}{2}, \quad A_1 \geq A_2, \quad \vert T \vert^2 \geq 1, \non \\ &&A_1^2 + \frac{L^3}{T_2} \geq 1, \quad A_2^2 +\frac{L^3}{T_2} \vert T\vert^2 \geq 1, \quad (A_1 - A_2)^2 +\frac{L^3}{T_2} \vert T-1 \vert^2 \geq 1.  \eea
To these we can also add the relation
\be \label{rule2} (A_1 - A_2 - 1)^2 +\frac{L^3}{T_2} \vert T-1 \vert^2 \geq 1\ee   
which trivially follows from figure 18. From \C{SL3}, it follows that \C{rule1} and \C{rule2} covers exactly $1/4$th of $\mathcal{F}_3$, which is the region $b$ in figure 20. We call this the identity transformation. 

\begin{figure}[ht]
\begin{center}
\[
\mbox{\begin{picture}(150,120)(0,0)
\includegraphics[scale=.7]{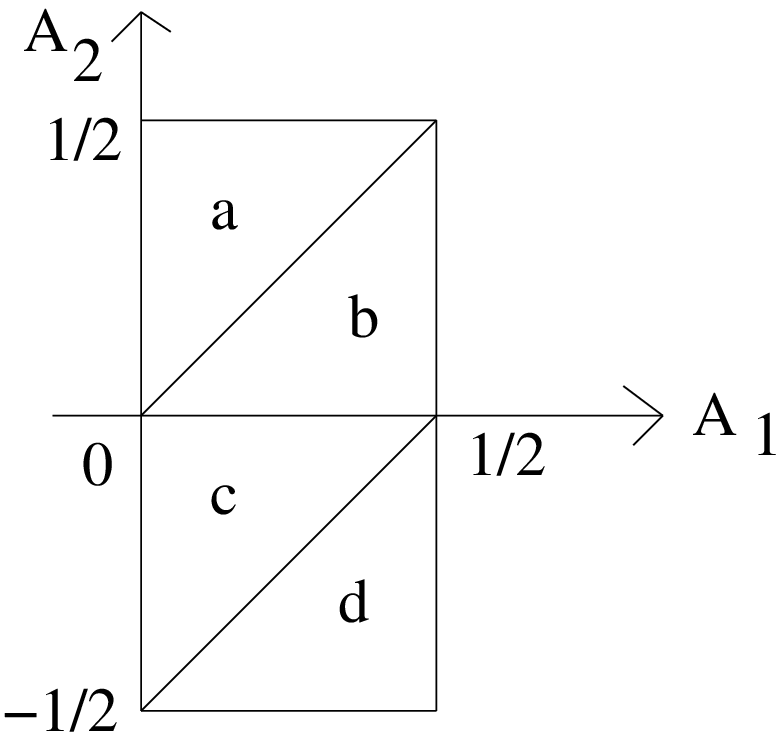}
\end{picture}}
\]
\caption{The $A_1$--$A_2$ region in \C{SL3}}
\end{center}
\end{figure}

We now demonstrate 4 non--trivial transformations which map 4 different regions of the space of the 6 Schwinger parameters into the fundamental domain $\mathcal{F}_3$. 

{\bf{1}}. Consider the region $A$ in figure 19, the region $h_1$ in figures 16, 17 and 18, and the region $g_1$ in figure 15. Now act with the $SL(3,\mathbb{Z})$ transformation
\be  \begin{pmatrix}
-1 &  0 & 0 \\
0 & 0 & -1  \\
0 & -1 & 0\\
\end{pmatrix}, \ee 
which acts on the moduli as 
\be \label{map}(L,A_1,A_2) \rightarrow (L,A_2,A_1), \quad T \rightarrow \bar{T}^{-1}. \ee
The transformed variables satisfy \C{SL3} in the region $b$ in figure 20. This covers the same patch of $\mathcal{F}_3$ as the identity transformation we have mentioned above. 

{\bf{2}}. Consider the region $B$ in figure 19, the region $h_1$ in figures 15 and 17, and the region $h_2$ in figures 16 and 18. Now act with the $SL(3,\mathbb{Z})$ transformation
\be  \label{trans1}\begin{pmatrix}
1 &  1 & 0 \\
0 & -1 & 0  \\
0 & 0 & -1\\
\end{pmatrix}, \ee 
which acts on the moduli as 
\be \label{change}(L,A_1,A_2,T) \rightarrow (L,1-A_1,-A_2,T). \ee
The transformed variables satisfy \C{SL3} in the region $c$ in figure 20.

{\bf{3}}. Consider the region $A$ in figure 19, the region $g_2$ in figure 15, and the region $h_1$ in figures 16, 17 and 18. Now act with the $SL(3,\mathbb{Z})$ transformation
\be  \begin{pmatrix}
1 &  0 & 0 \\
0 & 1 & 1  \\
0 & -1 & 0\\
\end{pmatrix}, \ee 
which acts on the moduli as 
\be (L,A_1,A_2) \rightarrow (L,A_1 -A_2,A_1), \quad T \rightarrow 1/(1-T). \ee

The transformed variables satisfy \C{SL3} in the region $a$ in figure 20.

{\bf{4}}. Consider the region $C$ in figure 19, the region $h_2$ in figure 16, and the region $h_1$ in figures 15, 17 and 18. Acting with the $SL(3,\mathbb{Z})$ transformation \C{trans1} which results in \C{change}, we see that the transformed variables satisfy \C{SL3} in the region $d$ in figure 20. 

Thus the above four transformations map the respective regions to cover exactly a single copy of $\mathcal{F}_3$. 

Of course, from the various diagrams it is clear that there are many regions we have to consider. In particular, we have to see whether each such region can be mapped into a patch of $\mathcal{F}_3$ or not. Finding appropriate $SL(3,\mathbb{Z})$ transformations to determine whether each region maps to some patch of $\mathcal{F}_3$ is not very simple. To see this, note that one can easily perform a  $GL(2,\mathbb{Z})_T$ transformation to put $T$ into the form of the defining equations involving only $T$ in $\mathcal{F}_3$, but then the other variables mix amongst each other. This is also true for $GL(2,\mathbb{Z})_\Theta$ and $GL(2,\mathbb{Z})_\Psi$ transformations as well. Thus one needs to perform more involved $SL(3,\mathbb{Z})$ transformations\footnote{In \C{listmat}, $A1$ and $A2$ simply shift $A_1$ and $A_2$ by 1 respectively; $T1, T2$ and $T3$ implement $GL(2,\mathbb{Z})_T$ transformations; and $U1$ sends $A_1 (A_2)$ to $-A_1 (-A_2)$. The involved $SL(3,\mathbb{Z})$ transformations involve $S1, S2, S3$ and $S4$,}. Given the complicated nature of these manipulations, we shall proceed with a simple assumption about the final structure. We assume that each region does map into a patch of $\mathcal{F}_3$. From diagrams 15, 16, 17, 18 and 19, we see that there are $3^3 \times 6 \times 4$ such regions. We assume that $\mathcal{F}_3$ is covered exactly once by 4 such regions. Thus we get $3^3 \times 6  = 162$ copies of $\mathcal{F}_3$. This assumption seems plausible to us because we started with an integral over the 6 Schwinger parameters which reduced to an $SL(3,\mathbb{Z})$ invariant measure on the shape moduli space of $SO(3) \backslash SL(3,\mathbb{R})$. Thus is is natural that the integral of the measure reduces to an integral of the measure over multiple copies of $\mathcal{F}_3$.  Also though finding explicit $SL(3,\mathbb{Z})$ transformations is difficult, we expect the assumption to be correct, because all the regions are related to each other by $SL(3,\mathbb{Z})$ transformations and we could choose 4 simple regions to cover $\mathcal{F}_3$ once.    

What if the assumption is not true? In that case, some of the regions map to $\mathcal{F}_3$, while the others do not, and $I_3$ in \C{impexp3} receives contributions from both. Then our analysis does give the partial contribution to the $D^6\mathcal{R}^4$ interaction (upto an overall undetermined factor which counts the number of times $\mathcal{F}_3$ is covered). This leaves undetermined the remaining contribution. Thus even in this case, we can partially fix the $D^6\mathcal{R}^4$ interaction.  However, the results we obtain later on support our assumption that the final integral is indeed over $\mathcal{F}_3$, though exactly how many times $\mathcal{F}_3$ is covered will not matter to us.

Proceeding with our assumption, we get that
\be \label{usemain} I_3 = \frac{5\cdot 27 \pi^{21/2}}{16 l_{11}^{15}} \s_3 \int_0^\infty d V_3 V_3^4 \int_{\mathcal{F}_3} d\mu   \sum_{\hat{k}^{\alpha I}} e^{-\pi^2 \mathcal{V}_2 V_3^{2/3} \hat{G}_{IJ}\hat{G}_{\alpha\beta} \hat{k}^{\alpha I} \hat{k}^{\beta J}} .\ee

\section{Renormalization of the three loop amplitude at $O(D^6\mathcal{R}^4)$}

We now consider the structure of the three loop amplitude that arises from \C{usemain}, in particular its ultraviolet divergences. Apart from the primitive three loop $\Lambda^{15}$ divergence, there are also one and two loop divergences to this amplitude. We first systematically isolate the one and two loop divergences, and then construct the one and two loop counterterms that cancel these divergences, which are easier to calculate than the three loop amplitude directly. These counterterms are multiplied by moduli dependent coefficients which depend on $\Omega$ and $\mathcal{V}_2$ in a precise way which we determine. Demanding the cancellation of these divergences upto a finite part, this uniquely fixes the moduli dependence of the corresponding part of the three loop amplitude.

\subsection{The one loop divergences and counterterms}

First let us analyze the one loop divergences that follow from figure 8, where the various numerators are given by \C{num}. In the expressions for $N^{(e)}, N^{(f)}, N^{(g)}, N^{(h)}$ and $N^{(i)}$ that are relevant for us, all but the last three terms in $N^{(i)}$ have the universal property that at $O(k^6)$, two powers of the external momenta are contracted with the loop momenta. For the last three terms in $N^{(i)}$, all powers of the external momenta have already factorized in each term out of the loop momenta. We first focus on the contributions coming from these universal terms.  

Consider the divergences that arise from figure $e$. There are three contributions coming from sending each loop momentum to infinity separately, and all of them diverge as $\Lambda^3$. From figure 8 and from \C{fige}, it follows that two of them have to be renormalized using a four point counterterm, and the third one by a five point counterterm. In 11 uncompactified dimensions, the total contribution from $I^{(e)}$ to the former divergence is given by
\be \label{div1} -8 S^2 \int \frac{d^{11}r}{r^8} \int d^{11} p \int d^{11} q \frac{(k_3 \cdot q)(k_4 \cdot q)}{q^6 p^4 (p+q)^2},\ee
and to the later divergence is given by
\be \label{div2} -4 S^2  \int d^{11}r \frac{(k_3 \cdot r)(k_4 \cdot r)}{r^{10}} \int d^{11} p \int d^{11} q \frac{1}{q^4 p^4 (p+q)^2}.\ee      
First let us consider \C{div1}. 
Compactifying on $T^2$, we renormalize this divergence by adding the one loop four point $\mathcal{R}^4$ counterterm\footnote{Two loop divergences have been cancelled in exactly this way in \C{ren2} and \C{ren3}. This is the generalization of this analysis to three loops.} as before. Thus from \C{totcont}, the total counterterm contribution to the amplitude is then given by 
\be \label{div3} \delta \mathcal{A}_4^{(3)} = \frac{(4\pi^2)^2 \kappa_{11}^8}{(2\pi)^{33}} \mathcal{K} \s_3 \cdot \frac{16}{9} \cdot \frac{\pi^3}{12 l_{11}^3}c_1 \cdot \mathcal{I}_1,\ee
where
\be \mathcal{I}_1 =-\frac{\pi^9}{2(4\pi^2 l_{11}^2 \mathcal{V}_2)^2} \sum_{m_I,n_I}\int_0^\infty d\lambda d\s d\rho \s^2 \lambda e^{-G^{IJ}\Big(\s m_I m_J +\lambda n_I n_J +\rho(m+n)_I (m+n)_J\Big)/l_{11}^2} \frac{\p \Delta^{-9/2}}{\p\s},\ee
where $\Delta$ is defined in \C{Delta}.
This integral and the ones below are done along the lines of earlier calculations and so we only give the answers.
This counterterm is depicted by $a$ in figure 21.  

\begin{figure}[ht]
\begin{center}
\[
\mbox{\begin{picture}(210,100)(0,0)
\includegraphics[scale=.6]{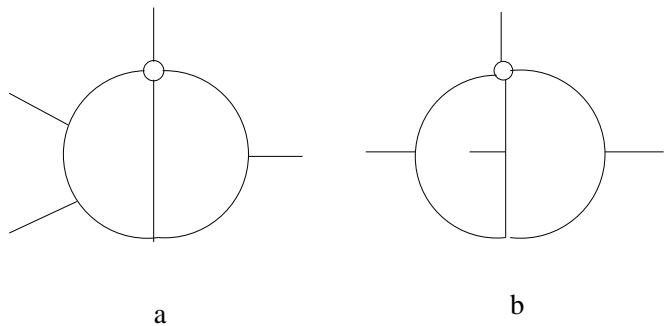}
\end{picture}}
\]
\caption{The planar and non--planar one loop four point counterterm diagrams}
\end{center}
\end{figure}

We next renormalize \C{div2} by adding a five point, one loop counterterm with coefficient proportional to $\hat{c}_1$,\footnote{Candidate counterterms, for example, are the $\mathcal{R}^3 F_5^2$ and $\mathcal{R}^3 G_3 G_3^*$ interactions. They should lie in the same supermultiplet as the $\mathcal{R}^4$ interaction, and hence have a finite piece proportional to $\zeta (2)$. This follows from the fact that these interactions have $E_{3/2} (\Omega,\bar\Omega)$ as their coefficient~\cite{Green:1998by} in 10 dimensions.} leading to the contribution 
\be \label{div4} \delta \mathcal{A}_4^{(3)} =  \frac{(4\pi^2)^2 \kappa_{11}^8}{(2\pi)^{33}} \mathcal{K} \s_3 \cdot \frac{\pi^3\hat{c}_1}{ l_{11}^3}  \cdot \mathcal{I}_2\ee
where
\be \mathcal{I}_2 = \frac{\pi^9}{(4\pi^2 l_{11}^2 \mathcal{V}_2)^2} \sum_{m_I,n_I}\int_0^\infty d\lambda d\s d\rho \s \lambda e^{-G^{IJ}\Big(\s m_I m_J +\lambda n_I n_J +\rho(m+n)_I (m+n)_J\Big)/l_{11}^2} \Delta^{-9/2}.\ee  
 This counterterm is depicted by figure 22. There are several other contributions from the other diagrams of this type. However, the overall coefficient will not be important to us, and we shall simply call the total coefficient $\hat{c}_1$.

\begin{figure}[ht]
\begin{center}
\[
\mbox{\begin{picture}(120,100)(0,0)
\includegraphics[scale=.6]{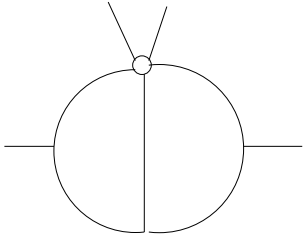}
\end{picture}}
\]
\caption{The one loop five point counterterm diagram}
\end{center}
\end{figure}

Proceeding similarly, from $f$ in figure 8 and \C{figf}, we see that the four point counterterm contributes
\be  \delta \mathcal{A}_4^{(3)} = \frac{(4\pi^2)^2 \kappa_{11}^8}{(2\pi)^{33}} \mathcal{K} \s_3 \cdot \frac{16}{9} \cdot \frac{\pi^3}{12 l_{11}^3}c_1 \cdot \mathcal{I}_1.\ee
There is another contribution which involves a six point one loop counterterm that diverges as $\Lambda$. Candidate counterterms are the $\mathcal{R}^4 F_5^2$ and $\mathcal{R}^4 G_3^2$ interactions, which lie in the same supermultiplet as the $\mathcal{R}^5$ interaction. We drop this contribution as it leaves no finite remainder on--shell. 

From $g$ in figure 8 and \C{figg}, we get the four point counterterm contribution to be
\be  \delta \mathcal{A}_4^{(3)} = \frac{(4\pi^2)^2 \kappa_{11}^8}{(2\pi)^{33}} \mathcal{K} \s_3\cdot \frac{16}{9} \cdot \frac{\pi^3}{12 l_{11}^3}c_1 \cdot \mathcal{I}_1,\ee
and the five point contribution of the form \C{div4}.

Also from $h$ in figure 8 and \C{figh}, we only get the five point counterterm contribution of the form \C{div4}.

Finally, from the universal terms in $i$ in \C{figi}, we get a five point counterterm of the form \C{div4}, and a four point counterterm contribution given by
\be \delta \mathcal{A}_4^{(3)} = \frac{(4\pi^2)^2 \kappa_{11}^8}{(2\pi)^{33}} \mathcal{K} \s_3 \cdot \frac{4}{3} \cdot \frac{\pi^3}{12 l_{11}^3}c_1  \cdot \mathcal{I}_3\ee
where
\be \mathcal{I}_3 = -\frac{\pi^9}{(4\pi^2 l_{11}^2 \mathcal{V}_2)^2} \sum_{m_I,n_I}\int_0^\infty d\lambda d\s d\rho \s \lambda \rho e^{-G^{IJ}\Big(\s m_I m_J +\lambda n_I n_J +\rho(m+n)_I (m+n)_J\Big)/l_{11}^2} \frac{\p \Delta^{-9/2}}{\p\s}.\ee
This non--planar counterterm is depicted by $b$ in figure 21. 

Thus adding the various contributions, the total one loop counterterm contribution is given by
\bea  \label{1loopfinal}\delta \mathcal{A}_4^{(3),1-loop} &=& \frac{(4\pi^2)^2 \kappa_{11}^8}{(2\pi)^{33}} \mathcal{K} \s_3 \Big[ \frac{\pi^3}{9 l_{11}^3}c_1 \Big(4 \mathcal{I}_1+ \mathcal{I}_3\Big)+ \frac{\pi^3 \hat{c}_1}{l_{11}^3} \mathcal{I}_2\Big] \non \\ &=& \frac{\kappa_{11}^8}{(2\pi)^{33}} \mathcal{K} \s_3 \cdot \frac{\pi^3}{3 l_{11}^3 }(c_1 + \hat{c}_1) \mathcal{I},\eea
where
\bea \label{calI}\mathcal{I} = 3 \mathcal{I}_2 =\frac{\pi^9}{(l_{11}^2 \mathcal{V}_2)^2}\sum_{m_I,n_I}\int_0^\infty \frac{d\lambda d\s d\rho}{\Delta^{7/2}}  e^{-G^{IJ}\Big(\s m_I m_J +\lambda n_I n_J +\rho(m+n)_I (m+n)_J\Big)/l_{11}^2} . \eea

Now \C{calI} can be evaluated along the lines of the two loop calculations mentioned earlier, and following~\cite{Dixon:1990pc,Green:1999pu} we get that 
\bea \label{defI}\mathcal{I} &=& \frac{6\pi^{11}}{l_{11}^{12}} \sum_{\hat{m}_I \hat{n}_I}\int_0^\infty dV_2 V_2^5 \int_{\mathcal{F}_2} \frac{d^2 \tau}{\tau_2^2}  e^{-\pi^2 G_{IJ} (\hat{m}+\hat{n}\tau)_I(\hat{m}+\hat{n}\bar\tau)_JV_2/\tau_2} \non \\ &=& \frac{1}{l_{11}^{12}} \Big[ f(\Lambda l_{11})^{12} + \frac{9}{4} \pi^{9/2} (\Lambda l_{11})^5 \mathcal{V}_2^{-7/2} E_{7/2} (\Omega,\bar\Omega) + 18 \zeta (5) \zeta (6) \mathcal{V}_2^{-6}\Big],\eea
where $f$ is an undetermined constant. Note that \C{defI} reduces to an $SL(2,\mathbb{Z})$ invariant integral over $\mathcal{F}_2$. While this is automatically true for $\mathcal{I}_2$, this is not the case for either $\mathcal{I}_1$ or $\mathcal{I}_3$. It is crucial that only the combination $4\mathcal{I}_1 +\mathcal{I}_3$ arises in \C{1loopfinal} which leads to an $SL(2,\mathbb{Z})$ invariant integral. 

The non--universal contributions from the last three terms in $N^{(i)}$ in \C{num} all give contributions of the form \C{div4}. Thus the complete one loop counterterm contribution is given by \C{1loopfinal} and \C{defI}.
Thus it immediately follows that there must be a divergent contribution to \C{usemain} of the form
\be \label{I31loop} I_3 \sim \s_3 \Lambda^{3} l_{11}^{-12} \mathcal{V}_2^{-6} \zeta(5) \zeta (6) .\ee

\subsection{The two loop divergences and counterterms}

The two loop divergences and counterterms can be analyzed like we did above. We now send two of the loop momenta to infinity simultaneously to get the divergences. From the diagrams in figure 8 and \C{num}, it is easy to see that there are two kinds of contributions with $\Lambda^8$ divergence, with the remaining loop momentum integral taking the form
\be \label{Div1} \int \frac{d^{11}p}{p^4} \ee
or 
\be \label{Div2} \int d^{11}p \frac{(k_3 \cdot p)(k_4 \cdot p)}{p^6}\ee
in 11 uncompactified dimensions. There is also a contribution which diverges as $\Lambda^6$, with the remaining momentum integral being
\be \label{Div3}\int \frac{d^{11}p}{p^2} \ee 
in 11 uncompactified dimensions. Compactifying on $T^2$, to renormalize \C{Div1} we add a five point two loop counterterm of the type
\be \label{Div12}\frac{t_1}{l_{11}^8} \cdot \frac{ \pi^{9/2}}{4\pi^2 l_{11}^2 \mathcal{V}_2} \sum_{l_I}\int_0^\infty  d\s \s^{-7/2} e^{-\s G^{IJ} l_I l_J/l_{ll}^2} \sim \frac{t_1}{l_{11}^{15}} \Big((\Lambda l_{11})^7 + \mathcal{V}_2^{-7/2} E_{7/2} (\Omega,\bar\Omega)\Big),\ee
and a four point counterterm of the type
\be \label{Div22}-\frac{t_2}{l_{11}^8} \cdot \frac{\pi^{9/2}}{8\pi^2 l_{11}^2 \mathcal{V}_2} \sum_{l_I} \int_0^\infty d\s \s^2 e^{-\s G^{IJ} l_I l_J/l_{ll}^2} \frac{\p \s^{-9/2}}{\p\s}\sim \frac{t_2}{l_{11}^{15}} \Big((\Lambda l_{11})^7 + \mathcal{V}_2^{-7/2} E_{7/2} (\Omega,\bar\Omega)\Big)\ee
to renormalize the divergence in \C{Div2}. These are denoted by $a$ and $b$ in figure 23 respectively. The counterterms in \C{Div12} and \C{Div22} must lie in the same supermultiplet as the $D^4 \mathcal{R}^4$ two loop primitive $\Lambda^8$ counterterm which does not leave any finite remainder\footnote{A candidate five point counterterm is the $D^3 {F}_5 \mathcal{R}^4$ term.}. We shall see this is consistent with our analysis. 

\begin{figure}[ht]
\begin{center}
\[
\mbox{\begin{picture}(370,100)(0,0)
\includegraphics[scale=.6]{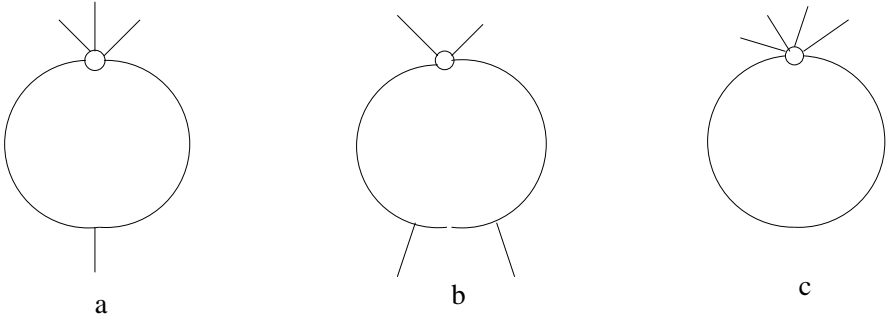}
\end{picture}}
\]
\caption{The two loop counterterms}
\end{center}
\end{figure}

Thus from \C{Div12} and \C{Div22} it immediately follows that there must be divergent contributions to \C{usemain} of the form
\be \label{I32loop1} I_3 \sim \s_3 \Lambda^{8} l_{11}^{-7} \mathcal{V}_2^{-7/2} E_{7/2} (\Omega,\bar\Omega).\ee

Compactifying on $T^2$, to renormalize the divergence in \C{Div3} we add a six point counterterm
\be \label{E9/2}\frac{t_3}{l_{11}^6} \cdot \frac{\pi^{9/2}}{4\pi^2 l_{11}^2 \mathcal{V}_2} \sum_{l_I} \int_0^\infty d\s \s^{-9/2}  e^{-\s G^{IJ} l_I l_J/l_{ll}^2} \sim \frac{t_3}{l_{11}^{15}} \Big((\Lambda l_{11})^9 + \mathcal{V}_2^{-9/2} E_{9/2} (\Omega,\bar\Omega)\Big)\ee
denoted $c$ in figure 23. This counterterm must lie in the same supermultiplet as the $D^6 \mathcal{R}^4$ two loop primitive $\Lambda^6$ counterterm which can have a finite remainder which follows from \C{2loopren}\footnote{A candidate six point counterterm is the $D^4 {F}_5^2 \mathcal{R}^4$ term.}. It follows that that there must be a divergent contribution to \C{usemain} of the form
\be \label{I32loop2}I_3 \sim \s_3 \Lambda^{6} l_{11}^{-9} \mathcal{V}_2^{-9/2} E_{9/2} (\Omega,\bar\Omega).\ee 

Thus the total two loop counterterm contribution is given by
\be \label{2loopfinal}\delta \mathcal{A}_4^{(3),2-loop} \sim \frac{\kappa_{11}^8}{(2\pi)^{33}} \cdot\frac{\mathcal{K}\s_3}{l_{11}^{15}} \Big[(t_1 +t_2) \Big((\Lambda l_{11})^7 + \mathcal{V}_2^{-7/2} E_{7/2}\Big)+ t_3 \Big((\Lambda l_{11})^9 +\mathcal{V}_2^{-9/2} E_{9/2} \Big)\Big].\ee

Again as in the one loop counterterm calculations, we see that the two loop divergent structure of the three loop amplitude is fixed. 

Thus to summarize, the one and two loop counterterm contributions are given by \C{1loopfinal} and \C{2loopfinal} respectively.
From the structure of these counterterms, we see that the three loop amplitude $I_3$ in \C{usemain} must have divergent contributions of the form
\be \label{matchdiv}\frac{\s_3}{l_{11}^{15}} \Big( (\Lambda l_{11})^{8}  \mathcal{V}_2^{-7/2} E_{7/2} (\Omega,\bar\Omega) + (\Lambda_{11})^{6}  \mathcal{V}_2^{-9/2} E_{9/2} (\Omega,\bar\Omega) + (\Lambda l_{11})^3  \zeta (5) \zeta (6) \mathcal{V}_2^{-6} \Big),\ee
where we have dropped irrelevant numerical factors. We now consider the three loop amplitude to see how these divergences arise. 

\subsection{The renormalized three loop amplitude}

We now analyze the structure of the renormalized three loop amplitude \C{usemain}, in particular the various ultraviolet divergences. We assume that these divergences arise from the various boundaries of the moduli space in \C{usemain}\footnote{As explained earlier, we have not constructed the explicit maps involved in going from the measure in \C{impexp3} to the measure in \C{usemain}, and so we do not prove this statement. However, this is a natural assumption given that this is exactly how the divergences arise at two loops.}. Hence they are completely determined from the boundary contributions. The leading divergence in \C{usemain} comes when all the KK momenta are vanishing. This divergence is cutoff at the boundary of the $V_3$ integral at $V^c = (\Lambda l_{11})^3$ which follows from \C{map2}. Thus the leading divergence is given by
\be I_3^{div} = \s_3 h \Lambda^{15}\ee  
where $h$ is an undetermined constant. This is the primitive three loop divergence which is independent of the $\mathcal{V}_2$ and $\Omega$ moduli. To evaluate the remaining subleading divergences, we shall make use of \C{Lapeqn} to evaluate the boundary contributions. This leads to
\be \label{defreg}\Big( \Delta^{SL(2,\mathbb{Z})} + \frac{4}{3} \mathcal{V}_2 \frac{\p}{\p \mathcal{V}_2} +\frac{1}{3} \mathcal{V}_2^2 \frac{\p^2}{\p \mathcal{V}_2^2}\Big) I_3^{div} = \frac{5\cdot 27 \pi^{21/2}}{16 l_{11}^{15}} \s_3\int_0^{V^c} dV_3 V_3^4 \int_{\mathcal{F}_3} d\mu \Delta^{SL(3,\mathbb{Z})} K_L\Big\vert_{bdy},\ee 
where $K_L$ is defined by \C{newdefF2}. In \C{defreg} it is understood that the integration over $\mathcal{F}_3$ is appropriately regularized. The (regularized) boundary of $\mathcal{F}_3$ consists of the union of the following two regions\footnote{The compact space $\mathcal{F}_3^*$, which is $\mathcal{F}_3$ along with its boundaries, has been described as a Satake compacitification in~\cite{Grenier}.}:

{\bf{(i)}} $L \rightarrow \infty$, in which one is left with the fundamental domain of $GL(2,\mathbb{Z})$ parametrized by $T$, which further has a boundary at $T_2 \rightarrow \infty$, and

{\bf{(ii)}} $T_2 \rightarrow \infty$, which must be accompanied by $L\rightarrow \infty$, or else the the defining relations 
\be \label{bound}0 \leq A_1 \leq \frac{1}{2}, \quad A_1^2 +\frac{L^3}{T_2} \geq 1\ee
in \C{SL3} are inconsistent for finite $L$. Thus in \C{defreg} we need to consider the contributions coming from the first two terms in \C{Lap3}.

\subsubsection{Boundary contributions as $L \rightarrow \infty$}

We first consider the contributions coming from ${\bf{(i)}}$. Thus we need to evaluate\footnote{Recall that the fundamental domain of $GL(2,\mathbb{Z})$ is half that of $SL(2,\mathbb{Z})$.}
\bea \label{subdiv}&&\mathcal{J}_1 (\mathcal{V}_2,\Omega,\bar\Omega)= \int_0^{V^c} dV_3 V_3^4 \int_{\mathcal{F}_3} d\mu L^4 \frac{\p}{\p L}\Big(\frac{1}{L^2} \frac{\p K_L}{\p L}\Big) \Big\vert_{L= L^c} \non \\ &&= \frac{1}{4} \int_0^{V^c} dV_3 V_3^4 \frac{1}{L^2 } \frac{\p}{\p L} \int_{\mathcal{F}_2} \frac{dT_1 dT_2}{T_2^2}\sum_{m_I,n_I} e^{-\pi^2 \mathcal{V}_2 V_3^{2/3}\hat{G}_{IJ} (m+n T)_I (m+n\bar{T})_J/L T_2}\Big\vert_{L= L^c},\eea 
where $K_L$ has simplified in the $L\rightarrow \infty$ limit on dropping terms that are exponentially suppressed, and $L$ has been cutoff at $L= L^c$. Now \C{subdiv} has a further nested divergence as $T_2 \rightarrow \infty$ coming from the boundary of $\mathcal{F}_2$. This boundary contribution $\mathcal{J}_1^{n-div}$ involving the nested divergence is given by
\bea \label{morediv}
&&\Delta_\Omega^{SL(2,\mathbb{Z})} \mathcal{J}_1^{n-div} \non \\ &&=  \frac{1}{4} \int_0^{V^c} dV_3 V_3^4 \frac{1}{L^2 }\frac{\p}{\p L} \int_{\mathcal{F}_2} \frac{dT_1 dT_2}{T_2^2}\Delta^{SL(2,\mathbb{Z})}_T\sum_{m_I,n_I}  e^{-\pi^2 \mathcal{V}_2 V_3^{2/3}\hat{G}_{IJ} (m+n T)_I (m+n\bar{T})_J/L T_2}\Big\vert_{L= L^c, T_2 = T_2^c}\non \\\eea  
where the subscript on $\Delta^{SL(2,\mathbb{Z})}$ specifies the moduli on which it acts, and we have used
\bea &&\Delta^{SL(2,\mathbb{Z})}_\Omega \sum_{m_I,n_I} e^{-\pi^2 \mathcal{V}_2 V_3^{2/3}\hat{G}_{IJ} (m+n T)_I (m+n\bar{T})_J/L T_2} \non \\ &&= \Delta^{SL(2,\mathbb{Z})}_T \sum_{m_I,n_I} e^{-\pi^2 \mathcal{V}_2 V_3^{2/3}\hat{G}_{IJ} (m+n T)_I (m+n\bar{T})_J/L T_2},\eea
and $T_2$ has been cutoff at $T_2 = T_2^c$. Thus from \C{morediv} we get that
\be \Delta_\Omega^{SL(2,\mathbb{Z})} \mathcal{J}_1^{n-div}= \frac{1}{4} \int_0^{V^c} \frac{dV_3 V_3^4}{L^2 }\frac{\p^2}{\p L\p T_2}\sum_{(m,n)\neq (0,0)}e^{-\pi^2 \mathcal{V}_2 V_3^{2/3} \vert m+n \Omega\vert^2/L T_2 \Omega_2}\Big\vert_{L= L^c, T_2 = T_2^c}\ee
where the exponential has further simplified in the limit $T_2 \rightarrow \infty$ on neglecting exponentially suppressed terms. This is precisely the contribution from the degenerate orbits of $SL(2,\mathbb{Z})_T$~\cite{Dixon:1990pc}. Thus as $L^c$ and $T_2^c \rightarrow \infty$, the leading divergent contribution is given by
\be \Delta_\Omega^{SL(2,\mathbb{Z})} \mathcal{J}_1^{n-div}= -\frac{\pi^2 \mathcal{V}_2}{4} \int_0^{V^c} dV_3 V_3^4 \cdot \frac{V_3^{2/3}}{(L^c)^4 (T_2^c)^2} \sum_{(m,n)\neq (0,0)}\frac{\vert m+n\Omega\vert^2}{\Omega_2}e^{-\frac{\pi^2 \mathcal{V}_2 V_3^{2/3}}{L^c T_2^c \Omega_2} \vert m+n \Omega\vert^2}.\ee 
We regularize the divergence by setting\footnote{We have regularized the divergences by choosing values for $L^c$ and $T_2^c$ in \C{valdiv} and \C{valdiv2}. We have not analyzed in detail how going to the boundary of the moduli space sends the various loop momenta to infinity, and hence these values of $L^c$ and $T_2^c$ have been put in by hand simply to get agreement with \C{matchdiv}. Thus these choices of renormalization are consistent with the S--duality of the type IIB theory. Other choices would correspond to different schemes, which must finally yield the $SL(2,\mathbb{Z})$ invariant answer in \C{matchdiv}. It would be interesting to understand the issue of renormalization in detail.}
\be \label{valdiv}L^c = (\Lambda l_{11}) V_3^{-1/3}, \quad T_2^c = (\Lambda l_{11})^3 V_3^{-1} , \ee 
leading to
\be \Delta_\Omega^{SL(2,\mathbb{Z})} \mathcal{J}_1^{n-div}= -\frac{105}{128 \pi^{13/2}} (\Lambda l_{11})^8 \mathcal{V}_2^{-7/2} E_{7/2} (\Omega,\bar\Omega),\ee
and thus
\be \label{bdy11}\mathcal{J}_1^{n-div} = -\frac{3}{32 \pi^{13/2}}(\Lambda l_{11})^8 \mathcal{V}_2^{-7/2} E_{7/2} (\Omega,\bar\Omega)\ee
upto moduli independent terms. 

We next evaluate the contribution to \C{subdiv} which involves the finite contribution coming from the integral over $\mathcal{F}_2$ as $L\rightarrow \infty$. This nested finite contribution $\mathcal{J}_1^{n-fin}$ involves the non--degenerate orbits of $SL(2,\mathbb{Z})_T$~\cite{Dixon:1990pc} and is given by
\be \label{fin}\mathcal{J}_1^{n-fin} = \frac{1}{2}  \int_0^{V^c} dV_3 V_3^4 \frac{1}{L^2 } \frac{\p}{\p L} \int_{-\infty}^\infty dT_1 \int_0^\infty \frac{dT_2}{T_2^2}\sum_{\substack{0< j \leq m-1\\ m>0, n \neq 0}} e^{-\pi^2 \mathcal{V}_2 V_3^{2/3} \Big( \frac{\vert mT+j+n\Omega\vert^2}{\Omega_2 T_2} -2mn\Big)/L}\Big\vert_{L= L^c},\ee
where the integral over $\mathcal{F}_2$ has been unfolded to twice the integral over the upper half plane. Now \C{fin} can also be written as
\bea \mathcal{J}_1^{n-fin} &=& -\frac{\mathcal{V}_2}{2} \frac{\p}{\p\mathcal{V}_2} \int_0^{V^c}  \frac{d V_3 V_3^4}{L^3 } \int_{-\infty}^\infty dT_1 \int_0^\infty \frac{dT_2}{T_2^2}\sum_{\substack{0< j \leq m-1 \\ m>0, n \neq 0}} e^{-\pi^2 \mathcal{V}_2 V_3^{2/3} \Big( \frac{\vert mT+j+n\Omega\vert^2}{\Omega_2 T_2} -2mn\Big)/L}\Big\vert_{L= L^c} \non \\ &=& -\frac{\mathcal{V}_2}{2} \frac{\p}{\p\mathcal{V}_2} \int_0^{V^c}  \frac{d V_3 V_3^5}{(\Lambda l_{11})^3 } \int_{-\infty}^\infty dT_1 \int_0^\infty \frac{dT_2}{T_2^2}\sum_{\substack{0< j \leq m-1 \\ m>0, n \neq 0}} e^{-\pi^2 \mathcal{V}_2 V_3 \Big( \frac{\vert mT+j+n\Omega\vert^2}{\Omega_2 T_2} -2mn\Big)/\Lambda l_{11}} \non \\ \eea  
on using \C{valdiv}. On rescaling $V_3$ by a factor of $\Lambda l_{11}$, this is equal to
\bea \mathcal{J}_1^{n-fin} &=& -(\Lambda l_{11})^3\frac{\mathcal{V}_2}{2} \frac{\p}{\p\mathcal{V}_2}\int_0^\infty  d V_3 V_3^5 \int_{-\infty}^\infty dT_1 \int_0^\infty \frac{dT_2}{T_2^2}\sum_{\substack{0< j \leq m-1,\\ m>0, n \neq 0}} e^{-\pi^2 \mathcal{V}_2 V_3 \Big( \frac{\vert mT+j+n\Omega\vert^2}{\Omega_2 T_2} -2mn\Big)} \non \\ &=& -(\Lambda l_{11})^3\frac{\mathcal{V}_2}{4} \frac{\p}{\p\mathcal{V}_2}\int_0^\infty  d V_3 V_3^5 \int_{\mathcal{F}_2} \frac{ d^2T }{T_2^2}\sum_{m_I, n_I} e^{-\pi^2  G_{IJ} (m+nT)_I (m+n\bar{T})_J V_3/T_2}\Big\vert_{non-degenerate}.\non \\ \eea
We have set $V^c \rightarrow \infty$ as the integral is finite.
This integral is precisely proportional to the one in \C{defI} which contributes to the non--degenerate orbits of $SL(2,\mathbb{Z})$ on identifying $T$ with $\tau$ and $V_3$ with $V_2$. Thus 
\be \label{bdy12}\mathcal{J}_1^{n-fin}= \frac{9}{2 \pi^{11}} (\Lambda l_{11})^3  \zeta (5) \zeta (6)\mathcal{V}_2^{-6}.\ee
Thus from \C{bdy11} and \C{bdy12}, the total boundary contribution from ${\bf{(i)}}$ is given by 
\be \mathcal{J}_1 (\mathcal{V}_2,\Omega,\bar\Omega)= \frac{9}{2 \pi^{11}} (\Lambda l_{11})^3  \zeta (5) \zeta (6)\mathcal{V}_2^{-6}-\frac{3}{32 \pi^{13/2}}(\Lambda l_{11})^8 \mathcal{V}_2^{-7/2} E_{7/2} (\Omega,\bar\Omega).\ee

Thus from \C{defreg}, we see that from the boundary of the moduli space as $L \rightarrow \infty$, we get divergent contributions to $I_3$ of the form
\be I_3 \sim \frac{\s_3}{l_{11}^{15}} \Big( \pi^{4}(\Lambda l_{11})^8 \mathcal{V}_2^{-7/2} E_{7/2} (\Omega,\bar\Omega) + \pi^{-1/2}(\Lambda l_{11})^3  \zeta (5) \zeta (6)\mathcal{V}_2^{-6}\Big)\ee
upto irrelevant numerical factors. 

\subsubsection{Boundary contributions as $T_2 \rightarrow \infty$}

We now consider the boundary contributions from ${\bf{(ii)}}$. We need to evaluate the contribution to the integral
\bea \mathcal{J}_2(\mathcal{V}_2,\Omega,\bar\Omega) &=& \int_0^{V^c} dV_3 V_3^4 \int_{\mathcal{F}_3} d\mu T_2^2 \frac{\p^2 K_L}{\p T_2^2 }  \Big\vert_{T_2 = T_2^c} \non \\ &=& \frac{1}{4}\int_0^{V^c} dV_3 V_3^4 \int_0^{L^c} \frac{dL}{L^4} \frac{\p}{\p T_2} \sum_{(m,n) \neq (0,0)}e^{-\pi^2 \mathcal{V}_2 V_3^{2/3} \vert m+n\Omega\vert^2/L T_2\Omega_2} \Big\vert_{T_2 = T_2^c},\non \\ \eea
from the boundary of $L$ as $L \rightarrow L^c$. The lattice factor has simplified in the $T_2 \rightarrow\infty$ limit on dropping exponentially suppressed terms. Thus we evaluate
\be \Big(\Delta^{SL(2,\mathbb{Z})}_\Omega + 2 \mathcal{V}_2 \frac{\p}{\p \mathcal{V}_2} \Big)\mathcal{J}_2 = \frac{1}{4}\int_0^{V^c} \frac{dV_3 V_3^4}{L^2}  \frac{\p^2}{\p L \p T_2}  \sum_{(m,n) \neq (0,0)}e^{-\pi^2 \mathcal{V}_2 V_3^{2/3} \vert m+n\Omega\vert^2/L T_2\Omega_2} \Big\vert_{T_2 = T_2^c,L= L^c},\ee
on using
\be \Big( \Delta^{SL(2,\mathbb{Z})}_\Omega + 2 \mathcal{V}_2 \frac{\p}{\p \mathcal{V}_2} \Big) e^{-\pi^2 \mathcal{V}_2 V_3^{2/3} \vert m+n\Omega\vert^2/L T_2\Omega_2}= L^4 \frac{\p}{\p L} \Big( \frac{1}{L^2} \frac{\p}{\p L}\Big)e^{-\pi^2 \mathcal{V}_2 V_3^{2/3} \vert m+n\Omega\vert^2/L T_2\Omega_2}.\ee
Thus as $T_2 \rightarrow \infty$ and $L \rightarrow \infty$, the leading divergent contribution is given by
\be \Big(\Delta^{SL(2,\mathbb{Z})}_\Omega + 2 \mathcal{V}_2 \frac{\p}{\p \mathcal{V}_2} \Big)\mathcal{J}_2 =  -\frac{\pi^2 \mathcal{V}_2}{4} \int_0^{V^c} dV_3 V_3^4 \cdot \frac{V_3^{2/3}}{(L^c)^4 (T_2^c)^2} \sum_{(m,n)\neq (0,0)}\frac{\vert m+n\Omega\vert^2}{\Omega_2}e^{-\frac{\pi^2 \mathcal{V}_2 V_3^{2/3}}{L^c T_2^c \Omega_2} \vert m+n \Omega\vert^2}.\ee
We regularize the divergence by setting 
\be \label{valdiv2}T_2^c = \Lambda l_{11}, \quad L^c = (\Lambda l_{11})^{5/3} V_3^{-4/3},\ee
leading to
\be \Big(\Delta^{SL(2,\mathbb{Z})}_\Omega + 2 \mathcal{V}_2 \frac{\p}{\p \mathcal{V}_2} \Big)\mathcal{J}_2 = -\frac{945}{256\pi^{17/2}} (\Lambda l_{11})^6 \mathcal{V}_2^{-9/2} E_{9/2} (\Omega,\bar\Omega),\ee
and thus
\be \mathcal{J}_2 = \frac{151}{448\pi^{17/2}}(\Lambda l_{11})^6 \mathcal{V}_2^{-9/2} E_{9/2} (\Omega,\bar\Omega).\ee
Note that to evaluate the boundary contribution as $T_2 \rightarrow \infty$ we have used the cutoff \C{valdiv2}, which is different from the cutoff \C{valdiv} used to evaluate the boundary contribution as $L \rightarrow \infty$. This is because 
\be \frac{(L^c)^3}{T_2^c} \sim (\Lambda l_{11})^4 V_3^{-4/3}\ee
which goes to infinity for fixed $V_3$ which is consistent with \C{bound}. However \C{valdiv} yields
\be \frac{(L^c)^3}{T_2^c} \sim O(1) \ee 
which need not be consistent with \C{bound} unless the constant if fixed to an appropriate value. 

Thus from \C{defreg}, we see that from the boundary of the moduli space as $T_2 \rightarrow \infty$, we get divergent contributions to $I_3$ of the form
\be I_3 \sim \frac{\s_3}{l_{11}^{15}}  \pi^2 (\Lambda l_{11})^6 \mathcal{V}_2^{-9/2} E_{9/2} (\Omega,\bar\Omega) \ee
upto irrelevant numerical factors.

Hence the complete ultraviolet divergent part of $I_3$ is given by
\bea \label{totdiv}I_3 &=& \frac{\s_3}{l_{11}^{15}} \Big( \hat{h}_1(\Lambda l_{11})^{15}+ \hat{h}_2(\Lambda l_{11})^8 \mathcal{V}_2^{-7/2} E_{7/2} (\Omega,\bar\Omega)+ \hat{h}_3(\Lambda l_{11})^6 \mathcal{V}_2^{-9/2} E_{9/2} (\Omega,\bar\Omega) \non \\ &&+ \pi^{-1/2}\hat{h}_4(\Lambda l_{11})^3 \zeta (5) \zeta (6) \mathcal{V}_2^{-6}\Big)\eea
where $\hat{h}_i$ are moduli independent constants that are not relevant for us. These include the primitive three loop divergence, as well as one and two loop subdivergences. This structure is completely determined by the boundary contributions of the moduli space. We have kept the explicit $\pi$ dependence of the term of $O(\mathcal{V}_2^{-6})$ for later use\footnote{Note that $\hat{h}_4$ has vanishing transcendentality and has no factors of $\pi$.}. 

We now argue that there are no more contributions to $I_3$ and \C{totdiv} is complete. Any finite contribution to $I_3$ would be of the form
\be I_3^{finite} =\frac{\s_3}{l_{11}^{15}} \mathcal{V}_2^{-15/2} f(\Omega,\bar\Omega)\ee 
where $f(\Omega,\bar\Omega)$ is $SL(2,\mathbb{Z})$ invariant. This
would lead to a term in the effective action given by
\be l_{11}^5 \int d^9 x \sqrt{-G^{(9)}} \mathcal{V}_2 D^6\mathcal{R}^4 \cdot \mathcal{V}_2^{-15/2}f(\Omega,\bar\Omega).\ee
In the type IIB theory, this leads to the term
\be \label{incon}l_s^5 \int d^9 x \sqrt{-g^B} r_B D^6\mathcal{R}^4 \cdot r_B^6 e^{-\phi_B/2}f(\Omega,\bar\Omega)\ee
in the effective action. Now the structure of $f(\Omega,\bar\Omega)$ is tightly constrained by supersymmetry. It must satisfy Poisson equation\footnote{It can also split into a sum of terms each of which satisfies the Poisson equation.} on the fundamental domain of $SL(2,\mathbb{Z})$ with possible source terms given by $E_{3/2}^2 (\Omega,\bar\Omega)$\cite{Basu:2008cf} involving the $\mathcal{R}^4$ interaction. If the source terms are non--vanishing, then using that
\be E_{3/2}^2 (\Omega,\bar\Omega) = 4\zeta(3)^2 e^{-3\phi_B} +16\zeta (2)\zeta (3) e^{-\phi_B}+ 16\zeta(2)^2 e^{\phi_B} +\ldots\ee
we see that
\be f(\Omega,\bar\Omega) \sim \zeta(3)^2 e^{-3\phi_B} +\zeta (2)\zeta (3) e^{-\phi_B}+ \zeta(2)^2 e^{\phi_B} +\ldots.\ee
Thus from \C{incon} we see it is inconsistent with perturbative string theory because of the fractional powers of $e^{-2\phi_B}$. Thus there cannot be source term contributions, and $f(\Omega,\bar\Omega)$ must satisfy Laplace equation, and hence\footnote{In general, a sum of various Eisenstein series is also allowed.}
\be f(\Omega,\bar\Omega) \sim E_s (\Omega,\bar\Omega) \sim \zeta (2s)e^{-s\phi_B} + \zeta (2s-1) e^{-(1-s)\phi_B} +\ldots.\ee 
From \C{incon} it follows that the only possibility consistent with perturbative string theory is $f(\Omega,\bar\Omega) = E_{3/2} (\Omega,\bar\Omega)$. However, that would give contributions at genus zero and one which violate perturbative string results. Hence
\be f(\Omega,\bar\Omega) =0\ee
and there are no  contributions beyond \C{totdiv}\footnote{One would expect such finite contributions to come from rank 3 non--degenerate orbits of $SL(3,\mathbb{Z})$. For toroidal compactifications, this would need a target space $T^d$ for $d\geq 3$, and hence this sector vanishes for our case. The classification of these orbits has been discussed in~\cite{Pioline:2004xq}, though in a different context. }.

Thus from \C{totdiv} we see that
\bea \label{finexp}\mathcal{A}_4^{(3)} &=& \frac{\kappa_{11}^8}{(2\pi)^{33}}\cdot \frac{\mathcal{K}\s_3}{l_{11}^{15}} \Big[ h_1(\Lambda l_{11} )^{15}+ h_2(\Lambda l_{11})^8 \mathcal{V}_2^{-7/2} E_{7/2} + h_3(\Lambda l_{11})^6 \mathcal{V}_2^{-9/2} E_{9/2}  \non \\ &&+ h_4\pi^{11/2}(\Lambda l_{11})^3 \zeta (5) \zeta (6) \mathcal{V}_2^{-6}\Big],\eea
where $h_4$ has vanishing transcendentality.

Let us first consider the terms involving $E_{7/2}$ in \C{2loopfinal} and \C{finexp}. They lead to terms in the effective action of the form\footnote{The precise numerical factors are not relevant for our analysis. }
\bea &&l_{11}^5 \int d^9 x \sqrt{-G^{(9)}} D^6\mathcal{R}^4 \Big(t_1 +t_2 + (\Lambda l_{11})^8 \Big)\mathcal{V}_2^{-5/2} E_{7/2} (\Omega,\bar\Omega) \non \\ &&\sim \zeta (7)l_{11}^5 \int d^9 x \sqrt{-G^{(9)}} D^6\mathcal{R}^4 \Big(t_1 + t_2 + (\Lambda l_{11})^8 \Big)\mathcal{V}_2^{-5/2} \Omega_2^{7/2} +\ldots,\eea
which in the type IIB theory gives
\be \zeta (7) l_s^5 \int d^9 x \sqrt{-g^{B}} D^6 \mathcal{R}^4 \Big(t_1 +t_2 + (\Lambda l_{11})^8 \Big) r_B^{5/3} e^{-8\phi_B/3}+\ldots\ee
which is inconsistent with the structure of string perturbation theory, and thus
\be \label{addfindiv}t_1 +t_2 + (\Lambda l_{11})^8 = 0\ee 
leaving no finite remainder. Hence the moduli independent term in \C{2loopfinal} simply changes the coefficient of the primitive three loop divergent term $\sim \Lambda^{15}$ in \C{finexp}.  

Similarly the terms involving $E_{9/2}$ in \C{2loopfinal} and \C{finexp} lead to terms in the effective action 
\bea &&l_{11}^5 \int d^9 x \sqrt{-G^{(9)}} D^6\mathcal{R}^4 \Big(t_3 + (\Lambda l_{11})^6 \Big)\mathcal{V}_2^{-7/2} E_{9/2} (\Omega,\bar\Omega) \non \\ &&\sim \zeta (9)l_{11}^5 \int d^9 x \sqrt{-G^{(9)}} D^6\mathcal{R}^4 \Big(t_3 + (\Lambda l_{11})^6 \Big)\mathcal{V}_2^{-7/2} \Omega_2^{9/2} +\ldots,\eea
which in the type IIB picture gives
\be \zeta (9) l_s^5 \int d^9 x \sqrt{-g^{B}} D^6 \mathcal{R}^4 \Big(t_3 + (\Lambda l_{11})^6 \Big) r_B^{3} e^{-4\phi_B} +\ldots\ee
which is inconsistent with the structure of string perturbation theory, and thus
\be \label{vanish} t_3 + (\Lambda l_{11})^6 = 0\ee 
leaving no finite remainder. Once again the moduli independent term in \C{2loopfinal} simply changes the coefficient of the primitive three loop divergent term $\sim \Lambda^{15}$ in \C{finexp}. Thus the two loop divergences when renormalized in quantum supergravity completely vanish and leave no finite remainder. Now \C{vanish} implies that the two loop $\Lambda^6$ counterterm leaves no finite remainder. Since it is in the same supermultiplet as the $D^6\mathcal{R}^4$ counterterm, it follows that 
\be \label{fixval}\eta =1\ee
 in \C{2loopren}.

Now in \C{1loopfinal} and \C{finexp} consider the terms which involve $\mathcal{V}_2^{-6}$. They give
\bea \mathcal{A}_4^{(3)} + \delta \mathcal{A}_4^{(3),1-loop} = \frac{\kappa_{11}^8}{(2\pi)^{33} l_{11}^{15}} \mathcal{K} \s_3 \Big[ \pi^{11/2} h_4 (\Lambda l_{11})^3 +6 \pi^3(c_1 +\hat{c}_1)\Big]\zeta (5) \zeta (6) \mathcal{V}_2^{-6} \non \\ = (2\pi^8 l_{11}^{15}) \mathcal{K} \s_3 \Big[ \frac{\pi^{5/2}h_4}{6} (\Lambda l_{11})^3 + c_1 +\hat{c}_1\Big]\frac{l_{11}^6\zeta (2)\zeta (5) \mathcal{V}_2^{-6}}{16\cdot 105 \pi^2}.\eea 
In the type IIA and IIB theories this gives
\bea \label{needadd}\mathcal{A}_4^{(3)} + \delta \mathcal{A}_4^{(3),1-loop} = (2\pi^8 l_{11}^{15} r_B) \mathcal{K} r_B \Big[ l_s^6 \s_3 r_B^4 \Big( \frac{\pi^{5/2}h_4}{6} (\Lambda l_{11})^3 + c_1 +\hat{c}_1 \Big) \frac{\zeta (2) \zeta (5)}{16 \cdot 105 \pi^2}+ O(k^8)\Big] \non \\ = (2\pi^8 l_{11}^{15} r_A^{-1}) \mathcal{K} r_A \Big[ \frac{l_s^6 \s_3}{r_A^6} \Big( \frac{\pi^{5/2}h_4}{6} (\Lambda l_{11})^3 + c_1 +\hat{c}_1 \Big) \frac{\zeta (2) \zeta (5)}{16 \cdot 105 \pi^2}+ O(k^8)\Big].\non \\ \eea
 
We now go back to the one loop amplitude \C{1loopfin} and consider the terms which did not obey the perturbative equality of the genus one amplitude in the IIA and IIB theories even after including two loop effects, as discussed after \C{2loopfin}.
Adding their contribution to \C{needadd}, we get a total contribution to the amplitude given by
\bea &&(2\pi^8 l_{11}^{15} r_B) \mathcal{K} r_B l_s^6 \s_3 \frac{\zeta(2) \zeta (5)}{16\cdot 21}\Big[ \frac{1}{3 r_B^6} + \frac{r_B^4}{5\pi^2} \Big( \frac{\pi^{5/2}h_4}{6} (\Lambda l_{11})^3 + c_1 +\hat{c}_1 \Big) \Big] \non \\ &&= (2\pi^8 l_{11}^{15} r_A^{-1}) \mathcal{K} r_A l_s^6 \s_3 \frac{\zeta(2) \zeta (5)}{16\cdot 21}\Big[ \frac{r_A^4}{3} + \frac{1}{5\pi^2 r_A^6} \Big( \frac{\pi^{5/2}h_4}{6} (\Lambda l_{11})^3 + c_1 +\hat{c}_1 \Big) \Big].\eea
Imposing the perturbative equality of the four graviton genus one amplitude in the IIA and IIB theories leads to
\be \frac{\pi^{5/2}h_4}{6} (\Lambda l_{11})^3 + c_1 +\hat{c}_1 = \frac{5\pi^2}{3}.\ee
This is precisely what is expected, and shows that $\hat{c}_1$ has a finite piece proportional to $\zeta (2)$, as expected on the basis of supersymmetry. Also the term involving $(\Lambda l_{11})^3$ has the correct structure since $h_4$ has vanishing transcendentality, which follows from comparing with \C{valc1}. 

The remaining terms in \C{1loopfinal} change the coefficient of the $\Lambda^{15}$ primitive divergent term, while the term involving $E_{7/2}$ merely adds to \C{addfindiv}. The $\Lambda^{15}$ term adds a term 
\be l_{11}^5 \int d^9 x \sqrt{-G^{(9)}} \mathcal{V}_2 D^6 \mathcal{R}_4 h_1 (\Lambda l_{11})^{15}\ee
to the effective action, which gives
\bea l_s^5 \int d^9 x \sqrt{-g^{A}} r_A D^6 \mathcal{R}^4 e^{2\phi_A} h_1 (\Lambda l_{11})^{15}\non \\ =  l_s^5 \int d^9 x \sqrt{-g^{A}} r_B D^6 \mathcal{R}^4 \frac{e^{2\phi_B}}{r_B^4} h_1 (\Lambda l_{11})^{15}\eea
in the IIA and IIB effective actions. To cancel this three loop primitive divergence we add a final moduli independent counterterm given by
\be \delta \mathcal{A}_4^{(3)} = \frac{\kappa_{11}^8}{(2\pi)^{33}} \cdot \mathcal{K} \s_3 z\ee 
sending 
\be h_1 (\Lambda l_{11})^{15} \rightarrow z + h_1 (\Lambda l_{11})^{15}\ee
in the renormalized amplitude, which contributes at genus two. From \C{2loopfin} and \C{fixval} it immediately follows that
\be  z + h_1 (\Lambda l_{11})^{15} = 24\zeta (4) ,\ee
 enforcing perturbative equality of the IIA and IIB theories at genus two.

Hence there are no more finite contributions, and the three loop amplitude adds a term of the form
\be l_{11}^5 \int d^9 x \sqrt{-G^{(9)}} \mathcal{V}_2 D^6 \mathcal{R}^4 \Big(\zeta (5) \zeta (6) \mathcal{V}_2^{-6} +\zeta (4) \eta \Big)\ee 
to the effective action. Putting in the various factors, we see that the renormalized three loop amplitude adds the contribution
\bea \mathcal{A}_4^{(3)} +\delta \mathcal{A}_4^{(3)}&=& (2\pi^8 l_{11}^{15} r_B) \mathcal{K} r_B  \Big[ l_s^6 \s_3 \Big( \frac{\zeta (2) \zeta (5)}{48 \cdot 21} r_B^4+\frac{\zeta (4)}{16} \eta e^{2\phi^B}r_B^{-4}\Big) + O(k^8)\Big]\non \\ &=& (2\pi^8 l_{11}^{15} r_A^{-1}) \mathcal{K} r_A \Big[ l_s^6 \s_3 \Big(\frac{\zeta (2) \zeta (5)}{48 \cdot 21 } r_A^{-6}  +\frac{\zeta (4)}{16} \eta\Big)+O(k^8) \Big]\eea
to the one and two loop amplitudes.

Thus, the renormalized total amplitude is given by
\bea &&\mathcal{A}_4= (2\pi^8 l_{11}^{15} r_B) \mathcal{K} r_B   \frac{l_s^6}{16\cdot 4!}\s_3\Big[ 4\zeta (3)^2 e^{-2\phi^B} +8\zeta (2) \zeta (3) (1+r_B^{-2})+ \frac{8}{21} \zeta (2) \zeta (5) (r_B^4 + r_B^{-6})\non \\ &&+ 24\zeta (4) e^{2\phi^B} (1+ \frac{5}{3} r_B^{-2} +r_B^{-4}) + \frac{8}{9} \zeta (6) e^{4\phi^B} (1+r_B^{-6})  \Big]\non \\ &&= (2\pi^8 l_{11}^{15} r_A^{-1}) \mathcal{K} r_A  \frac{l_s^6}{16\cdot 4!}\s_3\Big[ 4\zeta (3)^2 e^{-2\phi^A} +8\zeta (2) \zeta (3) (1+r_A^{-2}) + \frac{8}{21} \zeta(2) \zeta(5) (r_A^4 + r_A^{-6})\non \\ &&+ 24\zeta (4) e^{2\phi^A} (1 +\frac{5}{3} r_A^{-2}+r_A^{-4}) + \frac{8}{9} \zeta (6) e^{4\phi^A} (1+r_A^{-6})  \Big],\eea
which exhibits manifest T duality and perturbative equality of the IIA and IIB theories upto genus three. 

Thus to summarize, in nine dimensions we get a term in the effective action of the form
\be l_{11}^5 \int d^9 x \sqrt{-G^{(9)}} \mathcal{V}_2 D^6 \mathcal{R}^4 F(\Omega,\bar\Omega)\ee
where
\be F(\Omega,\bar\Omega) = \frac{4}{21}\zeta (2) E_{5/2} (\Omega,\bar\Omega) \mathcal{V}_2^{3/2} + \mathcal{E} (\Omega,\bar\Omega) \mathcal{V}_2^{-3}+ 4\zeta (2) E_{3/2} (\Omega,\bar\Omega)\mathcal{V}_2^{-3/2} + 24\zeta (4) + \frac{8}{21}\zeta (2) \zeta (5) \mathcal{V}_2^{-6},\ee
where $\mathcal{E}$ satisfies 
\be 4\Omega_2^2 \frac{\p^2 \mathcal{E}}{\p\Omega \p\bar\Omega} = 12 \mathcal{E} - 6 E_{3/2}^2.\ee 
In the type IIA and IIB theories, they yield perturbative contributions is given by
\bea \label{final}&&l_s^5 \int d^9 x \sqrt{-g^B} r_B D^6 \mathcal{R}^4 \Big[ 4\zeta (3)^2 e^{-2\phi^B} +8\zeta (2) \zeta (3) (1+r_B^{-2})+ \frac{8}{21} \zeta (2) \zeta (5) (r_B^4 + r_B^{-6})\non \\ &&+ 24\zeta (4) e^{2\phi^B} (1+ \frac{5}{3} r_B^{-2} +r_B^{-4}) + \frac{8}{9} \zeta (6) e^{4\phi^B} (1+r_B^{-6})  \Big] \non \\ &&=l_s^5 \int d^9 x \sqrt{-g^A} r_A D^6 \mathcal{R}^4\Big[ 4\zeta (3)^2 e^{-2\phi^A} +8\zeta (2) \zeta (3) (1+r_A^{-2}) + \frac{8}{21} \zeta(2) \zeta(5) (r_A^4 + r_A^{-6})\non \\ &&+ 24\zeta (4) e^{2\phi^A} (1 +\frac{5}{3} r_A^{-2}+r_A^{-4}) + \frac{8}{9} \zeta (6) e^{4\phi^A} (1+r_A^{-6})\Big].\eea
These results are in precise agreement with results obtained in~\cite{Basu:2007ck} based on U--duality in 8 dimensions and decompactifying to 9 dimensions. 

Recall that beyond three loops, the leading contribution in the low momentum expansion of the four graviton amplitude is of the form $D^8 \mathcal{R}^4$, and hence there are no more contributions to the $D^6 \mathcal{R}^4$ term~\cite{Bern:2009kd,Bjornsson:2010wm,Bjornsson:2010wu}. It is striking that the three loop amplitude when regularized yields a simple answer. This is a consequence of the fact that the $D^6\mathcal{R}^4$ interaction is BPS, and does not receive contributions beyond three loops in supergravity. Hence the expressions for the perturbative string amplitudes that have been obtained in \C{final} must match the 9 and 10 dimensional worldsheet calculations. This indeed is the case~\cite{Green:1999pv,Basu:2007ck,Green:2008uj,D'Hoker:2013eea,Gomez:2013sla,D'Hoker:2014gfa}.  

As discussed earlier, the supergravity calculations are valid only in a certain regime for either of the type II theories. For the IIA theory, this is the strongly coupled 10 dimensional theory, and for the IIB theory, the result is intrinsically 9 dimensional. However, the BPS interactions we have discussed do not receive any more contributions, and so are exact results. Hence they are valid at all values of the string coupling and also in the 10 dimensional decompactification limit of the IIB theory. This is unlike the non--BPS interactions which receive contributions from all loops in supergravity and hence have an infinite number of moduli dependent terms. These will generically give different values when expanded around small and large values of the moduli.      

\section{Going beyond BPS interactions}

In spite of the complications arising from the details of the three loop amplitude, the $D^6\mathcal{R}^4$ amplitude is simple. We do not expect such simplifications to occur at higher orders in the momentum expansion. In particular, these include non--BPS interactions of the form $D^{2k}\mathcal{R}^4$ for $k \geq 4$. For these interactions, all loop diagrams that arise from the Mercedes and ladder skeleton contribute. For the contributions that arise from the Mercedes skeleton, we expect the unrenormalized amplitude to be of the form
\be \int_0^\infty dV_3 V_3^a \int_{\mathcal{F}_3} d\mu F(L,T,\bar{T},A_1,A_2) K_L\ee  
generically, where $F(L,T,\bar{T},A_1,A_2)$ is not $SL(3,\mathbb{Z})$ invariant, and satisfies Poisson equation on $\mathcal{F}_3$, with source terms given by contact terms. This should lead to the amplitude satisfying Poisson equations on $\mathcal{F}_2$ for the complex structure $\Omega$ with source terms that are recursively determined by supersymmetry~\cite{Basu:2008cf}. For example, the coefficient of the $D^8\mathcal{R}^4$ interaction should satisfy Poisson equation with source terms involving $E_{1/2} (\Omega,\bar\Omega) E_{3/2} (\Omega,\bar\Omega)$~\cite{Green:2008bf}.     

For the diagrams that arise from the ladder skeleton, the amplitude has a qualitatively different structure. This is because the integrals are now over 5 Schwinger parameters. It would be interesting to determine the underlying auxiliary geometry for these diagrams that contribute to the amplitude. It is plausible that all amplitudes at all orders in the maximal supergravity loop expansion can be described in terms of some underlying auxiliary geometry $\mathcal{G}$. Hence one would have a geometric understanding of these multi--loop amplitudes, in particular the origin of ultraviolet divergences that arise from the boundaries of the moduli space of these geometries. Then the amplitudes would be described in terms of maps from $\mathcal{G}$ to the target space $T^d$. 

In general, it would be very interesting to understand non--BPS interactions in the effective action, about which very little is known. This would involve a detailed understanding of its perturbative structure as well as the various instanton contributions, using techniques of supergravity or otherwise. These issues have been recently discussed in specific contexts in~\cite{Basu:2013goa,Basu:2013oka}.

\section{Appendix}

\appendix

\section{Relevant facts about $SO(3)\backslash SL(3,\mathbb{R})$ and $SL(3,\mathbb{Z})$}

We mention certain facts about $SO(3)\backslash SL(3,\mathbb{R})$ and $SL(3,\mathbb{Z})$ which are relevant for our purposes. To start with, we consider the maximally symmetric coset space $SO(3)\backslash SL(3,\mathbb{R})$ which is parametrized by 5 moduli. Using the Iwasawa decomposition, we parametrize the inverse vielbein as\footnote{We use the notations of~\cite{Pioline:2004xq}.}  
\be \label{T1} E_a^{~\alpha}  = \begin{pmatrix}
1/L &0 &0  \\
0& \sqrt{L/T_2} &0 \\
0 & 0 & \sqrt{LT_2} 
\end{pmatrix}
\begin{pmatrix}
1 & A_1 & A_2  \\
0& 1 &T_1 \\
0 & 0 & 1 
\end{pmatrix}, \ee  
where we have fixed a gauge (corresponding to $SO(3)$ transformations). Any other gauge choice corresponds to left multiplication by an $SO(3)$ matrix. This leads to the inverse metric $g^{\alpha\beta} = \delta^{ab} E_a^{~\alpha} E_b^{~\beta}$, given by
\be \label{T2} g^{\alpha\beta} = \begin{pmatrix}
1/L^2 & A_1/L^2 & A_2/L^2  \\
A_1/L^2 & A_1^2/L^2 +L/T_2 & A_1A_2/L^2 +LT_1/T_2 \\
A_2/L^2 & A_1A_2/L^2 +LT_1/T_2 & A_2^2/L^2 + L\vert T \vert^2/T_2
\end{pmatrix}.\ee

We now calculate the metric on the 5 dimensional moduli space. The metric $\mathcal{G}_{\dot\alpha \dot\beta}$ ($\dot\alpha, \dot\beta = 1,\ldots,5$) is given by
\be -\frac{1}{2} {\rm Tr} (d g^{-1} dg) = \mathcal{G}_{\dot\alpha \dot\beta} dz^{\dot\alpha}dz^{\dot\beta},\ee
where $z^{\dot\alpha} = \{L,T,\bar{T},A_1,A_2\}$, where $T= T_1 +i T_2$. We get that
\be \label{Metric} \mathcal{G}_{\dot\alpha \dot\beta} dz^{\dot\alpha}dz^{\dot\beta} = \frac{3}{L^2} dL^2 + \frac{dTd\bar{T}}{T_2^2} + \frac{1}{L^3 T_2} \vert T dA_1 - d A_2\vert^2,\ee

We are interested in $SO(3)\backslash SL(3,\mathbb{R})/ SL(3,\mathbb{Z})$ after identifying by the discrete subgroup. The action of $SL(3,\mathbb{Z})$ is implemented by right multiplication. Thus under an $SL(3,\mathbb{Z})$ transformation by the matrix $S$, we have that $g^{\alpha\beta} \rightarrow (S^T g^{-1} S)^{\alpha\beta}$.

Note that from \C{Metric} we get that the $SL(3,\mathbb{Z})$ invariant volume element on moduli space is given by
\be \label{Measure} \frac{1}{L^4 T_2^2} dL dT_1 dT_2 dA_1 dA_2\ee
upto an irrelevant numerical factor. Also the $SL(3,\mathbb{Z})$ invariant Laplacian on the moduli space is
\be \label{Lap3}
\Delta^{SL(3,\mathbb{Z})} = \frac{1}{\sqrt{\mathcal{G}}} \p_{\dot\alpha} (\sqrt{\mathcal{G}} \mathcal{G}^{\dot\alpha\dot\beta} \p_{\dot\beta}) = \frac{L^4}{3} \frac{\p}{\p L} \Big( \frac{1}{L^2} \frac{\p}{\p L}\Big)+4 T_2^2 \frac{\p^2}{\p T\p\bar{T}}\ +\frac{L^3}{T_2} \Big\vert T\frac{\p}{\p A_2} + \frac{\p}{\p A_1}\Big\vert^2. \ee

The fundamental domain of $SL(3,\mathbb{Z})$, denoted $\mathcal{F}_3$, is given by~\cite{Gordon,Grenier2,Grenier}
\bea \label{SL3}   &&0 \leq A_1, T_1 \leq \frac{1}{2}, \quad \vert A_2 \vert \leq \frac{1}{2} , \quad \vert T \vert^2 \geq 1 \non \\ &&
A_1^2 + \frac{L^3}{T_2} \geq 1, \quad  A_2^2 + \frac{L^3}{T_2} \vert T \vert^2 \geq 1, \non \\ &&(A_1 - A_2)^2 +\frac{L^3}{T_2} \vert T -1 \vert^2 \geq 1, \non \\ &&(1-A_1 + A_2)^2 +\frac{L^3}{T_2} \vert T -1 \vert^2 \geq 1.\eea
Thus $T$ lies in the fundamental domain of $GL(2,\mathbb{Z})$. 

Under a non--trivial $SL(3,\mathbb{Z})$ transformation by a matrix $S$, we have that $g^{-1} \rightarrow  S^T g^{-1} S$. These $SL(3,\mathbb{Z})$ transformations are constructed out of the (overcomplete set of) matrices~\cite{Gordon}
\bea  \label{listmat}&&S1= \begin{pmatrix}
0 & 0 & 1  \\
1 & 0 & 0 \\
0 & 1 & 0
\end{pmatrix}, \quad 
S2= \begin{pmatrix}
0 & 1 & 0  \\
0 & 0 & 1 \\
1 & 0 & 0
\end{pmatrix},\quad
S3 = \begin{pmatrix}
0 & 1 & 0  \\
1 & 0 & -1 \\
-1 & 0 & 0
\end{pmatrix},\quad S4= \begin{pmatrix}
1 & 0 & 0  \\
-1 & 1 & 0 \\
1 & 0 & 1
\end{pmatrix},\non \\ &&A1 = \begin{pmatrix}
1 & 1 & 0  \\
0 & 1 & 0 \\
0 & 0 & 1
\end{pmatrix},\quad A2= \begin{pmatrix}
1 & 0 & 1  \\
0 & 1 & 0 \\
0 & 0 & 1
\end{pmatrix}, \quad T1= \begin{pmatrix}
1 & 0 & 0  \\
0 & 1 & 1 \\
0 & 0 & 1
\end{pmatrix}, \quad T2 = \begin{pmatrix}
1 & 0 & 0  \\
0 & 0 & 1 \\
0 & -1 & 0
\end{pmatrix},\non \\ &&T3 = \begin{pmatrix}
-1 & 0 & 0  \\
0 & -1 & 0 \\
0 & 0 & 1
\end{pmatrix}, \quad U1 = \begin{pmatrix}
1 & 0 & 0  \\
0 & -1 & 0 \\
0 & 0 & -1
\end{pmatrix}.\quad \quad \eea

\section{A relation among Laplacians of $SL(2,\mathbb{Z})$ and $SL(3,\mathbb{Z})$ acting on the lattice factor}

For the $T^2$ with complex structure $\Omega$, the $SL(2,\mathbb{Z})$ invariant Laplacian is given by
\be \label{Lap2}\Delta^{SL(2,\mathbb{Z})} = 4 \Omega_2^2 \frac{\p^2}{\p\Omega \p\bar\Omega}.\ee
We now prove that
\be \label{Lapeqn}\Delta^{SL(2,\mathbb{Z})} K_L = \Big(\Delta^{SL(3,\mathbb{Z})} - \frac{4}{3} \mathcal{V}_2 \frac{\p}{\p \mathcal{V}_2} - \frac{1}{3} \mathcal{V}_2^2 \frac{\p^2}{\p \mathcal{V}_2^2}\Big) K_L\ee
where $\Delta^{SL(3,\mathbb{Z})}$ is given by \C{Lap3}, 
\be \label{newdefF2}K_L = \sum_{\hat{k}^{\alpha I}} e^{-\pi^2 \mathcal{V}_2 V_3^{2/3} \hat{G}_{IJ} \hat{G}_{\alpha\beta} \hat{k}^{\alpha I} \hat{k}^{\beta J}} \ee
is the lattice factor\footnote{This is simply \C{defF2} with the overall factor removed.}, and $\hat{G}_{IJ}$ and $\hat{G}_{\alpha\beta}$ are defined in the main text in \C{measure}.

\subsection{Constraints based on symmetries}

To start with, let us see using elementary arguments what the constraints due to the symmetries are. From the definitions of $\Delta^{SL(2,\mathbb{Z})}$ (this is also true for $\Delta^{SL(3,\mathbb{Z})}$), it trivially follows that its action on the lattice factor $K_L$ will give two kinds of terms that depend on $\mathcal{V}_2$ and $V_3$ (apart from the exponential factors): $O(\mathcal{V}_2 V_3^{2/3})$ and $O((\mathcal{V}_2 V_3^{2/3})^2)$. There are only two other independent\footnote{Terms involving derivatives of $V_3$ are not independent of the terms involving derivatives of $\mathcal{V}_2$, as these two parameters only occur in the combination $\mathcal{V}_2 V_3^{2/3}$. Derivatives with respect to $\mathcal{V}_2$ are much more convenient for our purposes. } $SL(2,\mathbb{Z}) \times SL(3,\mathbb{Z})$ invariant possibilities which have this action on the lattice factor: (i) $\mathcal{V}_2 \p /\p \mathcal{V}_2$ for $O(\mathcal{V}_2 V_3^{2/3})$ and (ii) $\mathcal{V}_2^2 \p^2/\p^2 \mathcal{V}_2^2$ for $O((\mathcal{V}_2 V_3^{2/3})^2)$. Thus on general grounds, we have that 
\be \label{needint}\Delta^{SL(2,\mathbb{Z})} K_L= \Big(\alpha_1\Delta^{SL(3,\mathbb{Z})} +\alpha_2 \mathcal{V}_2 \frac{\p}{\p \mathcal{V}_2} +\alpha_3 \mathcal{V}_2^2 \frac{\p^2}{\p \mathcal{V}_2^2}\Big) K_L,\ee       
where $\alpha_1, \alpha_2$ and $\alpha_3$ are numbers. We now constrain them using only symmetries. We restrict ourselves to all $\hat{k}^{\alpha I} \neq 0$, as this case is trivially satisfied. For brevity, let us denote $V_3^{3/2} =\lambda$. Integrating over all $\lambda$, we get that ($s >0$) 
\be \label{integrate}\int_0^\infty d \lambda \lambda^{s-1} K_L = \Gamma(s) \sum_{\hat{k}^{\alpha I}} \frac{1}{(\pi^2 \mathcal{V}_2 \hat{G}_{\alpha\beta} \hat{G}_{IJ} \hat{k}^{\alpha I} \hat{k}^{\beta J})^s}.\ee 
Now consider the large $\Omega_2$ limit of \C{integrate}. From the expressions for $\hat{G}_{\alpha\beta}$ and $\hat{G}_{IJ}$ this is given by setting $\hat{l}_1 = \hat{m}_1 = \hat{n}_1 =0$ and thus
\be \pi^2 \mathcal{V}_2 \hat{G}_{\alpha\beta} \hat{G}_{IJ} \hat{k}^{\alpha I} \hat{k}^{\beta J} \rightarrow \frac{\pi^2\mathcal{V}_2}{\Omega_2} \hat{G}_{\alpha\beta} k^\alpha k^\beta,\ee
where $k^\alpha = (\hat{l}_2, \hat{m}_2,\hat{n}_2) \neq 0$.  Thus
\be \sum_{\hat{k}^{\alpha I}} \frac{1}{(\pi^2 \mathcal{V}_2 \hat{G}_{\alpha\beta} \hat{G}_{IJ} \hat{k}^{\alpha I} \hat{k}^{\beta J})^s} \rightarrow \frac{\Omega_2^s}{(\pi^2 \mathcal{V}_2)^s}E_s^{SL(3,\mathbb{Z})},\ee
where $E_s^{SL(3,\mathbb{Z})}$ is the Eisenstein series for $SL(3,\mathbb{Z})$ defined
 by
\be E_s^{SL(3,\mathbb{Z})} = \sum_{k^\alpha \neq 0} (\hat{G}_{\alpha\beta} k^\alpha k^\beta)^{-s}\ee 
which satisfies
\be \Delta^{SL(3,\mathbb{Z})} E_s^{SL(3,\mathbb{Z})} = \frac{2s}{3} (2s-3)E_s^{SL(3,\mathbb{Z})}.\ee
Thus in the large $\Omega_2$ limit, after integrating over $\lambda$, \C{needint} reduces to
\be \label{alls}s-1 = \frac{2}{3} \alpha_1 (2s-3) - \alpha_2 +(s+1)\alpha_3.\ee
Thus 
\be  \label{constraint}\frac{4}{3} \alpha_1 +\alpha_3 =1, \quad -\frac{2}{3} \alpha_1 -\alpha_2 +2\alpha_3 =0\ee
as \C{alls} is true for all $s$\footnote{Alternatively, one could have taken the large $T_2$ limit. Then $\hat{l}_1 = \hat{l}_2 = \hat{m}_1 = \hat{m}_2 =0$, and
\be \sum_{\hat{k}^{\alpha I}} \frac{1}{(\pi^2 \mathcal{V}_2 \hat{G}_{\alpha\beta} \hat{G}_{IJ} \hat{k}^{\alpha I} \hat{k}^{\beta J})^s} \rightarrow \Big(\frac{T_2 L}{\pi^2 \mathcal{V}_2} \Big)^s \sum_{(\hat{n}_1 , \hat{n}_2 )\neq (0,0)}\frac{\Omega_2^s}{\vert \hat{n}_1 \Omega - \hat{n}_2\vert^{2s}} = \Big(\frac{T_2 L}{\pi^2 \mathcal{V}_2} \Big)^s E_s^{SL(2,\mathbb{Z})} \ee
where the $SL(2,\mathbb{Z})$ invariant Eisenstein series satisfies
\be \Delta^{SL(2,\mathbb{Z})} E_s^{SL(2,\mathbb{Z})} = s(s-1) E_s^{SL(2,\mathbb{Z})}.\ee 
Again we reproduce \C{alls} on using \C{Lap3}.}. Thus these simple arguments yields two relations among the three coefficients.   

\subsection{Solving for the coefficients directly}

Note that $\Delta^{SL(2,\mathbb{Z})} K_L$, $\Delta^{SL(3,\mathbb{Z})} K_L,$ $\mathcal{V}_2 \p K_L/\p \mathcal{V}_2$ and $\mathcal{V}_2^2 \p^2 K_L/\p \mathcal{V}_2^2$ must each produce $SL(2,\mathbb{Z}) \times SL(3,\mathbb{Z})$ invariant expressions. We define the three independent invariants
\bea Z &=& \sum_{\hat{k}^{\alpha I}}\pi^2 \mathcal{V}_2 V_3^{2/3} \hat{G}_{\alpha\beta} \hat{G}_{IJ} \hat{k}^{\alpha I} \hat{k}^{\beta J} e^{-\pi^2 \mathcal{V}_2 V_3^{2/3} \hat{G}_{IJ} \hat{G}_{\alpha\beta} \hat{k}^{\alpha I} \hat{k}^{\beta J}}, \non \\ X &=&  \sum_{\hat{k}^{\alpha I}}(\pi^2 \mathcal{V}_2 V_3^{2/3} \hat{G}_{\alpha\beta} \hat{G}_{IJ} \hat{k}^{\alpha I} \hat{k}^{\beta J})^2 e^{-\pi^2 \mathcal{V}_2 V_3^{2/3} \hat{G}_{IJ} \hat{G}_{\alpha\beta} \hat{k}^{\alpha I} \hat{k}^{\beta J}}, \non \\ Y&=&\sum_{\hat{k}^{\alpha I}}\pi^4 \mathcal{V}_2^2 V_3^{4/3} (\hat{G}_{\alpha\beta}  \hat{k}^{\alpha I} \hat{k}^{\beta K})(\hat{G}_{\gamma\delta}  \hat{k}^{\gamma J} \hat{k}^{\delta L}) \hat{G}_{IJ} \hat{G}_{KL}e^{-\pi^2 \mathcal{V}_2 V_3^{2/3} \hat{G}_{IJ} \hat{G}_{\alpha\beta} \hat{k}^{\alpha I} \hat{k}^{\beta J}}.\eea
We find that
\bea &&\Delta^{SL(2,\mathbb{Z})} K_L = -2 Z + 2Y-X, \quad \Delta^{SL(3,\mathbb{Z})} K_L = -\frac{10}{3} Z - \frac{2}{3} X+ 2Y ,\non \\ &&\mathcal{V}_2 \frac{\p K_L}{\p \mathcal{V}_2} = -Z, \quad \mathcal{V}_2^2 \frac{\p^2 K_L}{\p \mathcal{V}_2^2} =  X.\eea
These calculations are considerably simplified using the $SL(2,\mathbb{Z}) \times SL(3,\mathbb{Z})$ invariance to perform the calculations at $L=1, A_1 = A_2 =0$ and then covariantizing. 

Thus substituting in \C{needint} we get that
\be \frac{10}{3} \alpha_1 +\alpha_2 = 2, \quad \frac{2}{3} \alpha_1 - \alpha_3 = 1, \quad \alpha_1 =1 .\ee
Hence $\alpha_1 =1, \alpha_2 = -4/3, \alpha_3 = -1/3$ proving \C{Lapeqn}. Note that \C{constraint} is automatically satisfied. The relation \C{Lapeqn} is crucial in regulating the ultraviolet divergences that arise from the boundary of moduli space.




\providecommand{\href}[2]{#2}\begingroup\raggedright\endgroup

\end{document}